\documentclass[twocolumn]{aastex}
\usepackage{amsmath}
\usepackage{amssymb}
\usepackage{appendix}
\usepackage{subcaption}
\usepackage{comment}

\providecommand{\edit}[1]{#1}
\usepackage{tabularx}
\usepackage{booktabs}
\usepackage{threeparttable}

\begin{document}

\author[0000-0003-4459-9054]{Yoav Rotman}
\affiliation{Space Science Institute, Lawrence Livermore National Laboratory, 7000 East Ave., Livermore, CA 94550, USA}
\affiliation{School of Earth and Space Exploration, Arizona State University, Tempe, AZ, USA}
\email{yrotman@asu.edu}

\author[0000-0002-1052-6749]{Peter McGill}
\affiliation{Space Science Institute, Lawrence Livermore National Laboratory, 7000 East Ave., Livermore, CA 94550, USA}
\email{X}

\author[0000-0003-0156-4564]{Luis Welbanks}
\affiliation{School of Earth and Space Exploration, Arizona State University, Tempe, AZ, USA}
\email{X}

\author[0000-0002-3627-1676]{Benjamin V.\ Rackham}
\affiliation{Department of Earth, Atmospheric and Planetary Sciences, Massachusetts Institute of Technology, Cambridge, MA 02139, USA}
\affiliation{Kavli Institute for Astrophysics and Space Research, Massachusetts Institute of Technology, Cambridge, MA 02139, USA}
\email{}

\author[0000-0003-0971-1709]{Aishwarya Iyer}
\altaffiliation{NASA Postdoctoral Fellow}
\affiliation{NASA Goddard Space Flight Center, 8800 Greenbelt Road, Greenbelt,
MD 20771, USA}
\email{}

\author[0000-0003-3714-5855]{D\'aniel Apai}
\affiliation{Steward Observatory, The University of Arizona, Tucson, AZ 85721, USA}
\affiliation{Lunar and Planetary Laboratory, The University of Arizona, Tucson, AZ 85721, USA}
\email{}

\author[0000-0001-6247-8323]{Michael R. Line}
\affiliation{School of Earth and Space Exploration, Arizona State University, Tempe, AZ, USA}
\email{}

\author[0000-0003-1309-2904]{Elisa V. Quintana}
\affiliation{NASA Goddard Space Flight Center, 8800 Greenbelt Road, Greenbelt, MD 20771, USA}
\email{}

\author[orcid=0000-0003-4206-5649]{Jessie L. Dotson}	
\affiliation{NASA Ames Research Center, MS 245-6, Moffett Field, CA 94035, USA}
\email{}

\author[0000-0001-8020-7121]{Knicole D. Col\'{o}n}
\affiliation{NASA Goddard Space Flight Center, 8800 Greenbelt Road, Greenbelt, MD 20771, USA}
\email{}

\author[0000-0001-7139-2724]{Thomas Barclay}
\affiliation{NASA Goddard Space Flight Center, 8800 Greenbelt Road, Greenbelt, MD 20771, USA}
\email{}

\author[0000-0002-3385-8391]{Christina Hedges} \affiliation{NASA Goddard Space Flight Center, 8800 Greenbelt Road, Greenbelt, MD 20771, USA} \affiliation{University of Maryland, Baltimore County, 1000 Hilltop Circle, Baltimore, Maryland, United States} 
\email{}

\author[0000-0002-5904-1865]{Jason F. Rowe}	\affiliation{Department of Physics and Astronomy, Bishops University, 2600 Rue College, Sherbrooke, QC J1M 1Z7, Canada}
\email{}

\author[0000-0002-0388-8004]{Emily A. Gilbert}
\affiliation{Jet Propulsion Laboratory, California Institute of Technology, 4800 Oak Grove Drive, Pasadena, CA 91109, USA}
\email{}

\author[0000-0003-2528-3409]{Brett M. Morris}
\affiliation{Space Telescope Science Institute, 3700 San Martin Drive, Baltimore, MD 21218, USA}
\email{}

\author[0000-0002-8035-4778]{Jessie L. Christiansen}	
\affiliation{California Institute of Technology/IPAC, 1200 California Blvd, MC 100-22. Pasadena, CA 91125, USA}
\email{}

\author[0000-0002-6276-1361]{Trevor O. Foote} \altaffiliation{NASA Postdoctoral Fellow} \affiliation{NASA Goddard Space Flight Center, 8800 Greenbelt Road, Greenbelt, MD 20771, USA}

\email{}

\author[0000-0001-9828-3229]{Aylin Garc{\'i}a Soto}
\affiliation{Department of Physics and Astronomy, Dartmouth College, Hanover NH 03755, USA}
\email{}

\author[0000-0002-8963-8056]{Thomas P. Greene}
\affiliation{California Institute of Technology/IPAC, 1200 California Blvd, MC 100-22. Pasadena, CA 91125, USA}
\email{}

\author[0000-0001-6541-0754]{Kelsey Hoffman}	
\affiliation{Department of Physics and Astronomy, Bishops University, 2600 Rue College, Sherbrooke, QC J1M 1Z7, Canada}
\email{}

\author[0000-0001-5084-4269]{Benjamin J. Hord}
\altaffiliation{NASA Postdoctoral Fellow}
\affiliation{NASA Goddard Space Flight Center, 8800 Greenbelt Road, Greenbelt, MD 20771, USA}
\email{}

\author[0000-0002-3239-5989]{Aurora Y. Kesseli}	
\affiliation{California Institute of Technology/IPAC, 1200 California Blvd, MC 100-22. Pasadena, CA 91125, USA}
\email{}

\author[0000-0001-9786-1031]{Veselin B. Kostov}	
\affiliation{NASA Goddard Space Flight Center, 8800 Greenbelt Road, Greenbelt, MD 20771, USA}
\affiliation{SETI Institute, 189 Bernardo Ave, Suite 200, Mountain View, CA 94043, USA}
\email{}

\author[0000-0003-4241-7413]{Megan Weiner Mansfield}
\affiliation{Department of Astronomy, University of Maryland, 4296 Stadium Drive, College Park, MD 20742, USA}
\email{}

\author[0000-0002-3295-1279]{Lindsey S. Wiser}
\affiliation{Johns Hopkins University Applied Physics Laboratory, 11100 Johns Hopkins Rd, Laurel, MD 20723, USA}
\email{}

\title{NASA's \textit{Pandora SmallSat Mission}: Simulated Modeling and Retrieval of Near-Infrared Exoplanet Transmission Spectra}

\begin{abstract}
Pandora is a SmallSat mission dedicated to understanding exoplanets and their host stars by disentangling the impact of stellar heterogeneity on exoplanet transmission spectra. Selected as a NASA Astrophysics Pioneers mission in 2021, Pandora will provide simultaneous long-term visible photometric monitoring (0.4--0.7 $\mu$m) and low-resolution near-infrared (NIR) spectroscopy (0.9--1.6 $\mu$m) of transiting systems for the purposes of monitoring host star variability and characterizing exoplanetary atmospheres. \edit{Pandora's year-long prime mission from 2026 to 2027 coincides with the middle of a decade defined by targeted efforts for atmospheric characterization of exoplanets, offering a key opportunity to leverage this new resource to maximize science with JWST and other observatories. Here we investigate Pandora's anticipated performance for the general exoplanet population accessible to transit spectroscopy, from hot Jupiters to temperate sub-Neptunes. By modeling the atmospheres of five test cases broadly consistent with the bulk properties of HD~209458~b, HD~189733~b, WASP-80~b, HAT-P-18~b, and K2-18~b, we find that Pandora may provide abundance constraints as precise as $\sim$1.0\,dex for main atmospheric absorbers such as H$_2$O and CH$_4$. Then, we explore the synergies between Pandora and JWST. Our results suggest that targets with JWST data in the near-infrared can benefit from the addition of Pandora observations and result in more reliable abundance estimates than with JWST data alone. Moreover, Pandora can serve the community by providing precursory observations of targets of interest for JWST atmospheric characterization. We conclude by outlining strategies for the use of Pandora as a standalone observatory and in synergy with JWST.} \\

\end{abstract}


\section{Introduction} 
The compositions of exoplanets offer unique insights into their formation and evolution, making compositional inferences a key scientific goal over the last two decades. These constraints typically come from transmission spectroscopy: as a planet transits its host star, the absorption spectrum of its atmosphere is imprinted into the observed stellar spectrum \citep{seager_theoretical_2000}. Extracting the atmospheric spectrum then allows us to make quantitative inferences about its temperature structure, chemical inventory, and aerosols. These inferences, in turn, inform us about the planet's formation and evolutionary history \citep[e.g.,][]{oberg_effects_2011, madhusudhan_co_2012, mordasini_imprint_2016}. 

The \textit{James Webb Space Telescope} \citep[JWST;][]{gardner_james_2006} has revolutionized the field of transmission spectroscopy, providing spectra of exoplanet atmospheres with wider wavelength coverage ($\sim$0.6--12.5 $\mu$m) and a higher resolution than any previous space-based observatory. JWST observations have enabled new insights into exoplanet atmospheres \citep[see, e.g.,][]{espinoza_highlights_2025}, including the first detections of carbon-bearing species \citep[e.g.,][]{jwst_transiting_exoplanet_community_early_release_science_team_identification_2022, bell_methane_2023} and photochemical processes \citep[e.g.,][]{tsai_photochemically_2023}. JWST spectra have also ushered in higher-precision constraints on the atmospheric chemical inventory \citep[e.g.,][]{welbanks_high_2024, beatty_sulfur_2024} as well as clouds and hazes \citep[e.g.,][]{dyrek_so2_2024, inglis_quartz_2024}. 

JWST's increased precision, however, has also highlighted the effects of stellar contamination on the planetary spectrum \citep[e.g.,][]{fu_water_2022, moran_high_2023, fournier-tondreau_near-infrared_2024, fournier-tondreau_transmission_2024, canas_gems_2025}, where heterogeneity in the stellar surface imprints stellar spectral features into the observed transmission spectrum through the transit light source (TLS) effect \citep{rackham_transit_2018, rackham_transit_2019}. While occulted starspots can be identified and corrected in the transit light curve \citep[e.g.,][]{pont_prevalence_2013}, unocculted spot modulation is often harder to definitively identify in the host star. If uncorrected, this effect can cause significant biases in analyses of transmission spectra \citep{pinhas_retrieval_2018, iyer_influence_2020, rackham_sag21_2023, rackham_toward_2024}. This issue is particularly pronounced for observations of small, rocky planets around K and M dwarfs. While these low-mass host stars are the most accessible to transit spectroscopy of small planets \citep{lim_atmospheric_2023, radica_promise_2025}, they are known to have higher levels of magnetic activity \citep[e.g.,][]{muirhead_catalog_2018}, leading to more heterogeneities on the stellar surface. 
Although stellar contamination of transmission spectra is not unique to JWST observations \citep[e.g.,][]{pont_detection_2008, pont_prevalence_2013, mccullough_water_2014, rackham_gj1214b_2017, zhang_trappist-1_2018}, it has become the major limitation for interpreting JWST transmission spectra, and often prevents confident identification of exoplanetary atmospheres \citep[e.g.,][]{moran_high_2023}.  

Pandora is a SmallSat Mission selected by NASA in 2021 as part of the Astrophysics Pioneers Program \citep{quintana_pandora_2021, quintana_pandora_2024, barclay_pandora_2025}, and \edit{launched in January} 2026. It is designed to address the challenge of the stellar contamination by combining near-IR spectroscopy ($\sim$0.9--1.6\,$\mu$m; $R\gtrsim30$) with simultaneous visible-light photometry ($\sim$0.4--0.7 $\mu$m) for twenty transiting exoplanet systems \citep{foote_pandora_2022}. Pandora has a 0.45 meter diameter primary mirror, and allows both wavelength channels to operate simultaneously. The bandpass of Pandora's Near-Infrared Detector Assembly (NIRDA) instrument covers multiple absorption bands of H$_2$O, which is ubiquitous in exoplanet atmospheres, but can also trace star spots  \citep{rackham_transit_2018, rackham_transit_2019}. By combining time-resolved, long-baseline optical photometry and NIR spectroscopy, Pandora will enable the disentanglement of stellar variability from planetary spectra, allowing for accurate inferences from both the planets and their host stars \citep{hoffman_pandora_2022}. Pandora is expected to observe $\sim\!20$ primary mission targets, and capture a baseline of at least ten transits for each target, throughout its mission lifetime. 

In this work, we assess Pandora's capabilities for atmospheric characterization and examine how observations from Pandora and JWST combined can improve the accuracy and precision of atmospheric inferences. We simulate synthetic spectra of five benchmark exoplanets and perform atmospheric retrievals using both observatories to analyze the expected accuracy and precision for each target and instrument combination. In Section \ref{sec:simulating}, we describe our target selection, synthetic data simulation, and retrieval methodology. In Section \ref{sec:pandora}, we describe the observational capabilities of Pandora. In Section \ref{sec:jwst}, we analyze possible uses for joint observations between JWST and Pandora, and which instruments and targets benefit most from synergistic observations. Finally, in Section \ref{sec:summary}, we summarize our findings and recommend future considerations when observing exoplanets with both observatories.








\begin{table*}[t]
\small
\centering
\begin{threeparttable}

\begin{tabular*}{\textwidth}{@{\extracolsep{\fill}} l c c c c c c @{}}
\toprule
Planet & $R_p$ [$R_{\rm J}$] & $M_p$ [$M_{\rm J}$] & $T_{\rm eq}$ [K] & $R_\star$ [$R_\odot$] & $M_\star$ [$M_\odot$] & $T_{\rm eff}$ [K] \\
\midrule

HD~209458~b\tnote{a} & 1.380\,$\pm$\,0.015 & 0.714\,$\pm$\,0.014 & 1459\,$\pm$\,12 & 1.162\,$\pm$\,0.012 & 1.148\,$\pm$\,0.040 & 6117\,$\pm$\,50 \\ \addlinespace

HD~189733~b\tnote{b} & 1.142$_{-0.034}^{+0.036}$ & 1.130$_{-0.045}^{+0.047}$ & 1220\,$\pm$\,13 & 0.777\,$\pm$\,0.001 & 0.840$^{+0.099}_{-0.112}$ & 5053$^{+46}_{-45}$ \\ \addlinespace

WASP-80~b\tnote{c}   & 0.999$_{-0.031}^{+0.030}$ & 0.538$_{-0.036}^{+0.035}$ & 825\,$\pm$\,19 & 0.586$_{-0.018}^{+0.017}$ & 0.577$^{+0.051}_{-0.054}$ & 4143$_{-94}^{+92}$ \\ \addlinespace

HAT-P-18~b\tnote{d}  & 0.995\,$\pm$\,0.052 & 0.197\,$\pm$\,0.013 & 852\,$\pm$\,28 & 0.749\,$\pm$\,0.037 & 0.770\,$\pm$\,0.031 & 4803\,$\pm$\,80 \\ \addlinespace

K2-18~b\tnote{e,f}    & 0.232\,$\pm$\,0.008 & 0.027\,$\pm$\,0.004 & 255\,$\pm$\,4 & 0.445\,$\pm$\,0.015 & 0.495\,$\pm$\,0.004 & 3457\,$\pm$\,39 \\

\bottomrule
\end{tabular*}
\caption{System parameters for our five selected targets. We adopt the median of the reported system parameters for our models. \textbf{References:} a) \citet{southworth_homogeneous_2010}; b) \citet{addison_minerva-australis_2019}; c) \citet{triaud_wasp-80b_2015}; d) \citet{hartman_hat-p-18b_2010}; e) \citet{benneke_water_2019}; f) \citet{cloutier_confirmation_2019}.} \label{table:sys_params}
\end{threeparttable}
\end{table*}

\section{Simulating and Retrieving Transmission Spectra} \label{sec:simulating}

We simulate synthetic transmission spectra for five exoplanet targets with the Aurora framework \citep{welbanks_aurora_2021}, modeling observations \edit{for} the Pandora NIR detector \edit{and} \edit{the Near-Infrared Camera (NIRCam) instrument on JWST. We select NIRCam due to its complimentary wavelength range; while Pandora covers the 0.9--1.6 $\mu$m band, NIRCam's long-wavelength channel covers 2.4--5.1 $\mu$m, allowing for a wider wavelength coverage encompassing more absorption features}. We then use the Bayesian inference methodology known as ``atmospheric retrieval'' \citep[e.g.,][]{madhusudhan_temperature_2009} to assess inferences from Pandora alone and in combination with JWST. Throughout this work, we assume that the observed spectra have already been corrected for stellar contamination, so that the data represents a planetary spectrum in the case of an ideal star. Companion studies will examine Pandora's ability to constrain stellar photospheric heterogeneity and mitigate the TLS effect \edit{\citep[][]{Rackham2026}}. Here, we describe our target selection, forward models, and retrieval methodologies.

\subsection{Targets}\label{sec:targets}

\edit{Our study aims to explore Pandora's performance for the general exoplanet population currently accessible for atmospheric characterization with transit spectroscopy. We broadly model this population by simulating the atmospheres of five targets spanning the hot Jupiter to temperate sub-Neptune regime. The cases selected broadly follow the bulk properties of HD~209458~b, HD~189733~b, WASP-80~b, HAT-P-18~b, and K2-18~b (i.e., mass, radius, equilibrium temperature, and host star, see Table \ref{table:sys_params})\footnote[1]{We note that both WASP-80~b and HAT-P-18~b are among the Pandora primary mission targets, see \url{https://pandorasat.com/targets/}}. The atmospheric compositions adopted in Section \ref{sec:pandora} generally follow existing constraints from the literature as described below, whereas those in Section \ref{sec:jwst} are motivated by chemical equilibrium expectations for an enhanced metallicity atmosphere. Below we describe the context of each planet archetype adopted in our study.}


\subsubsection{HD~209458~b}
HD~209458~b is a hot Jupiter orbiting a G0V star. It holds a unique spot as the first discovered transiting exoplanet and the first to have its atmosphere characterized \citep{charbonneau_detection_2000, charbonneau_detection_2002, vidal-madjar_extended_2003, vidal-madjar_detection_2004}. It is a uniquely well-suited target for transmission spectroscopy studies, with HST \citep[e.g.,][]{deming_infrared_2013, sing_continuum_2016}, \textit{Spitzer} \citep[e.g.,][]{evans_uniform_2015}, JWST \citep[e.g.,][]{xue_jwst_2024}, and ground-based observations \citep[e.g.,][]{giacobbe_five_2021}. 
The atmosphere of HD~209458~b shows signs of water \citep{deming_infrared_2013}, clouds and hazes \citep{sing_continuum_2016}, and more recently, evidence of carbon and sulfur bearing molecules  \citep{giacobbe_five_2021, xue_jwst_2024}.

\subsubsection{HD~189733~b}
HD~189733~b is a hot Jupiter orbiting a K2 dwarf \citep{bouchy_elodie_2005}. Like HD~209458~b, it has been extensively studied, with observations from both ground- \citep[e.g.,][]{redfield_sodium_2008, brogi_retrieving_2019} and space-based observatories \citep[e.g.,][]{ charbonneau_broadband_2008, agol_climate_2010}. Recent JWST observations of the transmission spectrum have identified H$_2$O, CO, CO$_2$, H$_2$S, and SiO$_2$ \citep{fu_hydrogen_2024, inglis_quartz_2024}, while previous HST observations showed signs of H$_2$O, haze-induced scattering, and unocculted starspots \citep[e.g.,][]{pont_detection_2008, pont_prevalence_2013, lecavelier_des_etangs_rayleigh_2008,sing_transit_2009}. \edit{Moreover,} previous \edit{studies} have \edit{suggested} strong stellar activity from HD 189733 \edit{\citep[e.g.,][]{miller-ricci_most_2008, pont_detection_2008, pont_prevalence_2013, mccullough_water_2014, narrett_axisymmetric_2024}}, making it a strong candidate for Pandora observations. 

\subsubsection{WASP-80~b}
WASP-80~b is a warm Jupiter orbiting a low-mass late K dwarf  \citep{triaud_wasp-80b_2013}. It straddles the lower edge of the warm Jupiter temperature range ($T_{\rm eq} =825$ K), making it a unique testbed for atmospheric chemistry in cooler Jovians. 
WASP-80~b has been observed with HST \citep[e.g.,][]{wong_hubble_2022, jacobs_probing_2023} and JWST \citep[e.g.,][]{bell_methane_2023, wiser_precise_2025}, as well as various ground-based observations \citep[e.g.,][]{fukui_multi-band_2014, mancini_physical_2014, carleo_gaps_2022}, leading to to detections of H$_2$O and carbon-bearing species, 
alongside possible aerosol contributions from clouds.

\subsubsection{HAT-P-18~b}
HAT-P-18~b is a low-density warm Saturn orbiting a K2 dwarf. Like WASP-80~b, the low temperature of HAT-P-18~b ($T_{\rm eq} = 852$ K) makes it an enticing target for understanding giant planet chemistry. Since its original discovery \citep{hartman_hat-p-18b_2010}, HAT-P-18~b has been observed by both ground-based \citep[e.g.][]{kirk_rayleigh_2017} and space-based observatories \citep[e.g.,][]{tsiaras_population_2018}, leading to the detection of a Rayleigh scattering slope, water absorption, and possible cloud and haze contributions to the transmission spectrum. More recently, HAT-P-18~b was an Early Release Observation (ERO) target for JWST, with NIRISS observations leading to detections of H$_2$O, CO$_2$, He, and a high-altitude cloud deck, with \edit{signatures} of Na \citep{fu_water_2022, fournier-tondreau_near-infrared_2024}. 

\subsubsection{K2-18~b}
K2-18~b is a temperate sub-Neptune orbiting a K2 dwarf \citep{montet_stellar_2015}. K2-18~b has been observed with JWST across the near- to mid-IR, with NIRSpec, NIRISS, and MIRI observations. HST/WFC3 observations yielded an H$_2$O detection with negligible CH$_4$ \citep{benneke_water_2019, tsiaras_water_2019, welbanks_mass-metallicity_2019, madhusudhan_interior_2020}, though follow-up observations with JWST detected CH$_4$ and CO$_2$, with only upper limit constraints on H$_2$O, and a strong super-Rayleigh slope \citep{madhusudhan_carbon-bearing_2023, hu_water-rich_2025}. 


\subsection{Atmospheric Model}\label{sec:fwd_model}

\begin{figure*}[t]
    \centering
    \includegraphics[width=\linewidth]{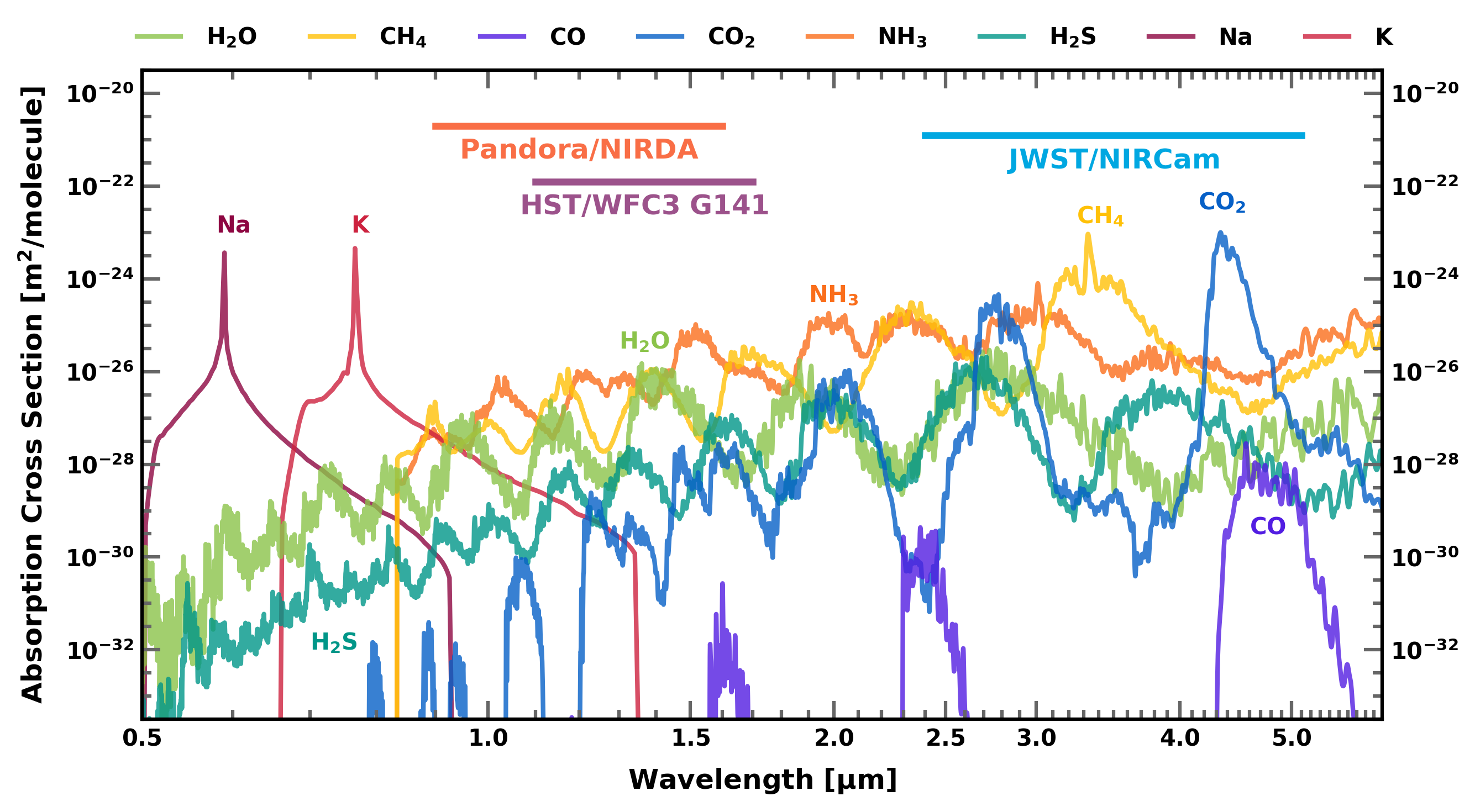}
    \caption{The absorption cross sections of common absorbers in exoplanet atmospheres considered in this work, shown at a pressure and temperature of 0.1 mbar and 1000 K. The wavelength coverage of Pandora's NIR detector (NIRDA), JWST's \edit{NIRCam F322W2 and F444W filters}, and HST's WFC3 instrument (for the G141 grism) are shown. Pandora/NIRDA covers absorption bands of H$_2$O, CH$_4$, NH$_3$ and the wing of the K doublet, making it most sensitive to these absorbers.} 
    \label{fig:wavelength_coverage}
\end{figure*}

We use Aurora \citep{welbanks_aurora_2021}, an exoplanet atmospheric spectra forward modeling and retrieval framework, to simulate the atmospheres of each target. Aurora solves radiative transfer assuming hydrostatic equilibrium and a plane-parallel one-dimensional atmosphere to produce the transmission spectrum of the planet. 
We assume an isothermal atmosphere at the equilibrium temperature of each planet in our models. \edit{The synthetic observations are the result of a planetary atmospheric model only; i.e., there is no spectroscopic contribution due to a heterogeneous stellar photosphere and the TLS effect.} 

\edit{In Section \ref{sec:pandora}, we present an investigation into the constraining power of Pandora's NIRDA detector. We build each planetary model based broadly on inferences from previous JWST observations and expectations from chemical equilibrium, including absorption contributions from H$_2$O, CH$_4$, CO, CO$_2$, H$_2$S, NH$_3$, Na, and K. Additionally, we account for inhomogeneous cloud coverage along the terminator using a linear model combination of cloudy and cloud-free models \citep{line_influence_2016}. We treat cloud coverage as an opaque gray slab at pressure $P_{\rm cloud}$ \citep[e.g.,][]{pinhas_retrieval_2018}, and treat any haze scattering slope as an enhancement to H$_2$ Rayleigh scattering \citep[e.g.,][]{lecavelier_des_etangs_rayleigh_2008}; this includes two parameters representing the scattering slope $\gamma$ (where $\gamma=-4$ for pure Rayleigh scattering) and a multiplicative enhancement term, $a$.}

\edit{In Section \ref{sec:jwst}, we compare observations with Pandora to those done in tandem with JWST/NIRCam. We model our selected planets, adopting the bulk parameters in Table \ref{table:sys_params} and assume an atmosphere with H$_2$O, CH$_4$, CO, and CO$_2$; we select abundances consistent with a solar C/O ratio and 10$\times$ solar metallicity at the equilibrium temperature of each of our targets \citep{moses2013}, and limit our absorbers to carbon- and oxygen-bearing species. Though these abundances may differ from the literature values, we choose these to homogenize our sample. This allows for considerations of future target selection and refinement, particularly for targets without current JWST observations. Additionally, we assume a cloud-free atmosphere with Rayleigh scattering for each target, as we are interested in the combined ability of Pandora and JWST to provide more accurate constraints on the atmospheric C and O abundances.}

\edit{We adopt} line-lists for H$_2$O, CO, CO$_2$ \citep{rothman_hitemp_2010}, CH$_4$ \citep{yurchenko_exomol_2014}, NH$_3$ \citep{yurchenko_variationally_2011}, H$_2$S, SO$_2$ \citep{underwood_exomol_2016}, Na \citep{allard_new_2019}, and K \citep{allard_k-h2_2016}, as well as collision-induced absorption from H$_2$-H$_2$ and H$_2$-He scattering \citep{richard_new_2012}. The absorption cross sections, as well as the bandpasses of Pandora, HST/WFC3/G141, and \edit{the JWST/NIRCam long-wavelength filters} are shown in Figure \ref{fig:wavelength_coverage}.

We use Aurora to perform our atmospheric retrievals on the data. Retrievals combine Bayesian inference sampling \citep[via MultiNest;][]{feroz_multinest_2009, buchner_x-ray_2014} and forward models to quantitatively place constraints on atmospheric parameters from exoplanetary spectra. By identifying the region of parameter space that maximizes the Bayesian likelihood, retrievals provide posterior probability distributions for the model parameters that govern the transmission spectrum, giving insight into the thermochemical state of the atmosphere. For a review of atmospheric retrieval techniques and models, see, e.g., \citet{madhusudhan_atmospheric_2018}. \edit{Here, we use 500 live points for our sampling, with a 100-layer atmospheric pressure grid and a spectral resolution of 10,000.}  A full table of the assumed absorbers, atmospheric temperatures, and cloud and haze profiles, as well as the Bayesian priors used throughout this work, are given in Tables \ref{table:params_priors} \edit{(for Section \ref{sec:pandora}) and \ref{table:params_priors2} (for Section \ref{sec:jwst})} in Appendix \ref{appendix:priors_and_model}. 

\subsection{Synthetic Pandora and JWST Observations}\label{sec:noise_sim}

For each target, we simulate observations of the transmission spectrum with Pandora's NIR detector and \edit{JWST/NIRCam (for both the F322W2 and F444W filters)}. For planets with published JWST NIRCam observations (HD~209458~b, HD~189733~b, and WASP-80~b), we adopt the published bins, bin widths, and uncertainties of the JWST data. For those without existing observations (HAT-P-18~b, K2-18~b, and F444W observations of WASP-80~b), we use the \texttt{PandExo} precision calculator \citep{batalha_pandexo_2017}, and bin the observations to $R=200$. We simulate astrophysical and instrument noise in each observation by normally scattering each \edit{synthetic} data point. 

We simulate the estimated uncertainty for Pandora's NIR observations following \citet{greene_characterizing_2016}. 
To calculate the signal-to-noise ratio (SNR) in each spectral bin, we use the \texttt{pandora-sat}\footnote[2]{Package publicly available at \url{https://github.com/PandoraMission/pandora-sat/}} Python package, developed by the Pandora team to simulate Pandora's performance \citep{hedges_open-source_2024}. We first calculate the observed SED of a given host star, modeled using an interpolated \texttt{PHOENIX} \citep{husser_new_2013} model and the NIRDA instrument sensitivity and resolution (average $R\approx110$). We then integrate the product of the stellar SED and the instrument sensitivity along each wavelength bin to calculate the observed flux per bin. From this, we calculate the photon shot noise of the observation, as well as the detector read and dark noise (from \texttt{pandora-sat}), to estimate the SNR of each spectral bin, which we convert to an associated uncertainty. 

For each target, we model a total observation time of twice the ingress to egress time ($T_{14}$) of the planet. This corresponds to a full transit observation and an equal out-of-transit baseline observation of the star. Given Pandora's low Earth orbit, it will be difficult to obtain uninterrupted observations of the entire transit (similar to observations with HST). We therefore assume a 50\% duty cycle, approximating that \edit{an average target will be inaccessible} for roughly half of the observation time. We note that Pandora will observe each primary mission target for a 24 hour visit, which will provide a longer out-of-transit baseline and likely lead to better \edit{SNR}. We assume a wavelength-independent 5-pixel width for the PSF \edit{and} do not consider possible effects from spacecraft jitter or noise from signals of nearby background stars, which can increase the uncertainty and introduce correlated noise \citep[e.g.,][]{sing_hubble_2011}. 

\begin{figure}[t]
    \centering
    \includegraphics[width=\linewidth]{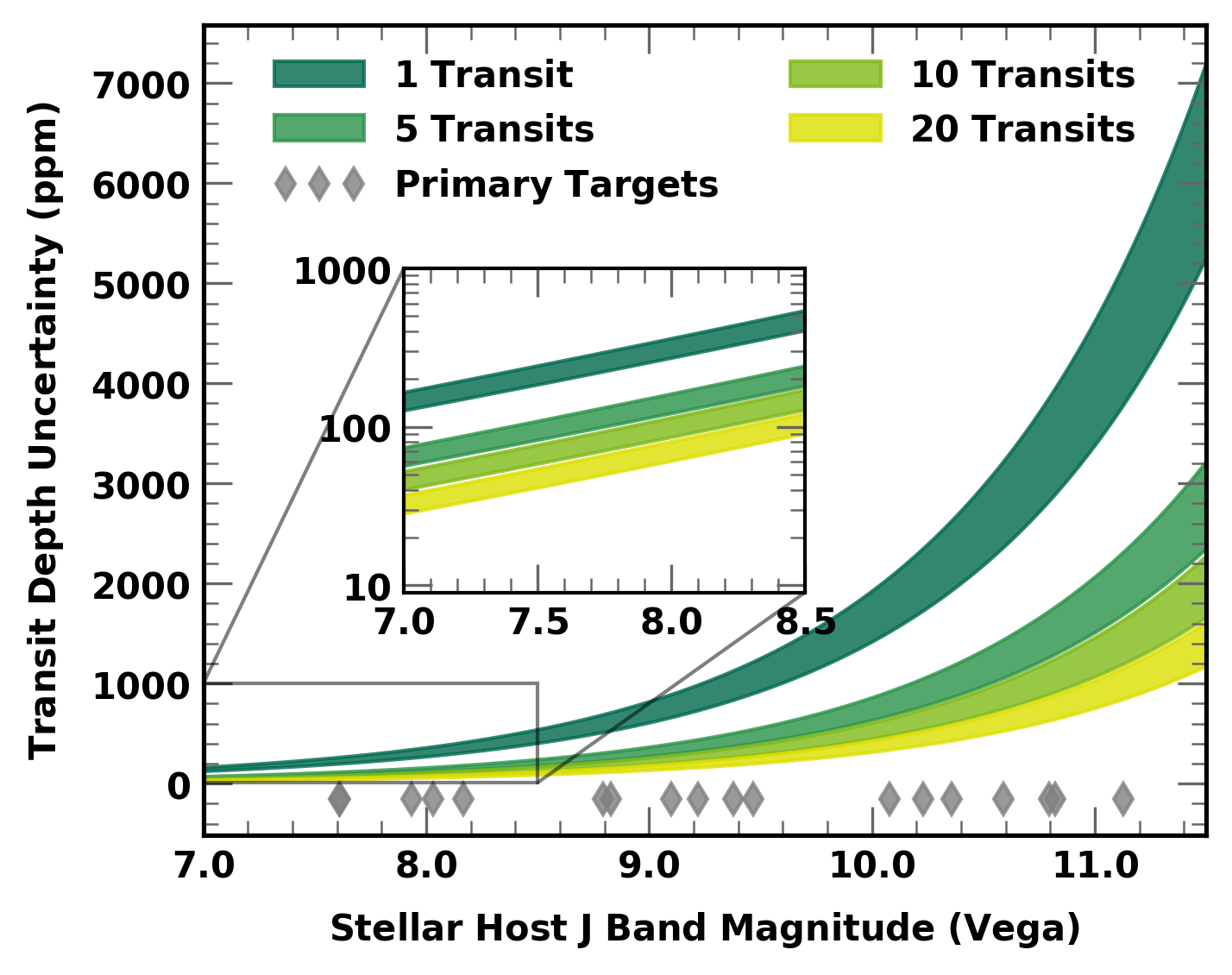}
    \caption{The range of uncertainties predicted by our calculations for Pandora observations of a warm Jupiter-like system (here, WASP-80~b) with varying stellar magnitude, after binning to $R\simeq30$. Different stellar fluxes and detector properties at different wavelengths will lead to different uncertainties throughout the bandpass; the bands here represent the range between the lowest and highest uncertainties. Stacking multiple observations will provide lower uncertainties, particularly for fainter targets. The inset shows the uncertainties for brighter stars, for which Pandora can achieve precisions as low as $\sim$30 ppm when multiple transits are observed. The magnitudes of stars included in the Pandora primary target list are shown as diamonds.}
    \label{fig:errorbars}
\end{figure}

We bin our synthetic Pandora observations to $R\approx30$ \citep[following][]{hoffman_pandora_2022}. For multiple observations, we assume that the uncertainties can be defined as decreasing by a factor of $\sqrt{N_{\rm trans}}$, representing the increase in SNR with multiple observations. This represents the optimistic scenario, in which our uncertainties are dominated by photon noise. In Figure \ref{fig:errorbars}, we show the estimated range of uncertainties for one, five, ten, and twenty stacked Pandora observations, each with a five hour total integration time, for a warm Jupiter-like system with varying J band stellar magnitude. 

\section{Pandora's Observational Capabilities}\label{sec:pandora}

Pandora's NIRDA instrument covers a similar wavelength range ($\sim$0.9--1.6\,$\mu$m) to the G141 grism of HST/WFC3 ($\sim$1.1--1.7\,$\mu$m); the latter has detected H$_2$O absorption in exoplanet atmospheres \citep[e.g.,][]{deming_infrared_2013, wakeford_hat-p-26b_2017, wakeford_complete_2018}, alongside terminator clouds and hazes \citep[e.g.,][]{pont_prevalence_2013,sing_continuum_2016}. 
With multiple dedicated observations per target, Pandora is expected to provide constraints similar to those of HST. Here we explore Pandora's observational capabilities with NIRDA, in order to identify the expected scientific output and the number of observations needed for atmospheric characterization.

\subsection{Constraining Atmospheric Composition of Primary Mission Targets}\label{sec:10transits}

Pandora is expected to observe ten transits as a preliminary baseline for each of its primary mission targets  \citep{hoffman_pandora_2022}. \edit{We generate} synthetic Pandora observations with uncertainties reflecting the expected precision from ten transit observations \edit{following the description in Section \ref{sec:fwd_model}}. As our targets represent the region of planetary parameter space probed by Pandora's primary mission targets, we aim to identify which absorbers and atmospheric properties will be best constrained by Pandora observations, and whether ten transits will be sufficient for primary mission targets.

\begin{figure}[t]
    \centering
    \includegraphics[width=\linewidth]{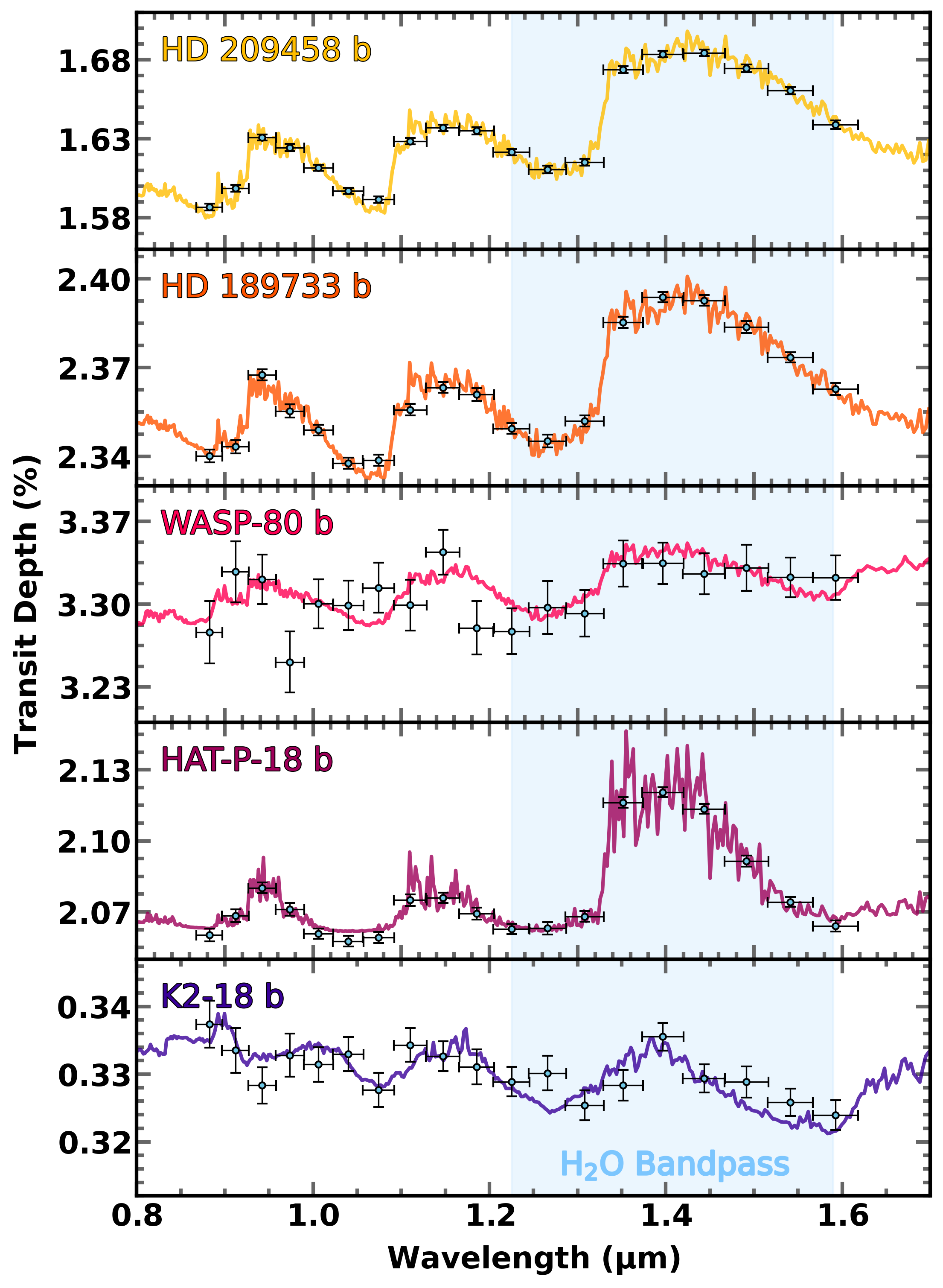}
    \caption{Synthetic forward models and observations with Pandora's NIRDA instrument of the five targets. The data is binned to a resolution of $R\simeq30$, and error bars are simulated for ten transit observations. The 1.4 $\mu$m water absorption band is highlighted \edit{for reference}, though K2-18~b shows a similar feature from CH$_4$ absorption.}
    \label{fig:simdata}
\end{figure}

For each target in Section \ref{sec:targets}, we generate a model of the atmospheric spectrum based on previous observations \edit{and Pandora's instrumental properties}. The full list of parameters used in generating the forward models is given in Table \ref{table:params_priors} in Appendix \ref{appendix:priors_and_model}. The forward models and synthetic data for ten transit observations \edit{with Pandora} are shown in Figure \ref{fig:simdata}. 

We analyze two forms of each forward model; a ``simple" cloud- and haze-free model comprised of absorbers in the Pandora bandpass (H$_2$O, CH$_4$, and NH$_3$) and a ``comprehensive" model that includes partial clouds, hazes, and species that do not have strong absorption lines in the Pandora bandpass (e.g., CO and CO$_2$). A full list of absorbers in the comprehensive model is provided in Table \ref{table:params_priors} in Appendix \ref{appendix:priors_and_model}.

\subsubsection{Molecular Abundances}

\begin{figure*}[]
    \centering
    \includegraphics[width=\linewidth]{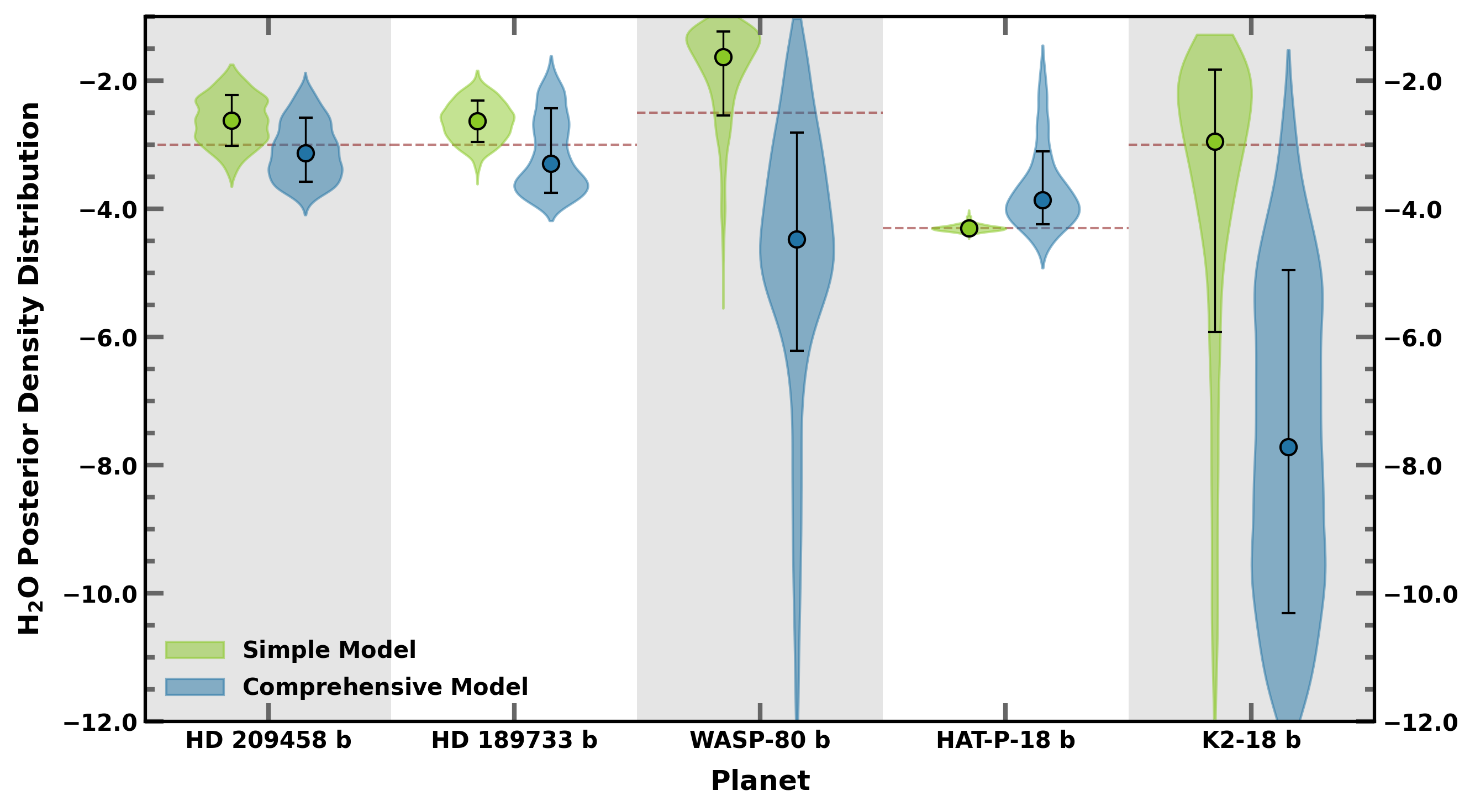}
    \caption{\edit{H$_2$O} posterior distributions for our five targets. The green distributions show the retrieved posteriors from the simple model, and the blue represents the \edit{comprehensive} model, which includes clouds and hazes (both described in Section \ref{sec:fwd_model}). Maroon lines indicate the injected true values from Table \ref{table:params_priors} in Appendix \ref{appendix:priors_and_model}, and the error bars represent the 16th to 84th percentile range. While the simple model is able to accurately retrieve the H$_2$O abundance with high precision, the inclusion of clouds and hazes leads to degeneracies with the H$_2$O abundance \citep[e.g.,][]{benneke_how_2013, welbanks_degeneracies_2019}, making it more difficult to constrain H$_2$O with the realistic model.}
    \label{fig:pandora_posteriors}
\end{figure*}

We find that, for most of our targets, Pandora is able to provide constraints on H$_2$O abundance of $\lesssim\!1$\,dex, comparable to HST constraints \citep[e.g.,][]{tsiaras_population_2018}. For K2-18~b, the water is unconstrained with only an upper limit, consistent with inferences from JWST\edit{, which found a CH$_4$-dominated atmosphere} \citep[e.g.,][]{madhusudhan_carbon-bearing_2023, hu_water-rich_2025}; additionally, WASP-80~b is the faintest target considered in this work and has a high altitude obscuring cloud deck \citep{bell_methane_2023}, leading to a wider and less informative H$_2$O posterior distribution. \edit{For} particularly cloudy planets like HAT-P-18~b and WASP-80~b, we find \edit{inferences broadly consistent with} the true H$_2$O abundance \edit{with precisions of $\sim\!2$ dex}. The posterior probability distributions of H$_2$O abundance in both the simple and comprehensive models are shown in Figure~\ref{fig:pandora_posteriors}.

For methane-dominated planets like K2-18~b, CH$_4$ can also be constrained to $\lesssim\!1$\,dex; in the case of an ideal cloud-free atmosphere, these constraints can be as low as $\sim\!0.5$\,dex, similar to the H$_2$O abundance. For the other planets in our sample, with lower CH$_4$ abundance, we find that \edit{inferences from} Pandora \edit{ result in} upper limits on the CH$_4$ abundance in the atmosphere. Similarly, we find that Pandora will be able to place upper limits on the abundance of NH$_3$. Finally, we find that for our targets, Pandora can successfully constrain the atmospheric temperature with uncertainties of $\sim\!100$\,K even for atmospheres with high cloud coverage (e.g., HAT-P-18~b). \edit{The CH$_4$ and NH$_3$ posterior distributions can be found in Appendix \ref{appendix:pandora_posteriors} \edit{(Figure \ref{fig:pandora_posteriors_CH4_NH3})}. }

The constraints on atmospheric composition \edit{may be related to} the atmospheric metallicity, an indicator of solid accretion during planet formation \citep[e.g.,][]{mordasini_imprint_2016}. We find that Pandora will be able to provide constraints on the atmospheric metallicity with uncertainties of $\lesssim1$ dex with ten transit observations, making population-level analysis \citep[e.g., ][]{welbanks_mass-metallicity_2019} with Pandora observations possible.

\subsubsection{Scattering Slopes}

We \edit{additionally} consider the constraints on scattering slopes in the atmospheres of our target planets. Rayleigh scattering predominantly impacts bluer wavelengths ($\lesssim1$\,$\mu$m), with a dependence on wavelength, scaling as $\lambda^{\gamma}$, where $\gamma=-4$ for pure H$_2$ Rayleigh scattering. Characterizing these slopes is crucial, as they both provide a baseline for the transit depth that ensures accurate measurements of absorption features \citep{benneke_atmospheric_2012} and contain information about aerosols \citep[e.g.,][]{wakeford_transmission_2015, pinhas_signatures_2017}. However, the shortwave predominance of scattering slopes makes them difficult to characterize with NIR instruments \edit{only}. Scattering slopes can be enhanced by condensates and hazes \citep[e.g.,][]{ohno_super-rayleigh_2020} \edit{but are broadly degenerate with} the TLS effect \citep[e.g.,][]{mccullough_water_2014, espinoza_wasp-19b_2019, pinhas_retrieval_2018}. Here, we consider two target cases with known Rayleigh or super-Rayleigh scattering (HD~189733~b and K2-18~b) and constraints \edit{with} Pandora/NIRDA observations. 

We find that in ten transits, observations with NIRDA will likely be sufficient to partially characterize the slope and amplitude enhancement of HD~189733~b's scattering slope, providing accurate ($<1\sigma$) constraints on the amplitude enhancement and slope, with precisions of $\sim$3\,dex and 5 respectively. However, we note that this slope may exhibit a degeneracy with the K abundance when only using NIRDA observations, due to the K doublet peaking at $\sim$0.77\,$\mu$m, off the blue edge of the NIRDA detector. A similar degeneracy has been noted in JWST/NIRISS observations, where the 0.6\,$\mu$m cutoff leads to observations of only the red wing of the Na doublet at $\sim$0.59\,$\mu$m \citep[e.g.,][]{taylor_awesome_2023}. Though we consider additional constraints provided by adding photometric data from Pandora/VISDA outside the scope of this work, \edit{leveraging such instruments} may be instrumental in breaking this degeneracy for planets with strong \edit{optical} slopes. 

\begin{figure}[]
    \centering
    \includegraphics[width=\linewidth]{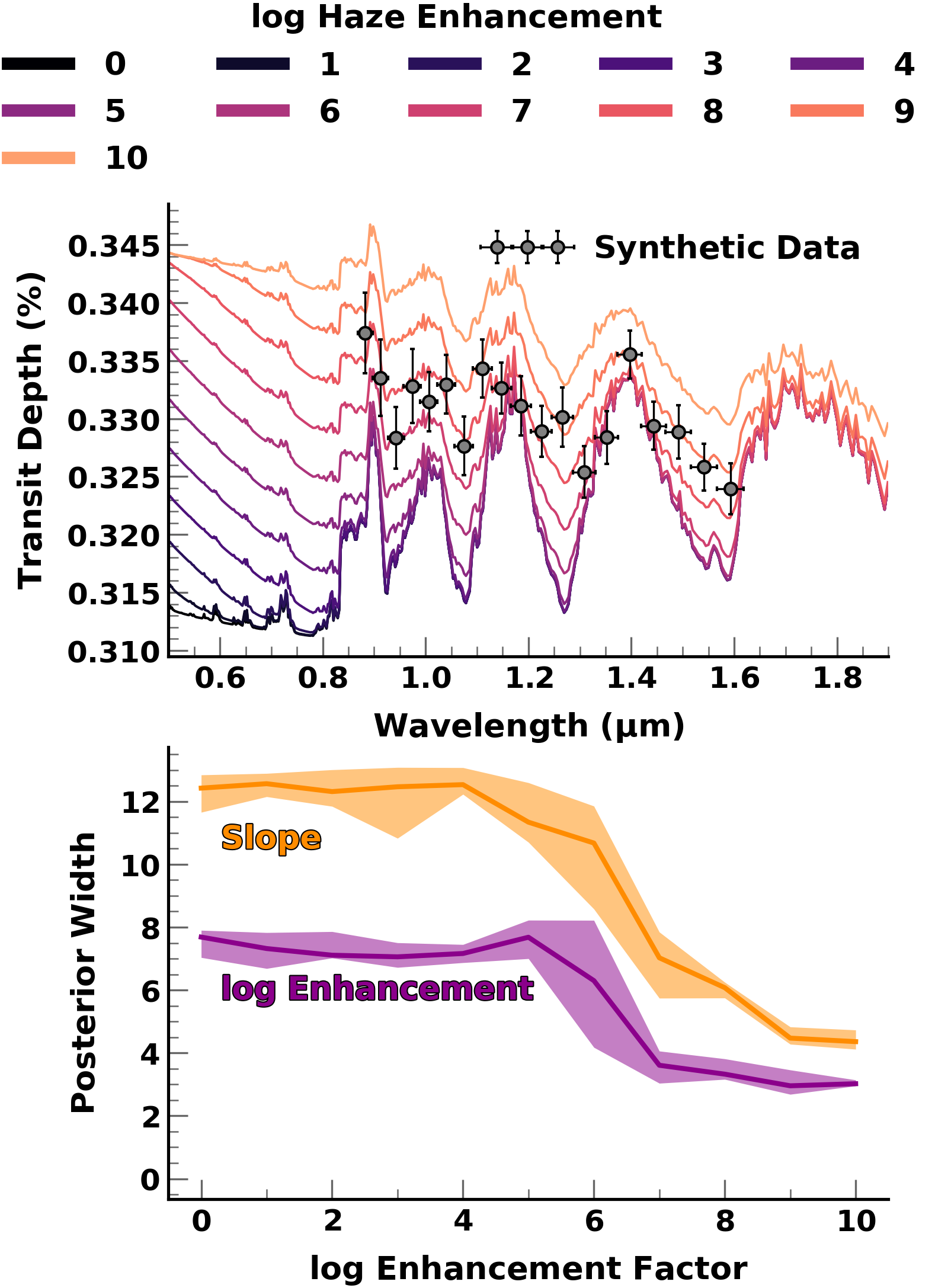}
    \caption{\textbf{Top:} The synthetic spectrum of K2-18~b overlaid with forward models with varying scattering slope enhancement amplitudes. Higher enhancement amplitudes affect the observed spectra further into the NIR, muting feature sizes in the $\sim$1.0 $\mu$m range. \textbf{Bottom:} The 68\% confidence interval of the posteriors of the enhancement ($\log(a)$; purple) and slope ($\gamma$; orange) for the different enhancements considered. At enhancements $\gtrsim10^6$, the scattering slope more strongly affects the Pandora bandpass, allowing for more precise constraints on both parameters.}
    \label{fig:K218b_haze_models}
\end{figure}

The analysis of K2-18~b \edit{results in constraints on} the scattering enhancement amplitude and slope with precisions of $1.8$\,dex and $2.3$, respectively. In order to identify the parameter space where aerosol scattering slopes will be constrainable with Pandora observations, we simulate new observations for K2-18~b, leaving all parameters constant aside from the enhancement factor, which we vary from 1 to 10$^{10}$. For each of these, we simulate ten instances of observations to estimate the precision of the \edit{parameter} posterior. We find that enhancements of $\gtrsim10^6$ will be observable using Pandora/NIRDA (Figure \ref{fig:K218b_haze_models}).

\subsubsection{Identification of Chemical Absorbers}

To analyze the ability of Pandora to identify the spectroscopic features of different chemical absorbers, we \edit{perform model comparisons} between \edit{models} that include or exclude a given absorber following the practices outlined in \citet{benneke_how_2013} and further described in \cite{welbanks_aurora_2021}. The Bayes factor, i.e., the ratio of evidences between the two models, is a measure of the model preference, with a log-Bayes factor of $|\ln B|>5.0$ corresponding to strong significance \edit{at best} \citep{trotta_bayes_2008}. For the model described in Section \ref{sec:fwd_model} and under the assumptions of our synthetic observations, we find that Pandora can yield strong model preferences for the presence of H$_2$O --- with values of $\ln B$ ranging from $-0.1$ for K2-18~b up to 544.3 for HD~209458~b. On the other hand, CH$_4$ and NH$_3$ \edit{absorption is} not significantly preferred, e.g.,  $\ln B=1.6$ for CH$_4$ K2-18~b and $\ln B<0$ for NH$_3$ in all five targets. These Bayesian model comparisons serve as an initial guideline for the capabilities of Pandora to identify spectral features and should be interpreted within the context of the atmospheric models considered as these comparisons do not represent unambiguous chemical detections \citep[see e.g.,][]{welbanks_application_2023, welbanks_challenges_2025}.



\subsection{Multi-Transit Observations}

Alongside observations of the 20 primary mission targets, Pandora will be executing auxiliary target observations for targets of interest. It is therefore prudent to identify the number of transits that would need to be observed with Pandora for suitable constraints on a given target. Here we compare constraints on key atmospheric parameters for observations simulated for multiple transits, ranging from a single transit to 20 transits. We use the same simple HD~209458~b forward model as those in Section \ref{sec:10transits}, limiting the absorbers to H$_2$O, CH$_4$, and NH$_3$, as others will be unconstrained from Pandora alone, and a cloud/haze free atmosphere. For each transit, we assume that the uncertainty in the observations can be scaled as $1/\sqrt{N}$, where $N$ is the number of transits observed. 

\begin{figure*}[t]
    \centering
    \includegraphics[width=\linewidth]{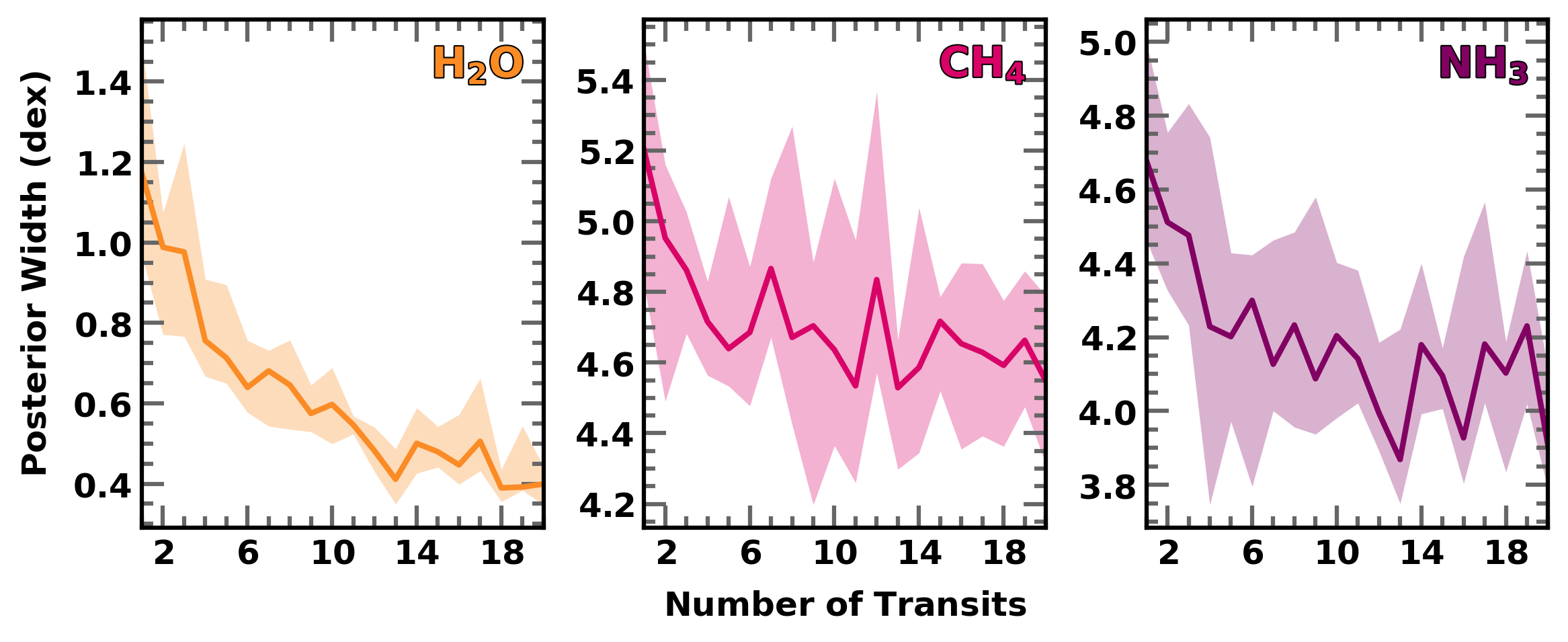}
    \caption{The width of the posterior (16$^{\rm th}$ to 84$^{\rm th}$ percentile) of the H$_2$O, CH$_4$, and NH$_3$ abundance for HD~209458~b as more transits are observed. The asymptotic nature of adding more transits can be clearly observed in the H$_2$O posterior; adding ten transits leads to a reduction of $\sim50\%$ in the H$_2$O posterior width for HD~209458~b, but there is a dropoff in relative improvement as $N_{\rm trans}$ becomes larger. CH$_4$ and NH$_3$ show some signs of tighter constraints when the number of transit observations is increased, but are more poorly constrained for HD~209458~b.}
    \label{fig:1_20_transit_improvement}
\end{figure*}

The constraints provided by Pandora observations improve with an increasing number of transits observed, particularly for the water abundance. \edit{The comparison of this for HD~209458~b is shown in Figure \ref{fig:1_20_transit_improvement}}. Combining transits yields diminishing returns as $N$ increases; \edit{going from 1 to 10} transit observations \edit{improves the H$_2$O posterior precision by a factor of $\sim\!2$,} but \edit{10} and \edit{20} transit observations provide roughly the same posterior width. 

\subsection{Observational Limits for Planetary Signals}

The uncertainty in the planetary \edit{transit depth} increases as a function of the stellar magnitude, making observations of planets even more challenging around \edit{fainter} stars. Given this challenge and the range of magnitudes in Pandora's primary target selection criteria ($m_J$ between 7 and 11.5), we aim to \edit{determine} the limit at which \edit{atmospheric spectra} will be distinguishable from a flat spectrum. 

\begin{figure}[t]
    \centering
    \includegraphics[width=\linewidth]{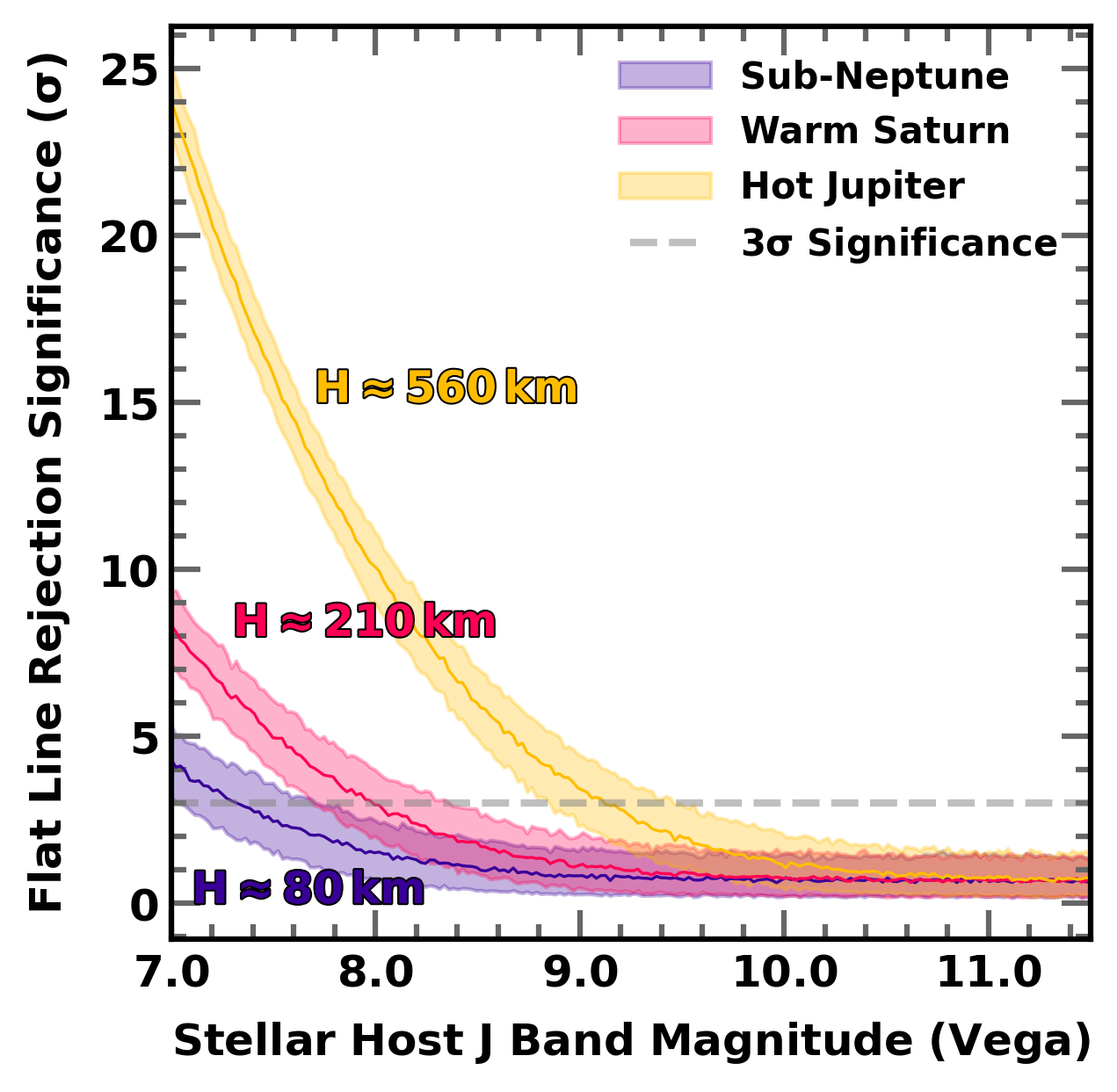}
    \caption{The rejection significance of a flat line for a \edit{hot Jupiter (HD 209458, yellow), warm Saturn (WASP-80, red), or sub-Neptune (K2-18, purple)} at a given J band magnitude. For each planet, we model the transit depth uncertainty assuming ten observations of five hours. The ability to reject a flat line is a function of the atmospheric scale height (labeled as $H$) versus the transit depth uncertainty. We show the 3$\sigma$ cutoff of a strong detection as a dashed gray line. For each planet, we show the median rejection significance and 68\% confidence interval from 100 random noise realizations. As the host star gets fainter, the ability to distinguish between a flat line and an atmospheric signal decreases. For stars with $m_J\gtrsim10$, most exoplanetary atmospheres will be require more than ten transits to distinguish from a featureless spectrum.}
    \label{fig:rejectiontest}
\end{figure}

\edit{We investigate these  limits for observations of Jupiter-, Saturn-, and Neptune-sized archetypes (modeling these after HD~209458~b, WASP-80~b, and K2-18~b)}. For each target, we simulate observations of \edit{spectra} across the range of Pandora's stellar magnitude limits, normalizing to a 5\edit{-}hour observation time for each transit. We consider 200 possible observations across the magnitude range for each target, with 100 noise instances at each magnitude. For each of these simulated observations, we use a chi-squared rejection test to calculate the probability of rejecting a flat line model. A rejection (e.g., $>\!3\sigma$) of a flat line implies a detection of an atmospheric signal. The results of these simulations are shown in Figure \ref{fig:rejectiontest}.

Largely, for target planets with stellar hosts of $m_J\gtrsim10$, we find that distinguishing between an atmospheric signal and a flat line will require over ten transits, \edit{even for high-SNR targets}. For smaller planets, \edit{a brighter host star is needed} for a $3 \sigma$ atmospheric detection ($m_J\lesssim7.5$). 

\subsection{\edit{Identifying H$_2$O Features with Pandora}}

For many smaller targets, \edit{a} major hindrance in HST and JWST observations has been the disentanglement of stellar contamination from the planetary spectrum \edit{\citep[e.g.,][]{zhang_trappist-1_2018, Wakeford2019, Garcia2022, lim_atmospheric_2023, moran_high_2023, banerjee_atmospheric_2024, espinoza2025, radica_promise_2025, rathcke_trappist1bc_2025}}. Pandora's unique capability to constrain the stellar and planetary spectra simultaneously and minimize the impact of stellar contamination presents a promising \edit{avenue for characterizing} planets with ambiguous detections or no previous observations. We explore Pandora's ability to \edit{determine} the presence of an atmosphere (or a deviation from a flat line transmission spectrum) via the H$_2$O absorption feature for each of our different planetary archetypes.

We first analytically calculate the expected signal-to-noise ratio (SNR) of a fully saturated H$_2$O absorption feature in the atmosphere of each of our planets, in order to identify a first-order approximation for the signal strength of the 1.4 $\mu$m water band. A fully saturated feature in a clear atmosphere covers about five scale heights in transit depth space \citep{brown_transmission_2001}. Following this, we calculate the expected transit depth signal of a fully saturated feature given the scale height of each of our planets. We then calculate the minimum and maximum uncertainty in the Pandora H$_2$O absorption band for each planet, and divide the two to calculate the expected range of SNRs of the water feature. We repeat this for a variety of stellar magnitudes. 

\begin{figure}[t]
    \centering
    \includegraphics[width=\linewidth]{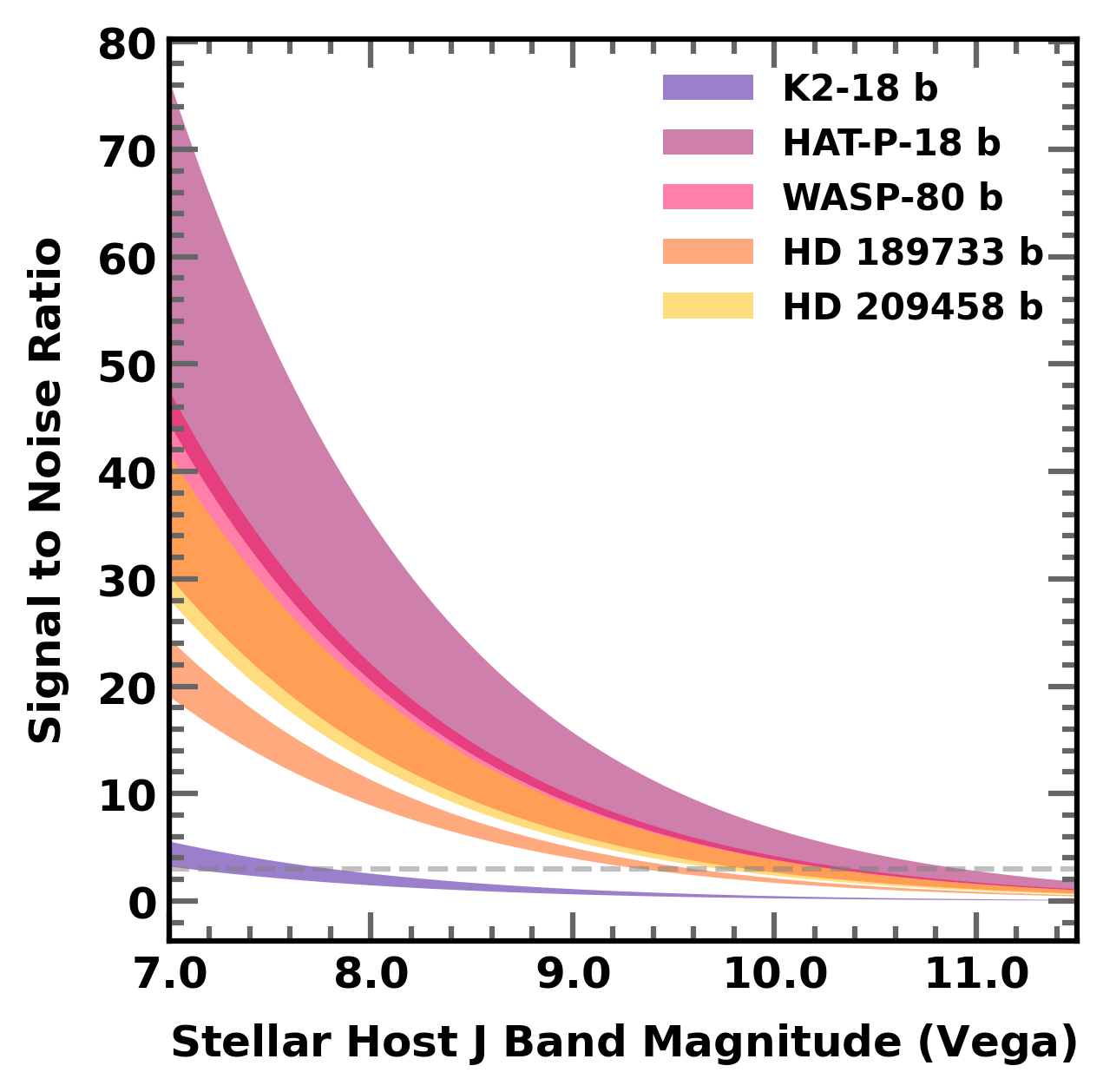}
    \caption{Estimated signal-to-noise ratios of the 1.4 $\mu$m H$_2$O absorption feature for each of our five targets at a given J band magnitude. For each planet, we model the transit depth uncertainty assuming ten observations of five hours. The cutoff for a signal (\edit{$SNR=3$}) is shown with a gray dotted line. The bands represent the maximum and minimum SNRs given the range of uncertainties in bins covering the absorption feature. As the host star becomes fainter, the uncertainty increases and the SNR of the absorption feature becomes smaller. For bright targets ($m_J\lesssim10$) Pandora will be able to produce SNRs of 10--80 for clear atmospheres with saturated water absorption features.}
    \label{fig:snr}
\end{figure}

Figure \ref{fig:snr} shows the expected SNR for all five of our targets, with values of $>5$ implying a strongly detectable feature. We find that Pandora will be able to provide SNRs of $>5$ for clear atmospheres of targets around stars brighter than $m_J\approx 10$. In particular, low-density planets with hot atmospheres and bright hosts can allow for SNRs of up to $\approx75$.

\section{Combining JWST and Pandora Observations}\label{sec:jwst}

The launch of Pandora \edit{in January of 2026} coincides well with JWST, which is now over halfway finished with its nominal 5-year primary mission lifetime. JWST provides a broader wavelength range and higher precision than either HST or Pandora, making it capable of constraining the abundances of multiple carbon- and sulfur-bearing molecules \citep[e.g.,][]{ jwst_transiting_exoplanet_community_early_release_science_team_identification_2022, bell_methane_2023}. Here we compare inferences from JWST observations alone to those from combined Pandora and JWST observations of the same target. For each target, we simulate  observations of a single transit \edit{with JWST/NIRCam, using the F322W2 and F444W filters}, as well as for ten transits with Pandora, as described in Section \ref{sec:simulating}. \edit{Because we aim to explore the consistency of Pandora across the parameter space of possible targets instead of its capability, we model each planet assuming an atmosphere in chemical equilibrium at a 10$\times$ solar metallicity, focusing on major carbon- and oxygen-bearing molecules (Section \ref{sec:fwd_model}). We assume an aerosol-free atmosphere at the equilibrium temperature of each target. We run three retrievals for each case: one with only the simulated Pandora observations, one with only simulated JWST observations, and one with both. For these optimistic scenarios, we analyze the ability to provide accurate constraints on the atmospheric enrichment, defined here as (C+O)/H.}

\begin{figure*}[t]
    \centering
    \includegraphics[width=\linewidth]{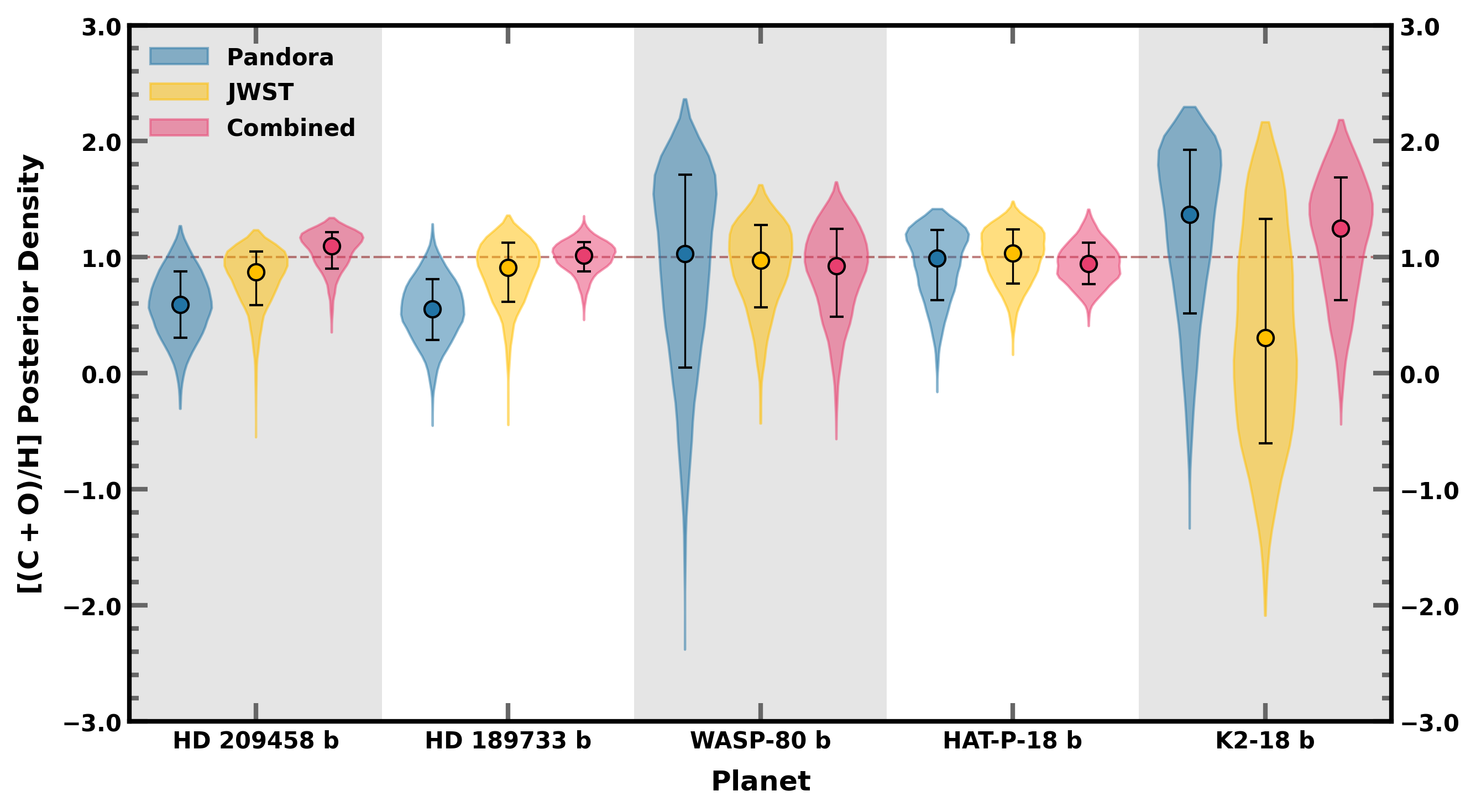}
    \caption{The posterior distributions of the \edit{metallicity (C+O/H) for each of our five targets inferred from Pandora, JWST, or combined observations. Metallicities are shown in logarithmic space and normalized to solar (e.g., (C+O/H)$_\odot=0$)}. Blue histograms represent the constraints from Pandora data only, with yellow representing the JWST-only and red representing the combined JWST and Pandora cases. Maroon lines indicate the \edit{true value of the models (10$\times$ solar)}, and the error bars represent the 16th to 84th percentile range. \edit{The combination of Pandora and JWST observations generally provides the most accurate and precise inference of the overall metallicity.}}
    \label{fig:jwst_posteriors}
\end{figure*}

We find that the combination of Pandora and JWST observations largely provides \edit {better constraints (e.g., higher accuracy, precision, or both)} than \edit{either the Pandora-only or} JWST-only analysis. The inclusion of Pandora provides a valuable baseline for the transit depth and extends the wavelength coverage \edit{of JWST's long-wavelength NIRCam observations (section \ref{sec:optical})}. The posterior distributions for \edit{the (C+O)/H measurements} for all five targets\edit{, normalized to their solar expectations \citep{moses2013},} are shown in Figure \ref{fig:jwst_posteriors}. Similar posteriors \edit{for individual species constrained in the retrieval} can be found in Appendix \ref{appendix:jwst_posteriors}. \edit{While Pandora excels at inferring the oxygen abundance (via H$_2$O, the dominant oxygen-carrying species), it is less able to accurately constrain carbon (via CO and CO$_2$), which is provided by the JWST observations.} 

Despite not having any strong absorption bands of CO or CO$_2$, \edit{we find that} the inclusion of Pandora data \edit{may} help improve the constraints on both species when combined with JWST. \edit{Notably, for} both HD~209458~b and HD~189733~b, which \edit{have the two highest CO abundances in our considered targets (Figure \ref{fig:jwst_posteriors_CO_CO2}), the inclusion of Pandora observations helps improve the precision of the inferred CO abundance. Similarly, despite Pandora's lack of CO$_2$ constraints, all targets with CO$_2$ except WASP-80~b show an improvement in the precision, accuracy, or both when Pandora is added to JWST/NIRCam observations. This improvement is likely due to Pandora observations providing a larger baseline for the transit depth, enabling a better estimate of the spectral continuum \citep[e.g.,][]{benneke_atmospheric_2012}.} 


\subsection{On the Importance of Optical and Near-IR Data from Pandora}\label{sec:optical}
To sufficiently characterize the atmosphere, observations in the optical and near-IR are often needed to complement data further towards the mid-IR wavelengths \citep[e.g.,][]{wakeford_complete_2018,fairman_importance_2024}. This allows for a transit depth baseline to be set \citep[e.g.,][]{benneke_atmospheric_2012}, \edit{breaking key degeneracies affecting transmission spectra \citep{welbanks_degeneracies_2019}, which may otherwise lead} to bimodal posterior distributions obtained with mid-IR observations alone \citep[e.g.,][]{schlawin_clear_2018}. Pandora's spectroscopic and photometric bandpasses can provide NIR spectra down to $\sim$0.9 $\mu$m and photometric data at $\sim\!0.5$ $\mu$m, while simultaneously observing the \edit{near-IR}. 

Here we \edit{explore synergies of} joint observations with Pandora and JWST. Motivated by previous analyses showing the importance of optical data \citep[e.g.,][]{wakeford_complete_2018}, we present an analysis of HD~209458~b with combined JWST and Pandora observations. We generate a spectrum \edit{of} HD~209458~b, \edit{following} Section \edit{\ref{sec:pandora}} but only including H$_2$O and CO$_2$, alongside a gray cloud opacity that covers 70$\%$ of the terminator. We consider observations with Pandora, JWST, or the combination of both, \edit{and vary the number of transit observations with Pandora} from one to twenty. We consider the additional observations with Pandora/VISDA outside the scope of this comparison, though we note that it would likely further improve constraints when compared to JWST or Pandora/NIRDA alone. For each scenario, we consider ten noise instances of the Pandora data, to ensure that any \edit{inferences} are not skewed by statistical outliers. 

\begin{figure}[t]
    \centering
    \includegraphics[width=\linewidth]{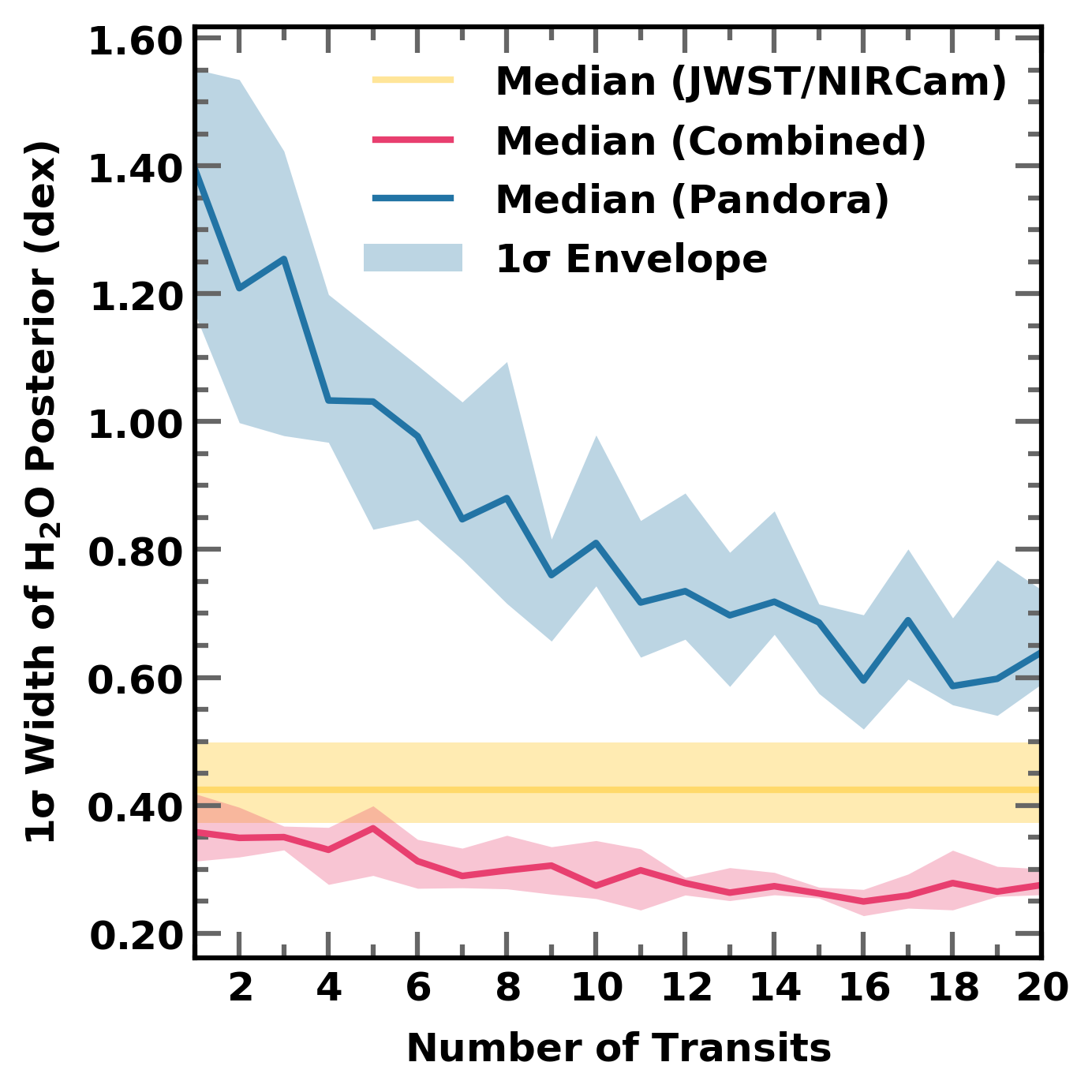}
    \caption{The width of posterior distributions of H$_2$O retrieved from synthetic observations of HD~209458~b. Inferences from Pandora observations are shown in light blue, with JWST/NIRCam/F322W+F444W in yellow. The inferences from combined observations of Pandora and NIRCam are shown in red. For each, we show the median and 68\% confidence interval of the posterior width for each number of Pandora transits. By combining both datasets, the inferences can provide better precision on the water abundance than either instrument alone due to the additional wavelength coverage.}
    \label{fig:hd209_water}
\end{figure}

We find that by combining observations with Pandora and NIRCam we obtain more reliable inferences on the H$_2$O abundance of HD~209458~b. In Figure \ref{fig:hd209_water}, we show the $1\sigma$ width (16$^{\rm th}$ to 84$^{\rm th}$ percentile range) of the H$_2$O abundance posteriors for each number of transits. Pandora observations by themselves have a larger $1\sigma$ range ($\sim\!0.6$--$1.4$ dex) than NIRCam ($\sim\!0.4$--$0.5$ dex)\edit{; however,} combining the observations allows for more precise constraints, down to $\sim\!0.3$ dex. In both cases, a plateau in the improvement is evident around the 10--12 transit mark, implying that \edit{for more than} 12 transits, Pandora observations provide diminishing returns with respect to the water abundance constraint.

\section{Discussion and Summary}\label{sec:summary}

The upcoming Pandora SmallSat Mission is a dedicated exoplanet observatory, which presents a unique opportunity to study novel or complex systems and complement spectroscopic observations with other facilities, such as HST and JWST. In this work, we simulated Pandora observations for five exoplanet targets, and analyzed the inferences that can be obtained from Pandora spectra, as well as combined Pandora and JWST spectra. We aimed to characterize Pandora's capabilities for exoplanet targets and as a tool for improving JWST observations of planets around active stars. Here, we summarize our main findings in this work.

\subsection{Detecting Atmospheric Signals}

For bright targets ($m_J{\approx}7$), Pandora will be able to obtain spectra with precisions of 30--100\,ppm after multiple transit observations. These precisions rival those of spectra obtained with HST's WFC/G141 grism, a crucial instrument for constraining water abundances in exoplanet atmospheres for over a decade \citep[e.g.,][]{deming_infrared_2013, kreidberg_detection_2015, sing_continuum_2016}. These precisions will allow for meaningful detections of an atmospheric signal (\edit{rejecting a flat line ``no atmosphere" case} at $>3\sigma$ confidence) for planets as small as sub-Neptunes, provided their host star is sufficiently bright \edit{and their atmospheres are aerosol-free and low-metallicity enough to provide large absorption features}. However, for host stars with $m_J\gtrsim10$, the uncertainty for spectra obtained from ten stacked transits will be of order $\sim\!500$ ppm or higher. At those magnitudes, more than ten transit observations may be required to rule out flat spectra, even for those with large scale heights.

\subsection{Constraining Atmospheric Spectra}

With ten transits observed and stellar heterogeneity corrections, Pandora's NIRDA spectra \edit{can} constrain H$_2$O with precisions of $\lesssim1$\,dex and atmospheric temperatures to within $\lesssim100$\,K \edit{for the scenarios considered here}. Pandora will also be able to observe the 1.4~$\mu$m H$_2$O absorption feature with a signal-to-noise ratio of up to $SNR=75$ for fully saturated, cloud-free atmospheres. Additionally, Pandora \edit{has the potential} to place accurate upper limits on the abundance of CH$_4$ and NH$_3$ in exoplanet atmospheres, and may be able to provide constraints of $\approx1$\,dex on the CH$_4$ abundance for methane-dominated atmospheres of sub-Neptunes \citep[e.g., K2-18~b;][]{madhusudhan_carbon-bearing_2023}. The latter will depend on the capability to distinguish between CH$_4$ and H$_2$O absorption, which has presented an issue for previous observations with HST/WFC3 G141 \citep[e.g.,][]{murphy_hst_2025}. This, however, \edit{may} be overcome with observations of CH$_4$ absorption at longer wavelengths with JWST \citep[e.g.,][]{bell_methane_2023, welbanks_high_2024}, presenting a key synergistic avenue between the two telescopes. Additionally, \edit{constraints} on the chemical inventory of the planetary atmosphere \edit{from Pandora observations alone may} allow for accurate constraints of $\lesssim1.3$ dex on the atmospheric metallicity \edit{for atmospheres with large absorption features}, allowing for population-level analysis of the bulk composition. \edit{We note that, particularly for sub-Neptunes, this is dependent on the atmospheric temperature, metallicity, and aerosol content of the atmosphere, as they may have high-metallicity or aerosol-obscured atmospheres indistinguishable with JWST from atmospheric non-detections \citep[e.g.,][]{wallack_jwst_2024, schlawin_possible_2024}.}

Furthermore, observations with Pandora/NIRDA \edit{may} be able to place constraints on the scattering slope for particularly hazy planets. Rayleigh scattering slopes in transmission spectra  are often enhanced by haze scattering in the upper atmosphere \citep[e.g.,][]{lecavelier_des_etangs_rayleigh_2008,pont_prevalence_2013}, leading to large slopes in optical wavelengths that may extend out towards NIR absorption features. Constraining this slope is crucial for analysis of the transmission spectrum. Firstly, the angle and enhancement of the slope provide key information about haze composition in the atmosphere \citep[e.g.,][]{pinhas_signatures_2017,ohno_super-rayleigh_2020}. \edit{Secondly, it} provides a key measurement of the mean molecular weight of the atmosphere \citep{benneke_atmospheric_2012}, allowing for disentanglement of the degeneracy between temperature and mean molecular weight that arises from only measuring the scale height via absorption features \edit{\citep[e.g.,][]{welbanks_degeneracies_2019}}. Most importantly, the scattering slope is often degenerate with TLS features, which preferentially affect \edit{shorter} wavelengths \citep[e.g.,][]{rackham_transit_2019,iyer_influence_2020}. This can lead to artificially large super-Rayleigh slopes \citep[e.g.,][]{canas_gems_2025} or \edit{negative} slopes \citep[e.g.,][]{cadieux_transmission_2024}. Pandora \edit{may} be able to constrain haze scattering slopes and enhancement factors for Rayleigh scattering with enhancements of more than $\sim10^6$. \edit{We note that we do not consider the effects of uncorrected TLS features in our spectra, and the ability to simultaneously constrain stellar and planetary properties; a companion study \citep[][]{Rackham2026} explores Pandora's constraining capabilities for the host stars and its impacts at the data and model levels.}

\subsection{Synergy with JWST}

We further consider the synergies of Pandora and JWST when observing absorption features in the planetary spectrum. \edit{Combining} Pandora and JWST observations can provide constraints on molecular abundances, particularly H$_2$O, that are both \edit{more precise} and more accurate than achievable with either telescope separately (Figure \ref{fig:jwst_posteriors}). The inclusion of Pandora provides a valuable wavelength baseline for observations \edit{at longer wavelengths with JWST (e.g., NIRCam)}, as well as an independent measurement of the water absorption feature, allowing for relative feature sizes in the JWST observations to be converted into absolute abundance measurements. \edit{This improvement with the addition of lower-wavelength data has been previously noted for HST \citep[e.g.,][]{fairman_importance_2024} and JWST; for example, \citet{batalha2017} noted the preference for broadband observations over repeated observations when aiming for accurate inferences from the planetary atmosphere}. 

Furthermore, we find that the inclusion of Pandora observations can help improve constraints on gases not accessible with Pandora alone, such as CO$_2$ (Figure \ref{fig:jwst_posteriors_CO_CO2}). Although Pandora can only place upper limit\edit{s} on the CO$_2$ abundance by itself, the wavelength coverage of Pandora provides a key baseline for comparing to the CO$_2$ feature observed with JWST. As such, we find that the CO$_2$ constraint\edit{s} acquired from joint JWST and Pandora observations \edit{can be} both more precise and more accurate than achievable with JWST alone.

\edit{We further explore the effect of multiple Pandora transit observations in combination with JWST. We find that combining multiple Pandora transits will provide more precise inferences of absorber abundances, though the effect of increasing the number of observations is lessened for absorbers with only upper limit characterization. In particular, the improvement due to an increased number of transits is most effective up until $\sim$10--12 transits, after which the precision does not improve significantly. Moreover, we find that the combination of Pandora observations with JWST will provide more precise inferences than possible with either instrument alone (leading to inferences as precise as $\sim\!0.25$ dex with the combined instrumentation at 10 transits).}

\subsection{Looking Forward with JWST and Pandora}\label{sec:conclusion}

With JWST now \edit{in operation for} more than three years of exoplanet observations, the field of exoplanet science is accelerating at a rapid pace. At the forefront of the field are questions about planet formation and evolution, including the composition of small planet atmospheres and the effect of underlying star--planet connections on both our inferences and the planet itself. Often, active host stars present barriers for JWST observations of their planets, including disagreement across multiple observations or biased estimates of abundances \citep[e.g.,][]{fu_water_2022, fournier-tondreau_near-infrared_2024}\edit{. While modeling methodologies have been proposed to overcome unknowns introduced in the observations by stellar activity \citep[e.g.,][]{rotman_enabling_2025, espinoza2025}, an approach at the data level is needed to ensure reliable observations}. Pandora is a mission designed specifically to better understand and \edit{overcome} these \edit{challenges}. 

In addition to improved characterization of individual planet atmospheres, the combination of JWST and Pandora offer\edit{s} an opportunity to further the study of populations of planet atmospheres. A significant, growing body of work concerns the understanding of different classes of planet atmospheres, \edit{the} processes \edit{that} drive \edit{their} evolution, and their relative frequencies \cite[see, e.g.][and references therein]{kempton_knutson_review_2024}. One roadblock to these analyses is the difficulty interpreting the depth of NIR spectral features in the absence of shorter wavelength information constraining the scattering slope. By providing these crucial complementary data for planets over a wide range of sizes and temperatures, Pandora will contribute significantly to population-level analyses of planets observed with JWST. 

\edit{Alongside the opportunities explored throughout this work, Pandora provides a unique opportunity to help overcome many hurdles inherent in observations with current instruments. Namely, multi-epoch observations of exoplanet atmospheres with JWST are often observed at different points in the stellar rotation, making them difficult to fit jointly \citep[e.g.,][]{may_double_2023}. By providing information about the stellar activity and allowing for corrections at the data level, joint Pandora--JWST programs will unlock new insights into planets around active hosts. Additionally, these same insights will inform the presence of offsets between different instruments and epochs, which have been difficult to constrain thus far with JWST alone \citep[e.g.,][]{carter_benchmark_2024}.}

Pandora presents a new opportunity to provide important constraints on the atmospheric composition, temperature, and metallicity of exoplanets, particularly in the presence of stellar activity. Furthermore, SmallSats like Pandora provide \edit{important avenues} for synergizing with current flagship missions. Leveraging Pandora's capabilities and synergies with JWST over its 1 year prime mission will help ensure that we maximize the science information available to us both now and over the coming decades of exoplanet science.

\begin{acknowledgments}
\edit{The authors would like to thank the reviewer for the comments that significantly improved the manuscript.} Pandora is supported by NASA’s Astrophysics Pioneers Program. This work was performed under the auspices of the U.S. Department of Energy by Lawrence Livermore National Laboratory under Contract DE-AC52-07NA27344. The document number is LLNL-JRNL-2009856. Y.R. and P.M were supported under the LLNL Space Science Institute’s Institutional Scientific Capability Portfolio. \edit{Computing
support for this work came from the Lawrence Livermore National Laboratory 19th Institutional Computing Grand Challenge program.}  L.W. \edit{and YR} acknowledge support from NASA XRP Grant [80NSSC24K0160]. L.W. thanks the Heising-Simons Foundation for support under the 51 Pegasi b Fellowship. 
\edit{This material is based upon work supported by NASA under award No.\ 80NSSC24K0197.
This material is partly based upon work supported by the National Aeronautics and Space Administration under Agreement No.\ 80NSSC21K0593 for the program ``Alien Earths''.
The results reported herein benefited from collaborations and/or information exchange within NASA’s Nexus for Exoplanet System Science (NExSS) research coordination network sponsored by NASA’s Science Mission Directorate.
This material is based upon work supported by the European Research Council (ERC) Synergy Grant under the European Union’s Horizon 2020 research and innovation program (grant No.\ 101118581---project REVEAL).}
T.G. acknowledges NASA Pandora funding through WBS 344310.01.02.01.02. B.J.H. acknowledges funding support from grant ORAU - 80HQTR21CA005. T.O.F. acknowledges funding support through the NASA Postdoctoral Program Grant Number: ORAU - 80HQTR21CA005. J.F.R. acknowledges financial support from the NSERC Discovery Program and the Canadian Space Agency ROSS program. K. H. acknowledges the support of the Canadian Space Agency (CSA) [22EXPROSS1].
\end{acknowledgments}


\software{NumPy \citep{harris_array_2020}, AstroPy \citep{the_astropy_collaboration_astropy_2022}, PandoraSat and PandoraSim \citep{hedges_open-source_2024}, Matplotlib \citep{hunter_matplotlib_2007}, MultiNest \citep{feroz_multinest_2009} via PyMultiNest \citep{buchner_x-ray_2014}.}

\bibliography{pandorabib, pandorabib2}{}

@ARTICLE{kempton_knutson_review_2024,
       author = {{Kempton}, Eliza M. -R. and {Knutson}, Heather A.},
        title = "{Transiting Exoplanet Atmospheres in the Era of JWST}",
      journal = {Reviews in Mineralogy and Geochemistry},
         year = 2024,
        month = jul,
       volume = {90},
       number = {1},
        pages = {411-464},
          doi = {10.2138/rmg.2024.90.12},
archivePrefix = {arXiv},
       eprint = {2404.15430},
 primaryClass = {astro-ph.EP},
       adsurl = {https://ui.adsabs.harvard.edu/abs/2024RvMG...90..411K}
}

@article{charbonneau_detection_2002,
	title = {Detection of an {Extrasolar} {Planet} {Atmosphere}*},
	volume = {568},
	issn = {0004-637X},
	url = {https://iopscience.iop.org/article/10.1086/338770/meta},
	doi = {10.1086/338770},
	language = {en},
	number = {1},
	urldate = {2022-03-14},
	journal = {The Astrophysical Journal},
	author = {Charbonneau, David and Brown, Timothy M. and Noyes, Robert W. and Gilliland, Ronald L.},
	month = mar,
	year = {2002},
	note = {Publisher: IOP Publishing},
	pages = {377},
}

@article{greene_characterizing_2016,
	title = {Characterizing {Transiting} {Exoplanet} {Atmospheres} with {JWST}},
	volume = {817},
	issn = {0004-637X},
	url = {https://ui.adsabs.harvard.edu/abs/2016ApJ...817...17G},
	doi = {10.3847/0004-637X/817/1/17},
	urldate = {2022-04-03},
	journal = {The Astrophysical Journal},
	author = {Greene, Thomas P. and Line, Michael R. and Montero, Cezar and Fortney, Jonathan J. and Lustig-Yaeger, Jacob and Luther, Kyle},
	month = jan,
	year = {2016},
	note = {ADS Bibcode: 2016ApJ...817...17G},
	pages = {17},
}

@misc{jwst_transiting_exoplanet_community_early_release_science_team_identification_2022,
	title = {Identification of carbon dioxide in an exoplanet atmosphere},
	url = {http://arxiv.org/abs/2208.11692},
	doi = {10.48550/arXiv.2208.11692},
	urldate = {2022-11-18},
	publisher = {arXiv},
	author = {{JWST Transiting Exoplanet Community Early Release Science Team} and Ahrer, Eva-Maria and Alderson, Lili and Batalha, Natalie M. and Batalha, Natasha E. and Bean, Jacob L. and Beatty, Thomas G. and Bell, Taylor J. and Benneke, Björn and Berta-Thompson, Zachory K. and Carter, Aarynn L. and Crossfield, Ian J. M. and Espinoza, Néstor and Feinstein, Adina D. and Fortney, Jonathan J. and Gibson, Neale P. and Goyal, Jayesh M. and Kempton, Eliza M.-R. and Kirk, James and Kreidberg, Laura and López-Morales, Mercedes and Line, Michael R. and Lothringer, Joshua D. and Moran, Sarah E. and Mukherjee, Sagnick and Ohno, Kazumasa and Parmentier, Vivien and Piaulet, Caroline and Rustamkulov, Zafar and Schlawin, Everett and Sing, David K. and Stevenson, Kevin B. and Wakeford, Hannah R. and Allen, Natalie H. and Birkmann, Stephan M. and Brande, Jonathan and Crouzet, Nicolas and Cubillos, Patricio E. and Damiano, Mario and Désert, Jean-Michel and Gao, Peter and Harrington, Joseph and Hu, Renyu and Kendrew, Sarah and Knutson, Heather A. and Lagage, Pierre-Olivier and Leconte, Jérémy and Lendl, Monika and MacDonald, Ryan J. and May, E. M. and Miguel, Yamila and Molaverdikhani, Karan and Moses, Julianne I. and Murray, Catriona Anne and Nehring, Molly and Nikolov, Nikolay K. and de la Roche, D. J. M. Petit dit and Radica, Michael and Roy, Pierre-Alexis and Stassun, Keivan G. and Taylor, Jake and Waalkes, William C. and Wachiraphan, Patcharapol and Welbanks, Luis and Wheatley, Peter J. and Aggarwal, Keshav and Alam, Munazza K. and Banerjee, Agnibha and Barstow, Joanna K. and Blecic, Jasmina and Casewell, S. L. and Changeat, Quentin and Chubb, K. L. and Colón, Knicole D. and Coulombe, Louis-Philippe and Daylan, Tansu and de Val-Borro, Miguel and Decin, Leen and Santos, Leonardo A. Dos and Flagg, Laura and France, Kevin and Fu, Guangwei and Muñoz, A. García and Gizis, John E. and Glidden, Ana and Grant, David and Heng, Kevin and Henning, Thomas and Hong, Yu-Cian and Inglis, Julie and Iro, Nicolas and Kataria, Tiffany and Komacek, Thaddeus D. and Krick, Jessica E. and Lee, Elspeth K. H. and Lewis, Nikole K. and Lillo-Box, Jorge and Lustig-Yaeger, Jacob and Mancini, Luigi and Mandell, Avi M. and Mansfield, Megan and Marley, Mark S. and Mikal-Evans, Thomas and Morello, Giuseppe and Nixon, Matthew C. and Ceballos, Kevin Ortiz and Piette, Anjali A. A. and Powell, Diana and Rackham, Benjamin V. and Ramos-Rosado, Lakeisha and Rauscher, Emily and Redfield, Seth and Rogers, Laura K. and Roman, Michael T. and Roudier, Gael M. and Scarsdale, Nicholas and Shkolnik, Evgenya L. and Southworth, John and Spake, Jessica J. and Steinrueck, Maria E. and Tan, Xianyu and Teske, Johanna K. and Tremblin, Pascal and Tsai, Shang-Min and Tucker, Gregory S. and Turner, Jake D. and Valenti, Jeff A. and Venot, Olivia and Waldmann, Ingo P. and Wallack, Nicole L. and Zhang, Xi and Zieba, Sebastian},
	month = aug,
	year = {2022},
	note = {arXiv:2208.11692 [astro-ph]},
}

@article{wakeford_complete_2018,
	title = {The {Complete} {Transmission} {Spectrum} of {WASP}-39b with a {Precise} {Water} {Constraint}},
	volume = {155},
	issn = {0004-6256},
	url = {https://ui.adsabs.harvard.edu/abs/2018AJ....155...29W},
	doi = {10.3847/1538-3881/aa9e4e},
	urldate = {2022-11-18},
	journal = {The Astronomical Journal},
	author = {Wakeford, H. R. and Sing, D. K. and Deming, D. and Lewis, N. K. and Goyal, J. and Wilson, T. J. and Barstow, J. and Kataria, T. and Drummond, B. and Evans, T. M. and Carter, A. L. and Nikolov, N. and Knutson, H. A. and Ballester, G. E. and Mandell, A. M.},
	month = jan,
	year = {2018},
	note = {ADS Bibcode: 2018AJ....155...29W},
	pages = {29},
}

@article{sing_continuum_2016,
	title = {A continuum from clear to cloudy hot-{Jupiter} exoplanets without primordial water depletion},
	volume = {529},
	issn = {0028-0836},
	url = {https://ui.adsabs.harvard.edu/abs/2016Natur.529...59S},
	doi = {10.1038/nature16068},
	urldate = {2022-11-18},
	journal = {Nature},
	author = {Sing, David K. and Fortney, Jonathan J. and Nikolov, Nikolay and Wakeford, Hannah R. and Kataria, Tiffany and Evans, Thomas M. and Aigrain, Suzanne and Ballester, Gilda E. and Burrows, Adam S. and Deming, Drake and Désert, Jean-Michel and Gibson, Neale P. and Henry, Gregory W. and Huitson, Catherine M. and Knutson, Heather A. and Lecavelier Des Etangs, Alain and Pont, Frederic and Showman, Adam P. and Vidal-Madjar, Alfred and Williamson, Michael H. and Wilson, Paul A.},
	month = jan,
	year = {2016},
	note = {ADS Bibcode: 2016Natur.529...59S},
	pages = {59--62},
}

@article{welbanks_degeneracies_2019,
	title = {On {Degeneracies} in {Retrievals} of {Exoplanetary} {Transmission} {Spectra}},
	volume = {157},
	issn = {0004-6256},
	url = {https://ui.adsabs.harvard.edu/abs/2019AJ....157..206W},
	doi = {10.3847/1538-3881/ab14de},
	urldate = {2022-11-18},
	journal = {The Astronomical Journal},
	author = {Welbanks, Luis and Madhusudhan, Nikku},
	month = may,
	year = {2019},
	note = {ADS Bibcode: 2019AJ....157..206W},
	pages = {206},
}

@article{madhusudhan_temperature_2009,
	title = {A {Temperature} and {Abundance} {Retrieval} {Method} for {Exoplanet} {Atmospheres}},
	volume = {707},
	issn = {0004-637X},
	url = {https://ui.adsabs.harvard.edu/abs/2009ApJ...707...24M},
	doi = {10.1088/0004-637X/707/1/24},
	urldate = {2022-11-18},
	journal = {The Astrophysical Journal},
	author = {Madhusudhan, N. and Seager, S.},
	month = dec,
	year = {2009},
	note = {ADS Bibcode: 2009ApJ...707...24M},
	pages = {24--39},
}

@article{welbanks_aurora_2021,
	title = {Aurora: {A} {Generalized} {Retrieval} {Framework} for {Exoplanetary} {Transmission} {Spectra}},
	volume = {913},
	issn = {0004-637X},
	shorttitle = {Aurora},
	url = {https://dx.doi.org/10.3847/1538-4357/abee94},
	doi = {10.3847/1538-4357/abee94},
	language = {en},
	number = {2},
	urldate = {2023-01-20},
	journal = {The Astrophysical Journal},
	author = {Welbanks, Luis and Madhusudhan, Nikku},
	month = jun,
	year = {2021},
	note = {Publisher: The American Astronomical Society},
	pages = {114},
}

@article{line_influence_2016,
	title = {The {Influence} of {Nonuniform} {Cloud} {Cover} on {Transit} {Transmission} {Spectra}},
	volume = {820},
	issn = {0004-637X},
	url = {https://dx.doi.org/10.3847/0004-637X/820/1/78},
	doi = {10.3847/0004-637X/820/1/78},
	language = {en},
	number = {1},
	urldate = {2023-01-20},
	journal = {The Astrophysical Journal},
	author = {Line, Michael R. and Parmentier, Vivien},
	month = mar,
	year = {2016},
	note = {Publisher: The American Astronomical Society},
	pages = {78},
}

@article{batalha_pandexo_2017,
	title = {{PandExo}: {A} {Community} {Tool} for {Transiting} {Exoplanet} {Science} with {JWST} \& {HST}},
	volume = {129},
	issn = {1538-3873},
	shorttitle = {{PandExo}},
	url = {https://dx.doi.org/10.1088/1538-3873/aa65b0},
	doi = {10.1088/1538-3873/aa65b0},
	language = {en},
	number = {976},
	urldate = {2023-02-01},
	journal = {Publications of the Astronomical Society of the Pacific},
	author = {Batalha, Natasha E. and Mandell, Avi and Pontoppidan, Klaus and Stevenson, Kevin B. and Lewis, Nikole K. and Kalirai, Jason and Earl, Nick and Greene, Thomas and Albert, Loïc and Nielsen, Louise D.},
	month = apr,
	year = {2017},
	note = {Publisher: The Astronomical Society of the Pacific},
	pages = {064501},
}

@article{madhusudhan_co_2012,
	title = {C/{O} {Ratio} as a {Dimension} for {Characterizing} {Exoplanetary} {Atmospheres}},
	volume = {758},
	issn = {0004-637X},
	url = {https://ui.adsabs.harvard.edu/abs/2012ApJ...758...36M},
	doi = {10.1088/0004-637X/758/1/36},
	urldate = {2023-02-01},
	journal = {The Astrophysical Journal},
	author = {Madhusudhan, Nikku},
	month = oct,
	year = {2012},
	note = {ADS Bibcode: 2012ApJ...758...36M},
	pages = {36},
}

@article{oberg_effects_2011,
	title = {The {Effects} of {Snowlines} on {C}/{O} in {Planetary} {Atmospheres}},
	volume = {743},
	issn = {0004-637X},
	url = {https://ui.adsabs.harvard.edu/abs/2011ApJ...743L..16O},
	doi = {10.1088/2041-8205/743/1/L16},
	urldate = {2023-02-01},
	journal = {The Astrophysical Journal},
	author = {Öberg, Karin I. and Murray-Clay, Ruth and Bergin, Edwin A.},
	month = dec,
	year = {2011},
	note = {ADS Bibcode: 2011ApJ...743L..16O},
	pages = {L16},
}

@article{feroz_multinest_2009,
	title = {{MultiNest}: an efficient and robust {Bayesian} inference tool for cosmology and particle physics},
	volume = {398},
	issn = {0035-8711},
	shorttitle = {{MultiNest}},
	url = {https://doi.org/10.1111/j.1365-2966.2009.14548.x},
	doi = {10.1111/j.1365-2966.2009.14548.x},
	number = {4},
	urldate = {2023-02-01},
	journal = {Monthly Notices of the Royal Astronomical Society},
	author = {Feroz, F. and Hobson, M. P. and Bridges, M.},
	month = oct,
	year = {2009},
	pages = {1601--1614},
}

@article{seager_theoretical_2000,
	title = {Theoretical {Transmission} {Spectra} during {Extrasolar} {Giant} {Planet} {Transits}},
	volume = {537},
	issn = {0004-637X},
	url = {https://ui.adsabs.harvard.edu/abs/2000ApJ...537..916S},
	doi = {10.1086/309088},
	urldate = {2023-02-02},
	journal = {The Astrophysical Journal},
	author = {Seager, S. and Sasselov, D. D.},
	month = jul,
	year = {2000},
	note = {ADS Bibcode: 2000ApJ...537..916S},
	pages = {916--921},
}

@article{pont_detection_2008,
	title = {Detection of atmospheric haze on an extrasolar planet: the 0.55-1.05 μm transmission spectrum of {HD} 189733b with the {HubbleSpaceTelescope}},
	volume = {385},
	issn = {0035-8711},
	shorttitle = {Detection of atmospheric haze on an extrasolar planet},
	url = {https://ui.adsabs.harvard.edu/abs/2008MNRAS.385..109P},
	doi = {10.1111/j.1365-2966.2008.12852.x},
	urldate = {2023-02-02},
	journal = {Monthly Notices of the Royal Astronomical Society},
	author = {Pont, F. and Knutson, H. and Gilliland, R. L. and Moutou, C. and Charbonneau, D.},
	month = mar,
	year = {2008},
	note = {ADS Bibcode: 2008MNRAS.385..109P},
	pages = {109--118},
}

@article{kreidberg_detection_2015,
	title = {A {Detection} of {Water} in the {Transmission} {Spectrum} of the {Hot} {Jupiter} {WASP}-12b and {Implications} for {Its} {Atmospheric} {Composition}},
	volume = {814},
	issn = {0004-637X},
	url = {https://ui.adsabs.harvard.edu/abs/2015ApJ...814...66K},
	doi = {10.1088/0004-637X/814/1/66},
	urldate = {2023-02-02},
	journal = {The Astrophysical Journal},
	author = {Kreidberg, Laura and Line, Michael R. and Bean, Jacob L. and Stevenson, Kevin B. and Désert, Jean-Michel and Madhusudhan, Nikku and Fortney, Jonathan J. and Barstow, Joanna K. and Henry, Gregory W. and Williamson, Michael H. and Showman, Adam P.},
	month = nov,
	year = {2015},
	note = {ADS Bibcode: 2015ApJ...814...66K},
	pages = {66},
}

@article{iyer_influence_2020,
	title = {The {Influence} of {Stellar} {Contamination} on the {Interpretation} of {Near}-infrared {Transmission} {Spectra} of {Sub}-{Neptune} {Worlds} around {M}-dwarfs},
	volume = {889},
	issn = {0004-637X},
	url = {https://ui.adsabs.harvard.edu/abs/2020ApJ...889...78I},
	doi = {10.3847/1538-4357/ab612e},
	urldate = {2023-02-09},
	journal = {The Astrophysical Journal},
	author = {Iyer, Aishwarya R. and Line, Michael R.},
	month = feb,
	year = {2020},
	note = {ADS Bibcode: 2020ApJ...889...78I},
	pages = {78},
}

@article{welbanks_mass-metallicity_2019,
	title = {Mass-{Metallicity} {Trends} in {Transiting} {Exoplanets} from {Atmospheric} {Abundances} of {H2O}, {Na}, and {K}},
	volume = {887},
	issn = {0004-637X},
	url = {https://ui.adsabs.harvard.edu/abs/2019ApJ...887L..20W},
	doi = {10.3847/2041-8213/ab5a89},
	urldate = {2023-05-18},
	journal = {The Astrophysical Journal},
	author = {Welbanks, Luis and Madhusudhan, Nikku and Allard, Nicole F. and Hubeny, Ivan and Spiegelman, Fernand and Leininger, Thierry},
	month = dec,
	year = {2019},
	note = {ADS Bibcode: 2019ApJ...887L..20W},
	pages = {L20},
}

@article{lecavelier_des_etangs_rayleigh_2008,
	title = {Rayleigh scattering in the transit spectrum of {HD} 189733b},
	volume = {481},
	copyright = {© ESO, 2008},
	issn = {0004-6361, 1432-0746},
	url = {https://www.aanda.org/articles/aa/abs/2008/14/aa09388-08/aa09388-08.html},
	doi = {10.1051/0004-6361:200809388},
	language = {en},
	number = {2},
	urldate = {2023-05-18},
	journal = {Astronomy \& Astrophysics},
	author = {Lecavelier des Etangs, A. and Pont, F. and Vidal-Madjar, A. and Sing, D.},
	month = apr,
	year = {2008},
	note = {Number: 2
Publisher: EDP Sciences},
	pages = {L83--L86},
}

@article{deming_infrared_2013,
	title = {Infrared {Transmission} {Spectroscopy} of the {Exoplanets} {HD} 209458b and {XO}-1b {Using} the {Wide} {Field} {Camera}-3 on the {Hubble} {Space} {Telescope}},
	volume = {774},
	issn = {0004-637X},
	url = {https://ui.adsabs.harvard.edu/abs/2013ApJ...774...95D},
	doi = {10.1088/0004-637X/774/2/95},
	urldate = {2023-05-22},
	journal = {The Astrophysical Journal},
	author = {Deming, Drake and Wilkins, Ashlee and McCullough, Peter and Burrows, Adam and Fortney, Jonathan J. and Agol, Eric and Dobbs-Dixon, Ian and Madhusudhan, Nikku and Crouzet, Nicolas and Desert, Jean-Michel and Gilliland, Ronald L. and Haynes, Korey and Knutson, Heather A. and Line, Michael and Magic, Zazralt and Mandell, Avi M. and Ranjan, Sukrit and Charbonneau, David and Clampin, Mark and Seager, Sara and Showman, Adam P.},
	month = sep,
	year = {2013},
	note = {ADS Bibcode: 2013ApJ...774...95D},
	pages = {95},
}

@article{wakeford_hat-p-26b_2017,
	title = {{HAT}-{P}-26b: {A} {Neptune}-mass exoplanet with a well-constrained heavy element abundance},
	volume = {356},
	issn = {0036-8075},
	shorttitle = {{HAT}-{P}-26b},
	url = {https://ui.adsabs.harvard.edu/abs/2017Sci...356..628W},
	doi = {10.1126/science.aah4668},
	urldate = {2023-05-22},
	journal = {Science},
	author = {Wakeford, Hannah R. and Sing, David K. and Kataria, Tiffany and Deming, Drake and Nikolov, Nikolay and Lopez, Eric D. and Tremblin, Pascal and Amundsen, David S. and Lewis, Nikole K. and Mandell, Avi M. and Fortney, Jonathan J. and Knutson, Heather and Benneke, Björn and Evans, Thomas M.},
	month = may,
	year = {2017},
	note = {ADS Bibcode: 2017Sci...356..628W},
	pages = {628--631},
}

@article{mccullough_water_2014,
	title = {Water {Vapor} in the {Spectrum} of the {Extrasolar} {Planet} {HD} 189733b. {I}. {The} {Transit}},
	volume = {791},
	issn = {0004-637X},
	url = {https://dx.doi.org/10.1088/0004-637X/791/1/55},
	doi = {10.1088/0004-637X/791/1/55},
	language = {en},
	number = {1},
	urldate = {2023-05-24},
	journal = {The Astrophysical Journal},
	author = {McCullough, P. R. and Crouzet, N. and Deming, D. and Madhusudhan, N.},
	month = jul,
	year = {2014},
	note = {Publisher: The American Astronomical Society},
	pages = {55},
}

@article{tsai_photochemically_2023,
	title = {Photochemically produced {SO2} in the atmosphere of {WASP}-39b},
	volume = {617},
	copyright = {2023 The Author(s)},
	issn = {1476-4687},
	url = {https://www.nature.com/articles/s41586-023-05902-2},
	doi = {10.1038/s41586-023-05902-2},
	language = {en},
	number = {7961},
	urldate = {2023-09-27},
	journal = {Nature},
	author = {Tsai, Shang-Min and Lee, Elspeth K. H. and Powell, Diana and Gao, Peter and Zhang, Xi and Moses, Julianne and Hébrard, Eric and Venot, Olivia and Parmentier, Vivien and Jordan, Sean and Hu, Renyu and Alam, Munazza K. and Alderson, Lili and Batalha, Natalie M. and Bean, Jacob L. and Benneke, Björn and Bierson, Carver J. and Brady, Ryan P. and Carone, Ludmila and Carter, Aarynn L. and Chubb, Katy L. and Inglis, Julie and Leconte, Jérémy and Line, Michael and López-Morales, Mercedes and Miguel, Yamila and Molaverdikhani, Karan and Rustamkulov, Zafar and Sing, David K. and Stevenson, Kevin B. and Wakeford, Hannah R. and Yang, Jeehyun and Aggarwal, Keshav and Baeyens, Robin and Barat, Saugata and de Val-Borro, Miguel and Daylan, Tansu and Fortney, Jonathan J. and France, Kevin and Goyal, Jayesh M. and Grant, David and Kirk, James and Kreidberg, Laura and Louca, Amy and Moran, Sarah E. and Mukherjee, Sagnick and Nasedkin, Evert and Ohno, Kazumasa and Rackham, Benjamin V. and Redfield, Seth and Taylor, Jake and Tremblin, Pascal and Visscher, Channon and Wallack, Nicole L. and Welbanks, Luis and Youngblood, Allison and Ahrer, Eva-Maria and Batalha, Natasha E. and Behr, Patrick and Berta-Thompson, Zachory K. and Blecic, Jasmina and Casewell, S. L. and Crossfield, Ian J. M. and Crouzet, Nicolas and Cubillos, Patricio E. and Decin, Leen and Désert, Jean-Michel and Feinstein, Adina D. and Gibson, Neale P. and Harrington, Joseph and Heng, Kevin and Henning, Thomas and Kempton, Eliza M.-R. and Krick, Jessica and Lagage, Pierre-Olivier and Lendl, Monika and Lothringer, Joshua D. and Mansfield, Megan and Mayne, N. J. and Mikal-Evans, Thomas and Palle, Enric and Schlawin, Everett and Shorttle, Oliver and Wheatley, Peter J. and Yurchenko, Sergei N.},
	month = may,
	year = {2023},
	note = {Number: 7961
Publisher: Nature Publishing Group},
	pages = {483--487},
}

@article{taylor_awesome_2023,
	title = {Awesome {SOSS}: atmospheric characterization of {WASP}-96 b using the {JWST} early release observations},
	volume = {524},
	issn = {0035-8711},
	shorttitle = {Awesome {SOSS}},
	url = {https://ui.adsabs.harvard.edu/abs/2023MNRAS.524..817T},
	doi = {10.1093/mnras/stad1547},
	urldate = {2023-10-10},
	journal = {Monthly Notices of the Royal Astronomical Society},
	author = {Taylor, Jake and Radica, Michael and Welbanks, Luis and MacDonald, Ryan J. and Blecic, Jasmina and Zamyatina, Maria and Roth, Alexander and Bean, Jacob L. and Parmentier, Vivien and Coulombe, Louis-Philippe and Feinstein, Adina D. and Espinoza, Néstor and Benneke, Björn and Lafrenière, David and Doyon, René and Ahrer, Eva-Maria},
	month = sep,
	year = {2023},
	note = {ADS Bibcode: 2023MNRAS.524..817T},
	pages = {817--834},
}

@article{fu_water_2022,
	title = {Water and an {Escaping} {Helium} {Tail} {Detected} in the {Hazy} and {Methane}-depleted {Atmosphere} of {HAT}-{P}-18b from {JWST} {NIRISS}/{SOSS}},
	volume = {940},
	issn = {2041-8205},
	url = {https://dx.doi.org/10.3847/2041-8213/ac9977},
	doi = {10.3847/2041-8213/ac9977},
	language = {en},
	number = {2},
	urldate = {2023-10-13},
	journal = {The Astrophysical Journal Letters},
	author = {Fu, Guangwei and Espinoza, Néstor and Sing, David K. and Lothringer, Joshua D. and Santos, Leonardo A. Dos and Rustamkulov, Zafar and Deming, Drake and Kempton, Eliza M.-R. and Komacek, Thaddeus D. and Knutson, Heather A. and Albert, Loïc and Pontoppidan, Klaus and Volk, Kevin and Filippazzo, Joseph},
	month = nov,
	year = {2022},
	note = {Publisher: The American Astronomical Society},
	pages = {L35},
}

@article{welbanks_application_2023,
	title = {On the {Application} of {Bayesian} {Leave}-one-out {Cross}-validation to {Exoplanet} {Atmospheric} {Analysis}},
	volume = {165},
	issn = {0004-6256, 1538-3881},
	url = {https://iopscience.iop.org/article/10.3847/1538-3881/acab67},
	doi = {10.3847/1538-3881/acab67},
	language = {en},
	number = {3},
	urldate = {2024-02-13},
	journal = {The Astronomical Journal},
	author = {Welbanks, Luis and McGill, Peter and Line, Michael and Madhusudhan, Nikku},
	month = mar,
	year = {2023},
	pages = {112},
}

@article{buchner_x-ray_2014,
	title = {X-ray spectral modelling of the {AGN} obscuring region in the {CDFS}: {Bayesian} model selection and catalogue},
	volume = {564},
	copyright = {© ESO, 2014},
	issn = {0004-6361, 1432-0746},
	shorttitle = {X-ray spectral modelling of the {AGN} obscuring region in the {CDFS}},
	url = {https://www.aanda.org/articles/aa/abs/2014/04/aa22971-13/aa22971-13.html},
	doi = {10.1051/0004-6361/201322971},
	language = {en},
	urldate = {2024-02-19},
	journal = {Astronomy \& Astrophysics},
	author = {Buchner, J. and Georgakakis, A. and Nandra, K. and Hsu, L. and Rangel, C. and Brightman, M. and Merloni, A. and Salvato, M. and Donley, J. and Kocevski, D.},
	month = apr,
	year = {2014},
	note = {Publisher: EDP Sciences},
	pages = {A125},
}

@article{allard_k-h2_2016,
	title = {K-{H2} line shapes for the spectra of cool brown dwarfs},
	volume = {589},
	issn = {0004-6361},
	url = {https://ui.adsabs.harvard.edu/abs/2016A&A...589A..21A},
	doi = {10.1051/0004-6361/201628270},
	urldate = {2024-02-29},
	journal = {Astronomy and Astrophysics},
	author = {Allard, N. F. and Spiegelman, F. and Kielkopf, J. F.},
	month = may,
	year = {2016},
	note = {ADS Bibcode: 2016A\&A...589A..21A},
	pages = {A21},
}

@misc{fairman_importance_2024,
	title = {The {Importance} of {Optical} {Wavelength} {Data} on {Atmospheric} {Retrievals} of {Exoplanet} {Transmission} {Spectra}},
	url = {https://arxiv.org/abs/2403.07801},
	language = {en},
	urldate = {2024-03-14},
	publisher = {arXiv},
	author = {Fairman, Charlotte and Wakeford, Hannah R. and MacDonald, Ryan J.},
	month = mar,
	year = {2024},
}

@incollection{madhusudhan_atmospheric_2018,
	title = {Atmospheric {Retrieval} of {Exoplanets}},
	url = {http://arxiv.org/abs/1808.04824},
    booktitle = {Springer Handbook of Exoplanets},
	urldate = {2024-03-14},
	author = {Madhusudhan, Nikku},
	year = {2018},
	doi = {10.1007/978-3-319-55333-7_104},
	note = {arXiv:1808.04824 [astro-ph]},
	pages = {2153--2182},
}

@article{sing_hubble_2011,
	title = {Hubble {Space} {Telescope} transmission spectroscopy of the exoplanet {HD} 189733b: high-altitude atmospheric haze in the optical and near-ultraviolet with {STIS}},
	volume = {416},
	issn = {0035-8711},
	shorttitle = {Hubble {Space} {Telescope} transmission spectroscopy of the exoplanet {HD} 189733b},
	url = {https://ui.adsabs.harvard.edu/abs/2011MNRAS.416.1443S},
	doi = {10.1111/j.1365-2966.2011.19142.x},
	urldate = {2024-03-18},
	journal = {Monthly Notices of the Royal Astronomical Society},
	author = {Sing, D. K. and Pont, F. and Aigrain, S. and Charbonneau, D. and Désert, J. -M. and Gibson, N. and Gilliland, R. and Hayek, W. and Henry, G. and Knutson, H. and Lecavelier Des Etangs, A. and Mazeh, T. and Shporer, A.},
	month = sep,
	year = {2011},
	note = {ADS Bibcode: 2011MNRAS.416.1443S},
	pages = {1443--1455},
}

@article{pinhas_retrieval_2018,
	title = {Retrieval of planetary and stellar properties in transmission spectroscopy with {Aura}},
	volume = {480},
	issn = {0035-8711},
	url = {https://doi.org/10.1093/mnras/sty2209},
	doi = {10.1093/mnras/sty2209},
	number = {4},
	urldate = {2024-03-25},
	journal = {Monthly Notices of the Royal Astronomical Society},
	author = {Pinhas, Arazi and Rackham, Benjamin V and Madhusudhan, Nikku and Apai, Dániel},
	month = nov,
	year = {2018},
	pages = {5314--5331},
}

@article{rothman_hitemp_2010,
	title = {{HITEMP}, the high-temperature molecular spectroscopic database},
	volume = {111},
	issn = {0022-4073},
	url = {https://ui.adsabs.harvard.edu/abs/2010JQSRT.111.2139R},
	doi = {10.1016/j.jqsrt.2010.05.001},
	urldate = {2024-04-02},
	journal = {Journal of Quantitative Spectroscopy and Radiative Transfer},
	author = {Rothman, L. S. and Gordon, I. E. and Barber, R. J. and Dothe, H. and Gamache, R. R. and Goldman, A. and Perevalov, V. I. and Tashkun, S. A. and Tennyson, J.},
	month = oct,
	year = {2010},
	note = {ADS Bibcode: 2010JQSRT.111.2139R},
	pages = {2139--2150},
}

@article{mordasini_imprint_2016,
	title = {The imprint of exoplanet formation history on observable present-day spectra of hot {Jupiters}},
	volume = {832},
	issn = {0004-637X},
	url = {https://dx.doi.org/10.3847/0004-637X/832/1/41},
	doi = {10.3847/0004-637X/832/1/41},
	language = {en},
	number = {1},
	urldate = {2024-04-09},
	journal = {The Astrophysical Journal},
	author = {Mordasini, C. and Boekel, R. van and Mollière, P. and Henning, Th and Benneke, Björn},
	month = nov,
	year = {2016},
	note = {Publisher: The American Astronomical Society},
	pages = {41},
}

@article{vidal-madjar_extended_2003,
	title = {An extended upper atmosphere around the extrasolar planet {HD209458b}},
	volume = {422},
	issn = {0028-0836},
	url = {https://ui.adsabs.harvard.edu/abs/2003Natur.422..143V},
	doi = {10.1038/nature01448},
	urldate = {2024-05-02},
	journal = {Nature},
	author = {Vidal-Madjar, A. and Lecavelier des Etangs, A. and Désert, J. -M. and Ballester, G. E. and Ferlet, R. and Hébrard, G. and Mayor, M.},
	month = mar,
	year = {2003},
	note = {ADS Bibcode: 2003Natur.422..143V},
	pages = {143--146},
}

@article{underwood_exomol_2016,
	title = {{ExoMol} molecular line lists – {XIV}. {The} rotation–vibration spectrum of hot {SO2}},
	volume = {459},
	issn = {0035-8711},
	url = {https://doi.org/10.1093/mnras/stw849},
	doi = {10.1093/mnras/stw849},
	number = {4},
	urldate = {2024-08-01},
	journal = {Monthly Notices of the Royal Astronomical Society},
	author = {Underwood, Daniel S. and Tennyson, Jonathan and Yurchenko, Sergei N. and Huang, Xinchuan and Schwenke, David W. and Lee, Timothy J. and Clausen, Sønnik and Fateev, Alexander},
	month = jul,
	year = {2016},
	pages = {3890--3899},
}

@article{bell_methane_2023,
	title = {Methane throughout the atmosphere of the warm exoplanet {WASP}-80b},
	volume = {623},
	copyright = {2023 The Author(s), under exclusive licence to Springer Nature Limited},
	issn = {1476-4687},
	url = {https://www.nature.com/articles/s41586-023-06687-0},
	doi = {10.1038/s41586-023-06687-0},
	language = {en},
	number = {7988},
	urldate = {2024-09-06},
	journal = {Nature},
	author = {Bell, Taylor J. and Welbanks, Luis and Schlawin, Everett and Line, Michael R. and Fortney, Jonathan J. and Greene, Thomas P. and Ohno, Kazumasa and Parmentier, Vivien and Rauscher, Emily and Beatty, Thomas G. and Mukherjee, Sagnick and Wiser, Lindsey S. and Boyer, Martha L. and Rieke, Marcia J. and Stansberry, John A.},
	month = nov,
	year = {2023},
	note = {Publisher: Nature Publishing Group},
	pages = {709--712},
}

@article{wakeford_transmission_2015,
	title = {Transmission spectral properties of clouds for hot {Jupiter} exoplanets},
	volume = {573},
	issn = {0004-6361},
	url = {https://ui.adsabs.harvard.edu/abs/2015A&A...573A.122W},
	doi = {10.1051/0004-6361/201424207},
	urldate = {2024-09-10},
	journal = {Astronomy and Astrophysics},
	author = {Wakeford, H. R. and Sing, D. K.},
	month = jan,
	year = {2015},
	note = {ADS Bibcode: 2015A\&A...573A.122W},
	pages = {A122},
}

@article{pinhas_signatures_2017,
	title = {On signatures of clouds in exoplanetary transit spectra},
	volume = {471},
	issn = {0035-8711},
	url = {https://ui.adsabs.harvard.edu/abs/2017MNRAS.471.4355P},
	doi = {10.1093/mnras/stx1849},
	urldate = {2024-09-10},
	journal = {Monthly Notices of the Royal Astronomical Society},
	author = {Pinhas, Arazi and Madhusudhan, Nikku},
	month = nov,
	year = {2017},
	note = {Publisher: OUP
ADS Bibcode: 2017MNRAS.471.4355P},
	pages = {4355--4373},
}

@misc{inglis_quartz_2024,
	title = {Quartz {Clouds} in the {Dayside} {Atmosphere} of the {Quintessential} {Hot} {Jupiter} {HD} 189733 b},
	url = {http://arxiv.org/abs/2409.11395},
	doi = {10.3847/2041-8213/ad725e},
	language = {en},
	urldate = {2024-09-18},
	author = {Inglis, Julie and Batalha, Natasha E. and Lewis, Nikole K. and Kataria, Tiffany and Knutson, Heather A. and Kilpatrick, Brian M. and Gagnebin, Anna and Mukherjee, Sagnick and Pettyjohn, Maria M. and Crossfield, Ian J. M. and Foote, Trevor O. and Grant, David and Henry, Gregory W. and Lally, Maura and McKemmish, Laura K. and Sing, David K. and Wakeford, Hannah R. and Trujillo, Juan C. Zapata and Zellem, Robert T.},
	month = sep,
	year = {2024},
	note = {arXiv:2409.11395 [astro-ph]},
}

@article{dyrek_so2_2024,
	title = {{SO2}, silicate clouds, but no {CH4} detected in a warm {Neptune}},
	volume = {625},
	issn = {0028-0836},
	url = {https://ui.adsabs.harvard.edu/abs/2024Natur.625...51D},
	doi = {10.1038/s41586-023-06849-0},
	urldate = {2024-10-12},
	journal = {Nature},
	author = {Dyrek, Achrène and Min, Michiel and Decin, Leen and Bouwman, Jeroen and Crouzet, Nicolas and Mollière, Paul and Lagage, Pierre-Olivier and Konings, Thomas and Tremblin, Pascal and Güdel, Manuel and Pye, John and Waters, Rens and Henning, Thomas and Vandenbussche, Bart and Ardevol Martinez, Francisco and Argyriou, Ioannis and Ducrot, Elsa and Heinke, Linus and van Looveren, Gwenael and Absil, Olivier and Barrado, David and Baudoz, Pierre and Boccaletti, Anthony and Cossou, Christophe and Coulais, Alain and Edwards, Billy and Gastaud, René and Glasse, Alistair and Glauser, Adrian and Greene, Thomas P. and Kendrew, Sarah and Krause, Oliver and Lahuis, Fred and Mueller, Michael and Olofsson, Goran and Patapis, Polychronis and Rouan, Daniel and Royer, Pierre and Scheithauer, Silvia and Waldmann, Ingo and Whiteford, Niall and Colina, Luis and van Dishoeck, Ewine F. and Östlin, Göran and Ray, Tom P. and Wright, Gillian},
	month = jan,
	year = {2024},
	note = {ADS Bibcode: 2024Natur.625...51D},
	pages = {51--54},
}

@article{welbanks_high_2024,
	title = {A high internal heat flux and large core in a warm {Neptune} exoplanet},
	volume = {630},
	copyright = {2024 The Author(s), under exclusive licence to Springer Nature Limited},
	issn = {1476-4687},
	url = {https://www.nature.com/articles/s41586-024-07514-w},
	doi = {10.1038/s41586-024-07514-w},
	language = {en},
	number = {8018},
	urldate = {2024-10-15},
	journal = {Nature},
	author = {Welbanks, Luis and Bell, Taylor J. and Beatty, Thomas G. and Line, Michael R. and Ohno, Kazumasa and Fortney, Jonathan J. and Schlawin, Everett and Greene, Thomas P. and Rauscher, Emily and McGill, Peter and Murphy, Matthew and Parmentier, Vivien and Tang, Yao and Edelman, Isaac and Mukherjee, Sagnick and Wiser, Lindsey S. and Lagage, Pierre-Olivier and Dyrek, Achrène and Arnold, Kenneth E.},
	month = jun,
	year = {2024},
	note = {Publisher: Nature Publishing Group},
	pages = {836--840},
}

@misc{quintana_pandora_2021,
	title = {The {Pandora} {SmallSat}: {Multiwavelength} {Characterization} of {Exoplanets} and their {Host} {Stars}},
	shorttitle = {The {Pandora} {SmallSat}},
	url = {http://arxiv.org/abs/2108.06438},
	doi = {10.48550/arXiv.2108.06438},
	urldate = {2024-12-03},
	publisher = {arXiv},
	author = {Quintana, Elisa V. and Colón, Knicole D. and Mosby, Gregory and Schlieder, Joshua E. and Supsinskas, Pete and Karburn, Jordan and Dotson, Jessie L. and Greene, Thomas P. and Hedges, Christina and Apai, Dániel and Barclay, Thomas and Christiansen, Jessie L. and Espinoza, Néstor and Mullally, Susan E. and Gilbert, Emily A. and Hoffman, Kelsey and Kostov, Veselin B. and Lewis, Nikole K. and Foote, Trevor O. and Mason, James and Youngblood, Allison and Morris, Brett M. and Newton, Elisabeth R. and Pepper, Joshua and Rackham, Benjamin V. and Rowe, Jason F. and Stevenson, Kevin},
	month = aug,
	year = {2021},
	note = {arXiv:2108.06438},
}

@article{rackham_transit_2018,
       author = {{Rackham}, Benjamin V. and {Apai}, D{\'a}niel and {Giampapa}, Mark S.},
        title = "{The Transit Light Source Effect: False Spectral Features and Incorrect Densities for M-dwarf Transiting Planets}",
      journal = {\apj},
         year = 2018,
        month = feb,
       volume = {853},
       number = {2},
          eid = {122},
        pages = {122},
          doi = {10.3847/1538-4357/aaa08c},
archivePrefix = {arXiv},
       eprint = {1711.05691},
 primaryClass = {astro-ph.EP},
       adsurl = {https://ui.adsabs.harvard.edu/abs/2018ApJ...853..122R}
}

@ARTICLE{rackham_transit_2019,
       author = {{Rackham}, Benjamin V. and {Apai}, D{\'a}niel and {Giampapa}, Mark S.},
        title = "{The Transit Light Source Effect. II. The Impact of Stellar Heterogeneity on Transmission Spectra of Planets Orbiting Broadly Sun-like Stars}",
      journal = {\aj},
         year = 2019,
        month = mar,
       volume = {157},
       number = {3},
          eid = {96},
        pages = {96},
          doi = {10.3847/1538-3881/aaf892},
archivePrefix = {arXiv},
       eprint = {1812.06184},
 primaryClass = {astro-ph.EP},
       adsurl = {https://ui.adsabs.harvard.edu/abs/2019AJ....157...96R}
}

@ARTICLE{rackham_sag21_2023,
       author = {{Rackham}, Benjamin V. and {Espinoza}, N{\'e}stor and {Berdyugina}, Svetlana V. and {Korhonen}, Heidi and {MacDonald}, Ryan J. and {Montet}, Benjamin T. and {Morris}, Brett M. and {Oshagh}, Mahmoudreza and {Shapiro}, Alexander I. and {Unruh}, Yvonne C. and {Quintana}, Elisa V. and {Zellem}, Robert T. and {Apai}, D{\'a}niel and {Barclay}, Thomas and {Barstow}, Joanna K. and {Bruno}, Giovanni and {Carone}, Ludmila and {Casewell}, Sarah L. and {Cegla}, Heather M. and {Criscuoli}, Serena and {Fischer}, Catherine and {Fournier}, Damien and {Giampapa}, Mark S. and {Giles}, Helen and {Iyer}, Aishwarya and {Kopp}, Greg and {Kostogryz}, Nadiia M. and {Krivova}, Natalie and {Mallonn}, Matthias and {McGruder}, Chima and {Molaverdikhani}, Karan and {Newton}, Elisabeth R. and {Panja}, Mayukh and {Peacock}, Sarah and {Reardon}, Kevin and {Roettenbacher}, Rachael M. and {Scandariato}, Gaetano and {Solanki}, Sami and {Stassun}, Keivan G. and {Steiner}, Oskar and {Stevenson}, Kevin B. and {Tregloan-Reed}, Jeremy and {Valio}, Adriana and {Wedemeyer}, Sven and {Welbanks}, Luis and {Yu}, Jie and {Alam}, Munazza K. and {Davenport}, James R.~A. and {Deming}, Drake and {Dong}, Chuanfei and {Ducrot}, Elsa and {Fisher}, Chloe and {Gilbert}, Emily and {Kostov}, Veselin and {L{\'o}pez-Morales}, Mercedes and {Line}, Mike and {Mo{\v{c}}nik}, Teo and {Mullally}, Susan and {Paudel}, Rishi R. and {Ribas}, Ignasi and {Valenti}, Jeff A.},
        title = "{The effect of stellar contamination on low-resolution transmission spectroscopy: needs identified by NASA's Exoplanet Exploration Program Study Analysis Group 21}",
      journal = {RAS Techniques and Instruments},
         year = 2023,
        month = jan,
       volume = {2},
       number = {1},
        pages = {148-206},
          doi = {10.1093/rasti/rzad009},
archivePrefix = {arXiv},
       eprint = {2201.09905},
 primaryClass = {astro-ph.IM},
       adsurl = {https://ui.adsabs.harvard.edu/abs/2023RASTI...2..148R}
}

@misc{fournier-tondreau_transmission_2024,
	title = {Transmission spectroscopy of {WASP}-52 b with {JWST} {NIRISS}: {Water} and helium atmospheric absorption, alongside prominent star-spot crossings},
	shorttitle = {Transmission spectroscopy of {WASP}-52 b with {JWST} {NIRISS}},
	url = {http://arxiv.org/abs/2412.17072},
	doi = {10.48550/arXiv.2412.17072},
	urldate = {2024-12-25},
	publisher = {arXiv},
	author = {Fournier-Tondreau, Marylou and Pan, Yanbo and Morel, Kim and Lafrenière, David and MacDonald, Ryan J. and Coulombe, Louis-Philippe and Allart, Romain and Albert, Loïc and Radica, Michael and Piaulet-Ghorayeb, Caroline and Roy, Pierre-Alexis and Pelletier, Stefan and Dang, Lisa and Doyon, René and Benneke, Björn and Cowan, Nicolas B. and Darveau-Bernier, Antoine and Lim, Olivia and Artigau, Étienne and Johnstone, Doug and Kaltenegger, Lisa and Taylor, Jake and Flagg, Laura},
	month = dec,
	year = {2024},
	note = {arXiv:2412.17072 [astro-ph]},
}

@article{moran_high_2023,
	title = {High {Tide} or {Riptide} on the {Cosmic} {Shoreline}? {A} {Water}-rich {Atmosphere} or {Stellar} {Contamination} for the {Warm} {Super}-{Earth} {GJ} 486b from {JWST} {Observations}},
	volume = {948},
	issn = {2041-8205},
	shorttitle = {High {Tide} or {Riptide} on the {Cosmic} {Shoreline}?},
	url = {https://dx.doi.org/10.3847/2041-8213/accb9c},
	doi = {10.3847/2041-8213/accb9c},
	language = {en},
	number = {1},
	urldate = {2024-12-27},
	journal = {The Astrophysical Journal Letters},
	author = {Moran, Sarah E. and Stevenson, Kevin B. and Sing, David K. and MacDonald, Ryan J. and Kirk, James and Lustig-Yaeger, Jacob and Peacock, Sarah and Mayorga, L. C. and Bennett, Katherine A. and López-Morales, Mercedes and May, E. M. and Rustamkulov, Zafar and Valenti, Jeff A. and Redai, Jéa I. Adams and Alam, Munazza K. and Batalha, Natasha E. and Fu, Guangwei and Gonzalez-Quiles, Junellie and Highland, Alicia N. and Kruse, Ethan and Lothringer, Joshua D. and Ceballos, Kevin N. Ortiz and Sotzen, Kristin S. and Wakeford, Hannah R.},
	month = may,
	year = {2023},
	note = {Publisher: The American Astronomical Society},
	pages = {L11},
}

@article{lim_atmospheric_2023,
	title = {Atmospheric {Reconnaissance} of {TRAPPIST}-1 b with {JWST}/{NIRISS}: {Evidence} for {Strong} {Stellar} {Contamination} in the {Transmission} {Spectra}},
	volume = {955},
	issn = {2041-8205},
	shorttitle = {Atmospheric {Reconnaissance} of {TRAPPIST}-1 b with {JWST}/{NIRISS}},
	url = {https://dx.doi.org/10.3847/2041-8213/acf7c4},
	doi = {10.3847/2041-8213/acf7c4},
	language = {en},
	number = {1},
	urldate = {2024-12-27},
	journal = {The Astrophysical Journal Letters},
	author = {Lim, Olivia and Benneke, Björn and Doyon, René and MacDonald, Ryan J. and Piaulet, Caroline and Artigau, Étienne and Coulombe, Louis-Philippe and Radica, Michael and L’Heureux, Alexandrine and Albert, Loïc and Rackham, Benjamin V. and Wit, Julien de and Salhi, Salma and Roy, Pierre-Alexis and Flagg, Laura and Fournier-Tondreau, Marylou and Taylor, Jake and Cook, Neil J. and Lafrenière, David and Cowan, Nicolas B. and Kaltenegger, Lisa and Rowe, Jason F. and Espinoza, Néstor and Dang, Lisa and Darveau-Bernier, Antoine},
	month = sep,
	year = {2023},
	note = {Publisher: The American Astronomical Society},
	pages = {L22},
}

@article{wallack_jwst_2024,
	title = {{JWST} {COMPASS}: {A} {NIRSpec}/{G395H} {Transmission} {Spectrum} of the {Sub}-{Neptune} {TOI}-836c},
	volume = {168},
	issn = {0004-6256},
	shorttitle = {{JWST} {COMPASS}},
	url = {https://ui.adsabs.harvard.edu/abs/2024AJ....168...77W},
	doi = {10.3847/1538-3881/ad3917},
	urldate = {2024-12-29},
	journal = {The Astronomical Journal},
	author = {Wallack, Nicole L. and Batalha, Natasha E. and Alderson, Lili and Scarsdale, Nicholas and Adams Redai, Jea I. and Aguichine, Artyom and Alam, Munazza K. and Gao, Peter and Wolfgang, Angie and Batalha, Natalie M. and Kirk, James and López-Morales, Mercedes and Moran, Sarah E. and Teske, Johanna and Wakeford, Hannah R. and Wogan, Nicholas F.},
	month = aug,
	year = {2024},
	note = {Publisher: IOP
ADS Bibcode: 2024AJ....168...77W},
	pages = {77},
}

@article{tsiaras_population_2018,
	title = {A {Population} {Study} of {Gaseous} {Exoplanets}},
	volume = {155},
	issn = {0004-6256},
	url = {https://ui.adsabs.harvard.edu/abs/2018AJ....155..156T},
	doi = {10.3847/1538-3881/aaaf75},
	urldate = {2024-12-30},
	journal = {The Astronomical Journal},
	author = {Tsiaras, A. and Waldmann, I. P. and Zingales, T. and Rocchetto, M. and Morello, G. and Damiano, M. and Karpouzas, K. and Tinetti, G. and McKemmish, L. K. and Tennyson, J. and Yurchenko, S. N.},
	month = apr,
	year = {2018},
	note = {Publisher: IOP
ADS Bibcode: 2018AJ....155..156T},
	pages = {156},
}

@article{fournier-tondreau_near-infrared_2024,
	title = {Near-infrared transmission spectroscopy of {HAT}-{P}-18 b with {NIRISS}: {Disentangling} planetary and stellar features in the era of {JWST}},
	volume = {528},
	issn = {0035-8711},
	shorttitle = {Near-infrared transmission spectroscopy of {HAT}-{P}-18 b with {NIRISS}},
	url = {https://ui.adsabs.harvard.edu/abs/2024MNRAS.528.3354F},
	doi = {10.1093/mnras/stad3813},
	urldate = {2025-01-14},
	journal = {Monthly Notices of the Royal Astronomical Society},
	author = {Fournier-Tondreau, Marylou and MacDonald, Ryan J. and Radica, Michael and Lafrenière, David and Welbanks, Luis and Piaulet, Caroline and Coulombe, Louis-Philippe and Allart, Romain and Morel, Kim and Artigau, Étienne and Albert, Loïc and Lim, Olivia and Doyon, René and Benneke, Björn and Rowe, Jason F. and Darveau-Bernier, Antoine and Cowan, Nicolas B. and Lewis, Nikole K. and Cook, Neil J. and Flagg, Laura and Genest, Frédéric and Pelletier, Stefan and Johnstone, Doug and Dang, Lisa and Kaltenegger, Lisa and Taylor, Jake and Turner, Jake D.},
	month = feb,
	year = {2024},
	note = {Publisher: OUP
ADS Bibcode: 2024MNRAS.528.3354F},
	pages = {3354--3377},
}

@article{cadieux_transmission_2024,
	title = {Transmission {Spectroscopy} of the {Habitable} {Zone} {Exoplanet} {LHS} 1140 b with {JWST}/{NIRISS}},
	volume = {970},
	issn = {0004-637X},
	url = {https://ui.adsabs.harvard.edu/abs/2024ApJ...970L...2C},
	doi = {10.3847/2041-8213/ad5afa},
	urldate = {2025-01-14},
	journal = {The Astrophysical Journal},
	author = {Cadieux, Charles and Doyon, René and MacDonald, Ryan J. and Turbet, Martin and Artigau, Étienne and Lim, Olivia and Radica, Michael and Fauchez, Thomas J. and Salhi, Salma and Dang, Lisa and Albert, Loïc and Coulombe, Louis-Philippe and Cowan, Nicolas B. and Lafrenière, David and L'Heureux, Alexandrine and Piaulet-Ghorayeb, Caroline and Benneke, Björn and Cloutier, Ryan and Charnay, Benjamin and Cook, Neil J. and Fournier-Tondreau, Marylou and Plotnykov, Mykhaylo and Valencia, Diana},
	month = jul,
	year = {2024},
	note = {Publisher: IOP
ADS Bibcode: 2024ApJ...970L...2C},
	pages = {L2},
}

@article{rackham_toward_2024,
	title = {Toward {Robust} {Corrections} for {Stellar} {Contamination} in {JWST} {Exoplanet} {Transmission} {Spectra}},
	volume = {168},
	issn = {0004-6256},
	url = {https://ui.adsabs.harvard.edu/abs/2024AJ....168...82R},
	doi = {10.3847/1538-3881/ad5833},
	urldate = {2025-01-14},
	journal = {The Astronomical Journal},
	author = {Rackham, Benjamin V. and de Wit, Julien},
	month = aug,
	year = {2024},
	note = {Publisher: IOP
ADS Bibcode: 2024AJ....168...82R},
	pages = {82},
}

@article{madhusudhan_carbon-bearing_2023,
	title = {Carbon-bearing {Molecules} in a {Possible} {Hycean} {Atmosphere}},
	volume = {956},
	issn = {0004-637X},
	url = {https://ui.adsabs.harvard.edu/abs/2023ApJ...956L..13M},
	doi = {10.3847/2041-8213/acf577},
	urldate = {2025-01-17},
	journal = {The Astrophysical Journal},
	author = {Madhusudhan, Nikku and Sarkar, Subhajit and Constantinou, Savvas and Holmberg, Måns and Piette, Anjali A. A. and Moses, Julianne I.},
	month = oct,
	year = {2023},
	note = {Publisher: IOP
ADS Bibcode: 2023ApJ...956L..13M},
	pages = {L13},
}

@article{husser_new_2013,
	title = {A new extensive library of {PHOENIX} stellar atmospheres and synthetic spectra},
	volume = {553},
	issn = {0004-6361},
	url = {https://ui.adsabs.harvard.edu/abs/2013A&A...553A...6H},
	doi = {10.1051/0004-6361/201219058},
	urldate = {2025-01-27},
	journal = {Astronomy and Astrophysics},
	author = {Husser, T. -O. and Wende-von Berg, S. and Dreizler, S. and Homeier, D. and Reiners, A. and Barman, T. and Hauschildt, P. H.},
	month = may,
	year = {2013},
	note = {ADS Bibcode: 2013A\&A...553A...6H},
	pages = {A6},
}

@article{benneke_how_2013,
	title = {How to {Distinguish} between {Cloudy} {Mini}-{Neptunes} and {Water}/{Volatile}-dominated {Super}-{Earths}},
	volume = {778},
	issn = {0004-637X},
	url = {https://ui.adsabs.harvard.edu/abs/2013ApJ...778..153B},
	doi = {10.1088/0004-637X/778/2/153},
	urldate = {2025-02-05},
	journal = {The Astrophysical Journal},
	author = {Benneke, Björn and Seager, Sara},
	month = dec,
	year = {2013},
	note = {Publisher: IOP
ADS Bibcode: 2013ApJ...778..153B},
	pages = {153},
}

@article{may_double_2023,
	title = {Double {Trouble}: {Two} {Transits} of the {Super}-{Earth} {GJ} 1132 b {Observed} with {JWST} {NIRSpec} {G395H}},
	volume = {959},
	issn = {2041-8205},
	shorttitle = {Double {Trouble}},
	url = {https://dx.doi.org/10.3847/2041-8213/ad054f},
	doi = {10.3847/2041-8213/ad054f},
	language = {en},
	number = {1},
	urldate = {2025-02-13},
	journal = {The Astrophysical Journal Letters},
	author = {May, E. M. and MacDonald, Ryan J. and Bennett, Katherine A. and Moran, Sarah E. and Wakeford, Hannah R. and Peacock, Sarah and Lustig-Yaeger, Jacob and Highland, Alicia N. and Stevenson, Kevin B. and Sing, David K. and Mayorga, L. C. and Batalha, Natasha E. and Kirk, James and López-Morales, Mercedes and Valenti, Jeff A. and Alam, Munazza K. and Alderson, Lili and Fu, Guangwei and Gonzalez-Quiles, Junellie and Lothringer, Joshua D. and Rustamkulov, Zafar and Sotzen, Kristin S.},
	month = dec,
	year = {2023},
	note = {Publisher: The American Astronomical Society},
	pages = {L9},
}

@article{carter_benchmark_2024,
	title = {A benchmark {JWST} near-infrared spectrum for the exoplanet {WASP}-39 b},
	volume = {8},
	copyright = {2024 The Author(s)},
	issn = {2397-3366},
	url = {https://www.nature.com/articles/s41550-024-02292-x},
	doi = {10.1038/s41550-024-02292-x},
	language = {en},
	number = {8},
	urldate = {2025-02-13},
	journal = {Nature Astronomy},
	author = {Carter, A. L. and May, E. M. and Espinoza, N. and Welbanks, L. and Ahrer, E. and Alderson, L. and Brahm, R. and Feinstein, A. D. and Grant, D. and Line, M. and Morello, G. and O’Steen, R. and Radica, M. and Rustamkulov, Z. and Stevenson, K. B. and Turner, J. D. and Alam, M. K. and Anderson, D. R. and Batalha, N. M. and Battley, M. P. and Bayliss, D. and Bean, J. L. and Benneke, B. and Berta-Thompson, Z. K. and Brande, J. and Bryant, E. M. and Burleigh, M. R. and Coulombe, L. and Crossfield, I. J. M. and Damiano, M. and Désert, J.-M. and Flagg, L. and Gill, S. and Inglis, J. and Kirk, J. and Knutson, H. and Kreidberg, L. and López Morales, M. and Mansfield, M. and Moran, S. E. and Murray, C. A. and Nixon, M. C. and Petit dit de la Roche, D. J. M. and Rackham, B. V. and Schlawin, E. and Sing, D. K. and Wakeford, H. R. and Wallack, N. L. and Wheatley, P. J. and Zieba, S. and Aggarwal, K. and Barstow, J. K. and Bell, T. J. and Blecic, J. and Caceres, C. and Crouzet, N. and Cubillos, P. E. and Daylan, T. and de Val-Borro, M. and Decin, L. and Fortney, J. J. and Gibson, N. P. and Heng, K. and Hu, R. and Kempton, E. M.-R. and Lagage, P. and Lothringer, J. D. and Lustig-Yaeger, J. and Mancini, L. and Mayne, N. J. and Mayorga, L. C. and Molaverdikhani, K. and Nasedkin, E. and Ohno, K. and Parmentier, V. and Powell, D. and Redfield, S. and Roy, P. and Taylor, J. and Zhang, X.},
	month = aug,
	year = {2024},
	note = {Publisher: Nature Publishing Group},
	pages = {1008--1019},
}

@article{benneke_atmospheric_2012,
	title = {Atmospheric {Retrieval} for {Super}-{Earths}: {Uniquely} {Constraining} the {Atmospheric} {Composition} with {Transmission} {Spectroscopy}},
	volume = {753},
	issn = {0004-637X},
	shorttitle = {Atmospheric {Retrieval} for {Super}-{Earths}},
	url = {https://ui.adsabs.harvard.edu/abs/2012ApJ...753..100B},
	doi = {10.1088/0004-637X/753/2/100},
	urldate = {2025-02-18},
	journal = {The Astrophysical Journal},
	author = {Benneke, Bjoern and Seager, Sara},
	month = jul,
	year = {2012},
	note = {Publisher: IOP
ADS Bibcode: 2012ApJ...753..100B},
	pages = {100},
}

@article{evans_uniform_2015,
	title = {A uniform analysis of {HD} 209458b {Spitzer}/{IRAC} light curves with {Gaussian} process models},
	volume = {451},
	issn = {0035-8711},
	url = {https://doi.org/10.1093/mnras/stv910},
	doi = {10.1093/mnras/stv910},
	number = {1},
	urldate = {2025-02-24},
	journal = {Monthly Notices of the Royal Astronomical Society},
	author = {Evans, Thomas M. and Aigrain, Suzanne and Gibson, Neale and Barstow, Joanna K. and Amundsen, David S. and Tremblin, Pascal and Mourier, Pierre},
	month = jul,
	year = {2015},
	pages = {680--694},
}

@article{radica_promise_2025,
	title = {Promise and {Peril}: {Stellar} {Contamination} and {Strict} {Limits} on the {Atmosphere} {Composition} of {TRAPPIST}-1 c from {JWST} {NIRISS} {Transmission} {Spectra}},
	volume = {979},
	issn = {0004-637X},
	shorttitle = {Promise and {Peril}},
	url = {https://ui.adsabs.harvard.edu/abs/2025ApJ...979L...5R},
	doi = {10.3847/2041-8213/ada381},
	urldate = {2025-03-10},
	journal = {The Astrophysical Journal},
	author = {Radica, Michael and Piaulet-Ghorayeb, Caroline and Taylor, Jake and Coulombe, Louis-Philippe and Benneke, Björn and Albert, Loic and Artigau, Étienne and Cowan, Nicolas B. and Doyon, René and Lafrenière, David and L'Heureux, Alexandrine and Lim, Olivia},
	month = jan,
	year = {2025},
	note = {Publisher: IOP
ADS Bibcode: 2025ApJ...979L...5R},
	pages = {L5},
}

@article{madhusudhan_interior_2020,
	title = {The {Interior} and {Atmosphere} of the {Habitable}-zone {Exoplanet} {K2}-18b},
	volume = {891},
	issn = {0004-637X},
	url = {https://ui.adsabs.harvard.edu/abs/2020ApJ...891L...7M},
	doi = {10.3847/2041-8213/ab7229},
	urldate = {2025-03-11},
	journal = {The Astrophysical Journal},
	author = {Madhusudhan, Nikku and Nixon, Matthew C. and Welbanks, Luis and Piette, Anjali A. A. and Booth, Richard A.},
	month = mar,
	year = {2020},
	note = {Publisher: IOP
ADS Bibcode: 2020ApJ...891L...7M},
	pages = {L7},
}

@article{sing_transit_2009,
	title = {Transit spectrophotometry of the exoplanet {HD} 189733b. {I}. {Searching} for water but finding haze with {HST} {NICMOS}},
	volume = {505},
	issn = {0004-6361},
	url = {https://ui.adsabs.harvard.edu/abs/2009A&A...505..891S},
	doi = {10.1051/0004-6361/200912776},
	urldate = {2025-03-11},
	journal = {Astronomy and Astrophysics},
	author = {Sing, D. K. and Désert, J. -M. and Lecavelier Des Etangs, A. and Ballester, G. E. and Vidal-Madjar, A. and Parmentier, V. and Hebrard, G. and Henry, G. W.},
	month = oct,
	year = {2009},
	note = {ADS Bibcode: 2009A\&A...505..891S},
	pages = {891--899},
}

@article{schlawin_possible_2024,
	title = {Possible {Carbon} {Dioxide} above the {Thick} {Aerosols} of {GJ} 1214 b},
	volume = {974},
	issn = {2041-8205},
	url = {https://dx.doi.org/10.3847/2041-8213/ad7fef},
	doi = {10.3847/2041-8213/ad7fef},
	language = {en},
	number = {2},
	urldate = {2025-03-27},
	journal = {The Astrophysical Journal Letters},
	author = {Schlawin, Everett and Ohno, Kazumasa and Bell, Taylor J. and Murphy, Matthew M. and Welbanks, Luis and Beatty, Thomas G. and Greene, Thomas P. and Fortney, Jonathan J. and Parmentier, Vivien and Edelman, Isaac R. and Gill, Samuel and Anderson, David R. and Wheatley, Peter J. and Henry, Gregory W. and Mehta, Nishil and Kreidberg, Laura and Rieke, Marcia J.},
	month = oct,
	year = {2024},
	note = {Publisher: The American Astronomical Society},
	pages = {L33},
}

@misc{rotman_enabling_2025,
	title = {Enabling {Robust} {Exoplanet} {Atmospheric} {Retrievals} with {Gaussian} {Processes}},
	url = {http://arxiv.org/abs/2503.21702},
	doi = {10.48550/arXiv.2503.21702},
	urldate = {2025-04-21},
	publisher = {arXiv},
	author = {Rotman, Yoav and Welbanks, Luis and Line, Michael R. and McGill, Peter and Radica, Michael and Nixon, Matthew C.},
	month = mar,
	year = {2025},
	note = {arXiv:2503.21702 [astro-ph]},
}

@article{charbonneau_detection_2000,
	title = {Detection of {Planetary} {Transits} {Across} a {Sun}-like {Star}},
	volume = {529},
	issn = {0004-637X},
	url = {https://ui.adsabs.harvard.edu/abs/2000ApJ...529L..45C/abstract},
	doi = {10.1086/312457},
	language = {en},
	number = {1},
	urldate = {2025-07-24},
	journal = {The Astrophysical Journal},
	author = {Charbonneau, David and Brown, Timothy M. and Latham, David W. and Mayor, Michel},
	month = jan,
	year = {2000},
	pages = {L45--L48},
}

@article{vidal-madjar_detection_2004,
	title = {Detection of {Oxygen} and {Carbon} in the {Hydrodynamically} {Escaping} {Atmosphere} of the {Extrasolar} {Planet} {HD} 209458b},
	volume = {604},
	issn = {0004-637X},
	url = {https://ui.adsabs.harvard.edu/abs/2004ApJ...604L..69V/abstract},
	doi = {10.1086/383347},
	language = {en},
	number = {1},
	urldate = {2025-07-24},
	journal = {The Astrophysical Journal},
	author = {Vidal-Madjar, A. and Désert, J.-M. and Lecavelier des Etangs, A. and Hébrard, G. and Ballester, G. E. and Ehrenreich, D. and Ferlet, R. and McConnell, J. C. and Mayor, M. and Parkinson, C. D.},
	month = mar,
	year = {2004},
	pages = {L69--L72},
}

@article{southworth_homogeneous_2010,
	title = {Homogeneous studies of transiting extrasolar planets – {III}. {Additional} planets and stellar models},
	volume = {408},
	issn = {0035-8711},
	url = {https://doi.org/10.1111/j.1365-2966.2010.17231.x},
	doi = {10.1111/j.1365-2966.2010.17231.x},
	number = {3},
	urldate = {2025-07-24},
	journal = {Monthly Notices of the Royal Astronomical Society},
	author = {Southworth, John},
	month = nov,
	year = {2010},
	pages = {1689--1713},
}

@article{xue_jwst_2024,
	title = {{JWST} {Transmission} {Spectroscopy} of {HD} 209458b: {A} {Supersolar} {Metallicity}, a {Very} {Low} {C}/{O}, and {No} {Evidence} of {CH4}, {HCN}, or {C2H2}},
	volume = {963},
	issn = {2041-8205},
	shorttitle = {{JWST} {Transmission} {Spectroscopy} of {HD} 209458b},
	url = {https://dx.doi.org/10.3847/2041-8213/ad2682},
	doi = {10.3847/2041-8213/ad2682},
	language = {en},
	number = {1},
	urldate = {2025-07-24},
	journal = {The Astrophysical Journal Letters},
	author = {Xue, Qiao and Bean, Jacob L. and Zhang, Michael and Welbanks, Luis and Lunine, Jonathan and August, Prune},
	month = feb,
	year = {2024},
	note = {Publisher: The American Astronomical Society},
	pages = {L5},
}

@misc{barclay_pandora_2025,
	title = {The {Pandora} {SmallSat}: {A} {Low}-{Cost}, {High} {Impact} {Mission} to {Study} {Exoplanets} and {Their} {Host} {Stars}},
	shorttitle = {The {Pandora} {SmallSat}},
	url = {http://arxiv.org/abs/2502.09730},
	doi = {10.48550/arXiv.2502.09730},
	urldate = {2025-07-24},
	publisher = {arXiv},
	author = {Barclay, Thomas and Quintana, Elisa V. and Colón, Knicole and Hord, Benjamin J. and Mosby, Gregory and Schlieder, Joshua E. and Zellem, Robert T. and Karburn, Jordan and Simms, Lance M. and Heatwole, Peter F. and Hedges, Christina L. and Dotson, Jessie L. and Greene, Thomas P. and Foote, Trevor O. and Lewis, Nikole K. and Rackham, Benjamin V. and Morris, Brett M. and Gilbert, Emily A. and Kostov, Veselin B. and Rowe, Jason F. and Wiser, Lindsay S.},
	month = feb,
	year = {2025},
	note = {arXiv:2502.09730 [astro-ph]},
}

@inproceedings{hedges_open-source_2024,
	title = {Open-source simulation tools for verifying {NASA} {Pandora} {SmallSat}'s scientific performance},
	volume = {13101},
	url = {https://www.spiedigitallibrary.org/conference-proceedings-of-spie/13101/131010F/Open-source-simulation-tools-for-verifying-NASA-Pandora-SmallSats-scientific/10.1117/12.3020326.full},
	doi = {10.1117/12.3020326},
	urldate = {2025-07-24},
	booktitle = {Software and {Cyberinfrastructure} for {Astronomy} {VIII}},
	publisher = {SPIE},
	author = {Hedges, Christina and Holcomb, Rae Jonah and Hord, Benjamin and Barclay, Thomas and Dotson, Jessie and Quintana, Elisa},
	month = jul,
	year = {2024},
	pages = {182--190},
}

@article{giacobbe_five_2021,
	title = {Five carbon- and nitrogen-bearing species in a hot giant planet’s atmosphere},
	volume = {592},
	copyright = {2021 The Author(s), under exclusive licence to Springer Nature Limited},
	issn = {1476-4687},
	url = {https://www.nature.com/articles/s41586-021-03381-x},
	doi = {10.1038/s41586-021-03381-x},
	language = {en},
	number = {7853},
	urldate = {2025-07-24},
	journal = {Nature},
	author = {Giacobbe, Paolo and Brogi, Matteo and Gandhi, Siddharth and Cubillos, Patricio E. and Bonomo, Aldo S. and Sozzetti, Alessandro and Fossati, Luca and Guilluy, Gloria and Carleo, Ilaria and Rainer, Monica and Harutyunyan, Avet and Borsa, Francesco and Pino, Lorenzo and Nascimbeni, Valerio and Benatti, Serena and Biazzo, Katia and Bignamini, Andrea and Chubb, Katy L. and Claudi, Riccardo and Cosentino, Rosario and Covino, Elvira and Damasso, Mario and Desidera, Silvano and Fiorenzano, Aldo F. M. and Ghedina, Adriano and Lanza, Antonino F. and Leto, Giuseppe and Maggio, Antonio and Malavolta, Luca and Maldonado, Jesus and Micela, Giuseppina and Molinari, Emilio and Pagano, Isabella and Pedani, Marco and Piotto, Giampaolo and Poretti, Ennio and Scandariato, Gaetano and Yurchenko, Sergei N. and Fantinel, Daniela and Galli, Alberto and Lodi, Marcello and Sanna, Nicoletta and Tozzi, Andrea},
	month = apr,
	year = {2021},
	note = {Publisher: Nature Publishing Group},
	pages = {205--208},
}

@article{trotta_bayes_2008,
	title = {Bayes in the sky: {Bayesian} inference and model selection in cosmology},
	volume = {49},
	issn = {0010-7514},
	shorttitle = {Bayes in the sky},
	url = {https://doi.org/10.1080/00107510802066753},
	doi = {10.1080/00107510802066753},
	number = {2},
	urldate = {2025-07-24},
	journal = {Contemporary Physics},
	author = {Trotta, Roberto},
	month = mar,
	year = {2008},
	note = {Publisher: Taylor \& Francis
\_eprint: https://doi.org/10.1080/00107510802066753},
	pages = {71--104},
}

@article{triaud_wasp-80b_2015,
	title = {{WASP}-80b has a dayside within the {T}-dwarf range},
	volume = {450},
	issn = {0035-8711},
	url = {https://ui.adsabs.harvard.edu/abs/2015MNRAS.450.2279T/abstract},
	doi = {10.1093/mnras/stv706},
	language = {en},
	number = {3},
	urldate = {2025-07-24},
	journal = {Monthly Notices of the Royal Astronomical Society},
	author = {Triaud, Amaury H. M. J. and Gillon, Michaël and Ehrenreich, David and Herrero, Enrique and Lendl, Monika and Anderson, David R. and Collier Cameron, Andrew and Delrez, Laetitia and Demory, Brice-Olivier and Hellier, Coel and Heng, Kevin and Jehin, Emmanuel and Maxted, Pierre F. L. and Pollacco, Don and Queloz, Didier and Ribas, Ignasi and Smalley, Barry and Smith, Alexis M. S. and Udry, Stéphane},
	month = jul,
	year = {2015},
	pages = {2279--2290},
}

@article{carleo_gaps_2022,
	title = {The {GAPS} {Programme} at {TNG} {XXXIX}. {Multiple} {Molecular} {Species} in the {Atmosphere} of the {Warm} {Giant} {Planet} {WASP}-80 b {Unveiled} at {High} {Resolution} with {GIANO}-{B}},
	volume = {164},
	issn = {0004-6256},
	url = {https://ui.adsabs.harvard.edu/abs/2022AJ....164..101C/abstract},
	doi = {10.3847/1538-3881/ac80bf},
	language = {en},
	number = {3},
	urldate = {2025-07-24},
	journal = {The Astronomical Journal},
	author = {Carleo, Ilaria and Giacobbe, Paolo and Guilluy, Gloria and Cubillos, Patricio E. and Bonomo, Aldo S. and Sozzetti, Alessandro and Brogi, Matteo and Gandhi, Siddharth and Fossati, Luca and Turrini, Diego and Biazzo, Katia and Borsa, Francesco and Lanza, Antonino F. and Malavolta, Luca and Maggio, Antonio and Mancini, Luigi and Micela, Giusi and Pino, Lorenzo and Poretti, Ennio and Rainer, Monica and Scandariato, Gaetano and Schisano, Eugenio and Andreuzzi, Gloria and Bignamini, Andrea and Cosentino, Rosario and Fiorenzano, Aldo and Harutyunyan, Avet and Molinari, Emilio and Pedani, Marco and Redfield, Seth and Stoev, Hristo},
	month = sep,
	year = {2022},
	pages = {101},
}

@misc{wiser_precise_2025,
	title = {A {Precise} {Metallicity} and {Carbon}-to-{Oxygen} {Ratio} for a {Warm} {Giant} {Exoplanet} from its {Panchromatic} {JWST} {Emission} {Spectrum}},
	url = {http://arxiv.org/abs/2506.01800},
	doi = {10.48550/arXiv.2506.01800},
	urldate = {2025-07-24},
	publisher = {arXiv},
	author = {Wiser, Lindsey S. and Bell, Taylor J. and Line, Michael R. and Schlawin, Everett and Beatty, Thomas G. and Welbanks, Luis and Greene, Thomas P. and Parmentier, Vivien and Murphy, Matthew M. and Fortney, Jonathan J. and Arnold, Kenny and Mehta, Nishil and Ohno, Kazumasa and Mukherjee, Sagnick},
	month = jun,
	year = {2025},
	note = {arXiv:2506.01800 [astro-ph]},
}

@article{jacobs_probing_2023,
	title = {Probing {Reflection} from {Aerosols} with the {Near}-infrared {Dayside} {Spectrum} of {WASP}-80b},
	volume = {956},
	issn = {0004-637X},
	url = {https://ui.adsabs.harvard.edu/abs/2023ApJ...956L..43J/abstract},
	doi = {10.3847/2041-8213/acfee9},
	language = {en},
	number = {2},
	urldate = {2025-07-24},
	journal = {The Astrophysical Journal},
	author = {Jacobs, Bob and Désert, Jean-Michel and Gao, Peter and Morley, Caroline V. and Arcangeli, Jacob and Barat, Saugata and Marley, Mark S. and Moses, Julianne I. and Fortney, Jonathan J. and Bean, Jacob L. and Stevenson, Kevin B. and Panwar, Vatsal},
	month = oct,
	year = {2023},
	pages = {L43},
}

@article{fukui_multi-band_2014,
	title = {Multi-band, {Multi}-epoch {Observations} of the {Transiting} {Warm} {Jupiter} {WASP}-80b},
	volume = {790},
	issn = {0004-637X},
	url = {https://ui.adsabs.harvard.edu/abs/2014ApJ...790..108F/abstract},
	doi = {10.1088/0004-637X/790/2/108},
	language = {en},
	number = {2},
	urldate = {2025-07-24},
	journal = {The Astrophysical Journal},
	author = {Fukui, Akihiko and Kawashima, Yui and Ikoma, Masahiro and Narita, Norio and Onitsuka, Masahiro and Ita, Yoshifusa and Onozato, Hiroki and Nishiyama, Shogo and Baba, Haruka and Ryu, Tsuguru and Hirano, Teruyuki and Hori, Yasunori and Kurosaki, Kenji and Kawauchi, Kiyoe and Takahashi, Yasuhiro H. and Nagayama, Takahiro and Tamura, Motohide and Kawai, Nobuyuki and Kuroda, Daisuke and Nagayama, Shogo and Ohta, Kouji and Shimizu, Yasuhiro and Yanagisawa, Kenshi and Yoshida, Michitoshi and Izumiura, Hideyuki},
	month = aug,
	year = {2014},
	pages = {108},
}

@article{mancini_physical_2014,
	title = {Physical properties and transmission spectrum of the {WASP}-80 planetary system from multi-colour photometry},
	volume = {562},
	issn = {0004-6361},
	url = {https://ui.adsabs.harvard.edu/abs/2014A&A...562A.126M/abstract},
	doi = {10.1051/0004-6361/201323265},
	language = {en},
	urldate = {2025-07-24},
	journal = {Astronomy and Astrophysics},
	author = {Mancini, L. and Southworth, J. and Ciceri, S. and Dominik, M. and Henning, Th and Jørgensen, U. G. and Lanza, A. F. and Rabus, M. and Snodgrass, C. and Vilela, C. and Alsubai, K. A. and Bozza, V. and Bramich, D. M. and Calchi Novati, S. and D'Ago, G. and Figuera Jaimes, R. and Galianni, P. and Gu, S.-H. and Harpsøe, K. and Hinse, T. and Hundertmark, M. and Juncher, D. and Kains, N. and Korhonen, H. and Popovas, A. and Rahvar, S. and Skottfelt, J. and Street, R. and Surdej, J. and Tsapras, Y. and Wang, X.-B. and Wertz, O.},
	month = feb,
	year = {2014},
	pages = {A126},
}

@article{triaud_wasp-80b_2013,
	title = {{WASP}-80b: a gas giant transiting a cool dwarf},
	volume = {551},
	issn = {0004-6361},
	shorttitle = {{WASP}-80b},
	url = {https://ui.adsabs.harvard.edu/abs/2013A%26A...551A..80T/abstract},
	doi = {10.1051/0004-6361/201220900},
	language = {en},
	urldate = {2025-07-24},
	journal = {Astronomy \&amp; Astrophysics, Volume 551, id.A80, {\textless}NUMPAGES{\textgreater}5{\textless}/NUMPAGES{\textgreater} pp.},
	author = {Triaud, A. H. M. J. and Anderson, D. R. and Collier Cameron, A. and Doyle, A. P. and Fumel, A. and Gillon, M. and Hellier, C. and Jehin, E. and Lendl, M. and Lovis, C. and Maxted, P. F. L. and Pepe, F. and Pollacco, D. and Queloz, D. and Ségransan, D. and Smalley, B. and Smith, A. M. S. and Udry, S. and West, R. G. and Wheatley, P. J.},
	month = mar,
	year = {2013},
	pages = {A80},
}

@article{wong_hubble_2022,
	title = {The {Hubble} {PanCET} {Program}: {A} {Featureless} {Transmission} {Spectrum} for {WASP}-29b and {Evidence} of {Enhanced} {Atmospheric} {Metallicity} on {WASP}-80b},
	volume = {164},
	issn = {0004-6256},
	shorttitle = {The {Hubble} {PanCET} {Program}},
	url = {https://ui.adsabs.harvard.edu/abs/2022AJ....164...30W/abstract},
	doi = {10.3847/1538-3881/ac7234},
	language = {en},
	number = {1},
	urldate = {2025-07-24},
	journal = {The Astronomical Journal},
	author = {Wong, Ian and Chachan, Yayaati and Knutson, Heather A. and Henry, Gregory W. and Adams, Danica and Kataria, Tiffany and Benneke, Björn and Gao, Peter and Deming, Drake and López-Morales, Mercedes and Sing, David K. and Alam, Munazza K. and Ballester, Gilda E. and Barstow, Joanna K. and Buchhave, Lars A. and dos Santos, Leonardo A. and Fu, Guangwei and García Muñoz, Antonio and MacDonald, Ryan J. and Mikal-Evans, Thomas and Sanz-Forcada, Jorge and Wakeford, Hannah R.},
	month = jul,
	year = {2022},
	pages = {30},
}

@article{redfield_sodium_2008,
	title = {Sodium {Absorption} from the {Exoplanetary} {Atmosphere} of {HD} 189733b {Detected} in the {Optical} {Transmission} {Spectrum}},
	volume = {673},
	issn = {0004-637X},
	url = {https://ui.adsabs.harvard.edu/abs/2008ApJ...673L..87R/abstract},
	doi = {10.1086/527475},
	language = {en},
	number = {1},
	urldate = {2025-07-25},
	journal = {The Astrophysical Journal},
	author = {Redfield, Seth and Endl, Michael and Cochran, William D. and Koesterke, Lars},
	month = jan,
	year = {2008},
	pages = {L87},
}

@misc{hu_water-rich_2025,
	title = {A water-rich interior in the temperate sub-{Neptune} {K2}-18 b revealed by {JWST}},
	url = {http://arxiv.org/abs/2507.12622},
	doi = {10.48550/arXiv.2507.12622},
	urldate = {2025-07-25},
	publisher = {arXiv},
	author = {Hu, Renyu and Bello-Arufe, Aaron and Tokadjian, Armen and Yang, Jeehyun and Damiano, Mario and Roy, Pierre-Alexis and Coulombe, Louis-Philippe and Madhusudhan, Nikku and Constantinou, Savvas and Benneke, Björn},
	month = jul,
	year = {2025},
	note = {arXiv:2507.12622 [astro-ph]},
}

@article{charbonneau_broadband_2008,
	title = {The {Broadband} {Infrared} {Emission} {Spectrum} of the {Exoplanet} {HD} 189733b},
	volume = {686},
	issn = {0004-637X},
	url = {https://iopscience.iop.org/article/10.1086/591635/meta},
	doi = {10.1086/591635},
	language = {en},
	number = {2},
	urldate = {2025-07-24},
	journal = {The Astrophysical Journal},
	author = {Charbonneau, David and Knutson, Heather A. and Barman, Travis and Allen, Lori E. and Mayor, Michel and Megeath, S. Thomas and Queloz, Didier and Udry, Stéphane},
	month = oct,
	year = {2008},
	note = {Publisher: IOP Publishing},
	pages = {1341},
}

@article{beatty_sulfur_2024,
	title = {Sulfur {Dioxide} and {Other} {Molecular} {Species} in the {Atmosphere} of the {Sub}-{Neptune} {GJ} 3470 b},
	volume = {970},
	issn = {0004-637X},
	url = {https://ui.adsabs.harvard.edu/abs/2024ApJ...970L..10B},
	doi = {10.3847/2041-8213/ad55e9},
	urldate = {2025-07-29},
	journal = {The Astrophysical Journal},
	author = {Beatty, Thomas G. and Welbanks, Luis and Schlawin, Everett and Bell, Taylor J. and Line, Michael R. and Murphy, Matthew and Edelman, Isaac and Greene, Thomas P. and Fortney, Jonathan J. and Henry, Gregory W. and Mukherjee, Sagnick and Ohno, Kazumasa and Parmentier, Vivien and Rauscher, Emily and Wiser, Lindsey S. and Arnold, Kenneth E.},
	month = jul,
	year = {2024},
	note = {Publisher: IOP
ADS Bibcode: 2024ApJ...970L..10B},
	pages = {L10},
}

@misc{canas_gems_2025,
	title = {{GEMS} {JWST}: {Transmission} spectroscopy of {TOI}-5205b reveals significant stellar contamination and a metal-poor atmosphere},
	shorttitle = {{GEMS} {JWST}},
	url = {https://ui.adsabs.harvard.edu/abs/2025arXiv250206966C},
	doi = {10.48550/arXiv.2502.06966},
	urldate = {2025-07-29},
	publisher = {arXiv},
	author = {Cañas, Caleb I. and Lustig-Yaeger, Jacob and Tsai, Shang-Min and Müller, Simon and Helled, Ravit and Louie, Dana R. and Guzmán Caloca, Giannina and Kanodia, Shubham and Gao, Peter and Libby-Roberts, Jessica and Hardegree-Ullman, Kevin K. and Colón, Knicole D. and Czekala, Ian and Delamer, Megan and Han, Te and Lin, Andrea S. J. and Mahadevan, Suvrath and May, Erin M. and Ninan, Joe P. and Piette, Anjali A. A. and Stefánsson, Guðmundur and Stevenson, Kevin B. and Teske, Johanna and Wallack, Nicole L.},
	month = feb,
	year = {2025},
	note = {ADS Bibcode: 2025arXiv250206966C},
}

@article{yurchenko_exomol_2014,
	title = {{ExoMol} line lists - {IV}. {The} rotation-vibration spectrum of methane up to 1500 {K}},
	volume = {440},
	issn = {0035-8711},
	url = {https://ui.adsabs.harvard.edu/abs/2014MNRAS.440.1649Y},
	doi = {10.1093/mnras/stu326},
	urldate = {2025-07-29},
	journal = {Monthly Notices of the Royal Astronomical Society},
	author = {Yurchenko, Sergei N. and Tennyson, Jonathan},
	month = may,
	year = {2014},
	note = {Publisher: OUP
ADS Bibcode: 2014MNRAS.440.1649Y},
	pages = {1649--1661},
}

@article{yurchenko_variationally_2011,
	title = {A variationally computed line list for hot {NH3}},
	volume = {413},
	issn = {0035-8711},
	url = {https://ui.adsabs.harvard.edu/abs/2011MNRAS.413.1828Y},
	doi = {10.1111/j.1365-2966.2011.18261.x},
	urldate = {2025-07-29},
	journal = {Monthly Notices of the Royal Astronomical Society},
	author = {Yurchenko, S. N. and Barber, R. J. and Tennyson, J.},
	month = may,
	year = {2011},
	note = {Publisher: OUP
ADS Bibcode: 2011MNRAS.413.1828Y},
	pages = {1828--1834},
}

@article{allard_new_2019,
	title = {New study of the line profiles of sodium perturbed by {H2}},
	volume = {628},
	issn = {0004-6361},
	url = {https://ui.adsabs.harvard.edu/abs/2019A&A...628A.120A},
	doi = {10.1051/0004-6361/201935593},
	urldate = {2025-07-29},
	journal = {Astronomy and Astrophysics},
	author = {Allard, N. F. and Spiegelman, F. and Leininger, T. and Molliere, P.},
	month = aug,
	year = {2019},
	note = {Publisher: EDP
ADS Bibcode: 2019A\&A...628A.120A},
	pages = {A120},
}

@article{richard_new_2012,
	title = {New section of the {HITRAN} database: {Collision}-induced absorption ({CIA})},
	volume = {113},
	issn = {0022-4073},
	shorttitle = {New section of the {HITRAN} database},
	url = {https://ui.adsabs.harvard.edu/abs/2012JQSRT.113.1276R},
	doi = {10.1016/j.jqsrt.2011.11.004},
	urldate = {2025-07-29},
	journal = {Journal of Quantitative Spectroscopy and Radiative Transfer},
	author = {Richard, C. and Gordon, I. E. and Rothman, L. S. and Abel, M. and Frommhold, L. and Gustafsson, M. and Hartmann, J. -M. and Hermans, C. and Lafferty, W. J. and Orton, G. S. and Smith, K. M. and Tran, H.},
	month = jul,
	year = {2012},
	note = {Publisher: Elsevier
ADS Bibcode: 2012JQSRT.113.1276R},
	pages = {1276--1285},
}

@article{the_astropy_collaboration_astropy_2022,
	title = {The {Astropy} {Project}: {Sustaining} and {Growing} a {Community}-oriented {Open}-source {Project} and the {Latest} {Major} {Release} (v5.0) of the {Core} {Package}*},
	volume = {935},
	issn = {0004-637X},
	shorttitle = {The {Astropy} {Project}},
	url = {https://dx.doi.org/10.3847/1538-4357/ac7c74},
	doi = {10.3847/1538-4357/ac7c74},
	language = {en},
	number = {2},
	urldate = {2025-07-29},
	journal = {The Astrophysical Journal},
	author = {{The Astropy Collaboration} and Price-Whelan, Adrian M. and Lim, Pey Lian and Earl, Nicholas and Starkman, Nathaniel and Bradley, Larry and Shupe, David L. and Patil, Aarya A. and Corrales, Lia and Brasseur, C. E. and Nöthe, Maximilian and Donath, Axel and Tollerud, Erik and Morris, Brett M. and Ginsburg, Adam and Vaher, Eero and Weaver, Benjamin A. and Tocknell, James and Jamieson, William and van Kerkwijk, Marten H. and Robitaille, Thomas P. and Merry, Bruce and Bachetti, Matteo and Günther, H. Moritz and Authors, Paper and Aldcroft, Thomas L. and Alvarado-Montes, Jaime A. and Archibald, Anne M. and Bódi, Attila and Bapat, Shreyas and Barentsen, Geert and Bazán, Juanjo and Biswas, Manish and Boquien, Médéric and Burke, D. J. and Cara, Daria and Cara, Mihai and Conroy, Kyle E and Conseil, Simon and Craig, Matthew W. and Cross, Robert M. and Cruz, Kelle L. and D’Eugenio, Francesco and Dencheva, Nadia and Devillepoix, Hadrien A. R. and Dietrich, Jörg P. and Eigenbrot, Arthur Davis and Erben, Thomas and Ferreira, Leonardo and Foreman-Mackey, Daniel and Fox, Ryan and Freij, Nabil and Garg, Suyog and Geda, Robel and Glattly, Lauren and Gondhalekar, Yash and Gordon, Karl D. and Grant, David and Greenfield, Perry and Groener, Austen M. and Guest, Steve and Gurovich, Sebastian and Handberg, Rasmus and Hart, Akeem and Hatfield-Dodds, Zac and Homeier, Derek and Hosseinzadeh, Griffin and Jenness, Tim and Jones, Craig K. and Joseph, Prajwel and Kalmbach, J. Bryce and Karamehmetoglu, Emir and Kałuszyński, Mikołaj and Kelley, Michael S. P. and Kern, Nicholas and Kerzendorf, Wolfgang E. and Koch, Eric W. and Kulumani, Shankar and Lee, Antony and Ly, Chun and Ma, Zhiyuan and MacBride, Conor and Maljaars, Jakob M. and Muna, Demitri and Murphy, N. A. and Norman, Henrik and O’Steen, Richard and Oman, Kyle A. and Pacifici, Camilla and Pascual, Sergio and Pascual-Granado, J. and Patil, Rohit R. and Perren, Gabriel I and Pickering, Timothy E. and Rastogi, Tanuj and Roulston, Benjamin R. and Ryan, Daniel F and Rykoff, Eli S. and Sabater, Jose and Sakurikar, Parikshit and Salgado, Jesús and Sanghi, Aniket and Saunders, Nicholas and Savchenko, Volodymyr and Schwardt, Ludwig and Seifert-Eckert, Michael and Shih, Albert Y. and Jain, Anany Shrey and Shukla, Gyanendra and Sick, Jonathan and Simpson, Chris and Singanamalla, Sudheesh and Singer, Leo P. and Singhal, Jaladh and Sinha, Manodeep and Sipőcz, Brigitta M. and Spitler, Lee R. and Stansby, David and Streicher, Ole and Šumak, Jani and Swinbank, John D. and Taranu, Dan S. and Tewary, Nikita and Tremblay, Grant R. and Val-Borro, Miguel de and Van Kooten, Samuel J. and Vasović, Zlatan and Verma, Shresth and de Miranda Cardoso, José Vinícius and Williams, Peter K. G. and Wilson, Tom J. and Winkel, Benjamin and Wood-Vasey, W. M. and Xue, Rui and Yoachim, Peter and Zhang, Chen and Zonca, Andrea and Contributors, Astropy Project},
	month = aug,
	year = {2022},
	note = {Publisher: The American Astronomical Society},
	pages = {167},
}

@article{harris_array_2020,
	title = {Array programming with {NumPy}},
	volume = {585},
	copyright = {2020 The Author(s)},
	issn = {1476-4687},
	url = {https://www.nature.com/articles/s41586-020-2649-2},
	doi = {10.1038/s41586-020-2649-2},
	language = {en},
	number = {7825},
	urldate = {2025-07-29},
	journal = {Nature},
	author = {Harris, Charles R. and Millman, K. Jarrod and van der Walt, Stéfan J. and Gommers, Ralf and Virtanen, Pauli and Cournapeau, David and Wieser, Eric and Taylor, Julian and Berg, Sebastian and Smith, Nathaniel J. and Kern, Robert and Picus, Matti and Hoyer, Stephan and van Kerkwijk, Marten H. and Brett, Matthew and Haldane, Allan and del Río, Jaime Fernández and Wiebe, Mark and Peterson, Pearu and Gérard-Marchant, Pierre and Sheppard, Kevin and Reddy, Tyler and Weckesser, Warren and Abbasi, Hameer and Gohlke, Christoph and Oliphant, Travis E.},
	month = sep,
	year = {2020},
	note = {Publisher: Nature Publishing Group},
	pages = {357--362},
}

@ARTICLE{rathcke_trappist1bc_2025,
       author = {{Rathcke}, Alexander D. and {Buchhave}, Lars A. and {Wit}, Julien de and {Rackham}, Benjamin V. and {August}, Prune C. and {Diamond-Lowe}, Hannah and {Mendon{\c{C}}a}, Jo{\~a}o M. and {Bello-Arufe}, Aaron and {L{\'o}pez-Morales}, Mercedes and {Kitzmann}, Daniel and {Heng}, Kevin},
        title = "{Stellar Contamination Correction Using Back-to-back Transits of TRAPPIST-1 b and c}",
      journal = {\apjl},
         year = 2025,
        month = jan,
       volume = {979},
       number = {1},
          eid = {L19},
        pages = {L19},
          doi = {10.3847/2041-8213/ada5c7},
archivePrefix = {arXiv},
       eprint = {2412.16541},
 primaryClass = {astro-ph.EP},
       adsurl = {https://ui.adsabs.harvard.edu/abs/2025ApJ...979L..19R}
}

@ARTICLE{rackham_gj1214b_2017,
       author = {{Rackham}, Benjamin and {Espinoza}, N{\'e}stor and {Apai}, D{\'a}niel and {L{\'o}pez-Morales}, Mercedes and {Jord{\'a}n}, Andr{\'e}s and {Osip}, David J. and {Lewis}, Nikole K. and {Rodler}, Florian and {Fraine}, Jonathan D. and {Morley}, Caroline V. and {Fortney}, Jonathan J.},
        title = "{ACCESS I: An Optical Transmission Spectrum of GJ 1214b Reveals a Heterogeneous Stellar Photosphere}",
      journal = {\apj},
         year = 2017,
        month = jan,
       volume = {834},
       number = {2},
          eid = {151},
        pages = {151},
          doi = {10.3847/1538-4357/aa4f6c},
archivePrefix = {arXiv},
       eprint = {1612.00228},
 primaryClass = {astro-ph.EP},
       adsurl = {https://ui.adsabs.harvard.edu/abs/2017ApJ...834..151R}
}
\bibliographystyle{aasjournalv7}

\appendix
\renewcommand{\thefigure}{B\arabic{figure}}
\setcounter{figure}{0}

\section{Model Parameters and Priors}\label{appendix:priors_and_model}

Table \ref{table:params_priors} lists the atmospheric model parameters used to generate synthetic observations and the priors adopted in the retrievals \edit{for the Pandora-only observations} (see Section \ref{sec:pandora}).

\begin{table*}[h!]
\centering
\small
\begin{tabular}{lccccc c}
\hline
Parameter & HD 209458b$^a$ & HD 189733b$^b$ & WASP-80b$^c$ & HAT-P-18b$^d$ & K2-18b$^e$ & Prior \\
\hline
$T_{\rm iso}$ [K] & 1459 & 1220 & 825 & 852 & 255 & $\mathcal{U}[0, 1800]$ \\
$\log$(H$_2$O) & -3.0 & -3.0 & -2.5 & -4.3 & -3.0 & $\mathcal{U}[-12, -1]$ \\
$\log$(CH$_4$) & -8.0 & -8.0 & -3.5 & -6.0 & -2.0 & $\mathcal{U}[-12, -1]$ \\
$\log$(CO)     & -4.0 & -3.5 & -5.0 & -4.3 & -3.5 & $\mathcal{U}[-12, -1]$ \\
$\log$(CO$_2$) & -6.0 & -5.5 & -6.0 & -5.0 & -4.0 & $\mathcal{U}[-12, -1]$ \\
$\log$(NH$_3$) & -7.0 & -6.5 & -5.5 & -7.0 & -5.5 & $\mathcal{U}[-12, -1]$ \\
$\log$(H$_2$S) & -4.5 & -4.0 & -4.0 & -5.5 & -- & $\mathcal{U}[-12, -1]$ \\
$\log$(Na)     & -5.5 & -5.5 & --   & --   & --  & $\mathcal{U}[-12, -1]$ \\
$\log$(K)      & -7.0 & -7.0 & --   & --   & --  & $\mathcal{U}[-12, -1]$ \\
$\phi_{\rm cloud}$ & 0.5 & 0.5 & 0.4 & 0.9 & 0.6 & $\mathcal{U}[0, 1]$ \\
$\log P_{\rm cloud}$ [bar] & 1.7 & 0.5 & -7.0 & -1.5 & -0.5 & $\mathcal{U}[-8, 2]$ \\
$\log(a)$ & 0.0 & 3.0 & 0.0 & 0.0 & 8.0 & $\mathcal{U}[-4, 10]$ \\
$\gamma$ & -4.0 & -8.0 & -4.0 & -4.0 & -11.0 & $\mathcal{U}[-20, 2]$ \\
\hline
\end{tabular}
\caption{Summary of forward-model parameters (columns 2–6) and retrieval priors (last column) \edit{used for the models in Section \ref{sec:pandora}}. All $\log(X_i)$ values are molecular abundances given as log$_{10}$ volume mixing ratios. The variables for the cloud model described in Section \ref{sec:fwd_model} are cloud coverage fraction ($\phi_{\rm cloud}$), the cloud-top pressure ($\log P_{\rm cloud}$ in bar), Rayleigh scattering enhancement factor ($\log(a)$), and scattering slope ($\gamma$). ``--'' indicates a species not included in the model. $\mathcal{U}$ denotes a uniform prior. Atmospheric abundances roughly follow those from existing studies of these planets with JWST data as a guideline: \textbf{a)} \citet{xue_jwst_2024}, \textbf{b)} \citet{fu_hydrogen_2024}, \textbf{c)} \citet{bell_methane_2023, wiser_precise_2025}, \textbf{d)} \citet{fournier-tondreau_near-infrared_2024}, \textbf{e)} \citet{madhusudhan_carbon-bearing_2023}.}
\label{table:params_priors}
\end{table*}

\edit{Table \ref{table:params_priors2} lists the atmospheric model parameters used to generate synthetic observations and the priors adopted in the retrievals for the combined Pandora and JWST observations (see Section \ref{sec:jwst}). These parameters are determined via equilibrium chemistry expectations at the equilibrium temperature of the planet \citep[e.g.,][]{moses2013}}.

\begin{table*}[h!]
\centering
\small
\begin{tabular}{lccccc c}
\hline
Parameter & HD 209458b & HD 189733b & WASP-80b & HAT-P-18b & K2-18b & Prior \\
\hline
$T_{\rm iso}$ [K] & 1459 & 1220 & 825 & 852 & 255 & $\mathcal{U}[0, 1800]$ \\
$\log$(H$_2$O) & -2.3 & -2.3 & -2.0 & -2.0 & -2.0 & $\mathcal{U}[-12, -1]$ \\
$\log$(CH$_4$) & -7.0 & -5.5 & -2.5 & -2.5 & -2.5 & $\mathcal{U}[-12, -1]$ \\
$\log$(CO)     & -2.5 & -2.5 & -4.0 & -4.0 & -- & $\mathcal{U}[-12, -1]$ \\
$\log$(CO$_2$) & -5.0 & -5.0 & -5.5 & -5.5 & -- & $\mathcal{U}[-12, -1]$ \\
\hline
\end{tabular}
\caption{\edit{Summary of forward-model parameters and retrieval priors used for the models in Section \ref{sec:jwst}. All $\log(X_i)$ values are molecular abundances given as log$_{10}$ volume mixing ratios. Atmospheric abundances are determined following equilibrium chemistry expectations \citep[e.g.,][]{moses2013}.}}
\label{table:params_priors2}
\end{table*}

\section{CH$_4$ and NH$_3$ Constraints from Pandora}\label{appendix:pandora_posteriors}

Here, we show the posterior distributions of CH$_4$ and NH$_3$ obtained with Pandora observations, for both the comprehensive and simple models. Both molecules can be accurately constrained with upper limits on their abundances, with the exception of a full abundance constraint of CH$_4$ on K2-18~b. NH$_3$ is additionally mostly constrained for HAT-P-18~b in the case of the simple model, although a tail extends towards the lower end of the prior. The comprehensive model, however, shows an unconstrained posterior that ends in an upper limit, though the upper limit is accurate. 

\begin{figure}[h]
    \centering
    \begin{subfigure}[t]{0.48\linewidth}
        \centering
        \includegraphics[width=\linewidth]{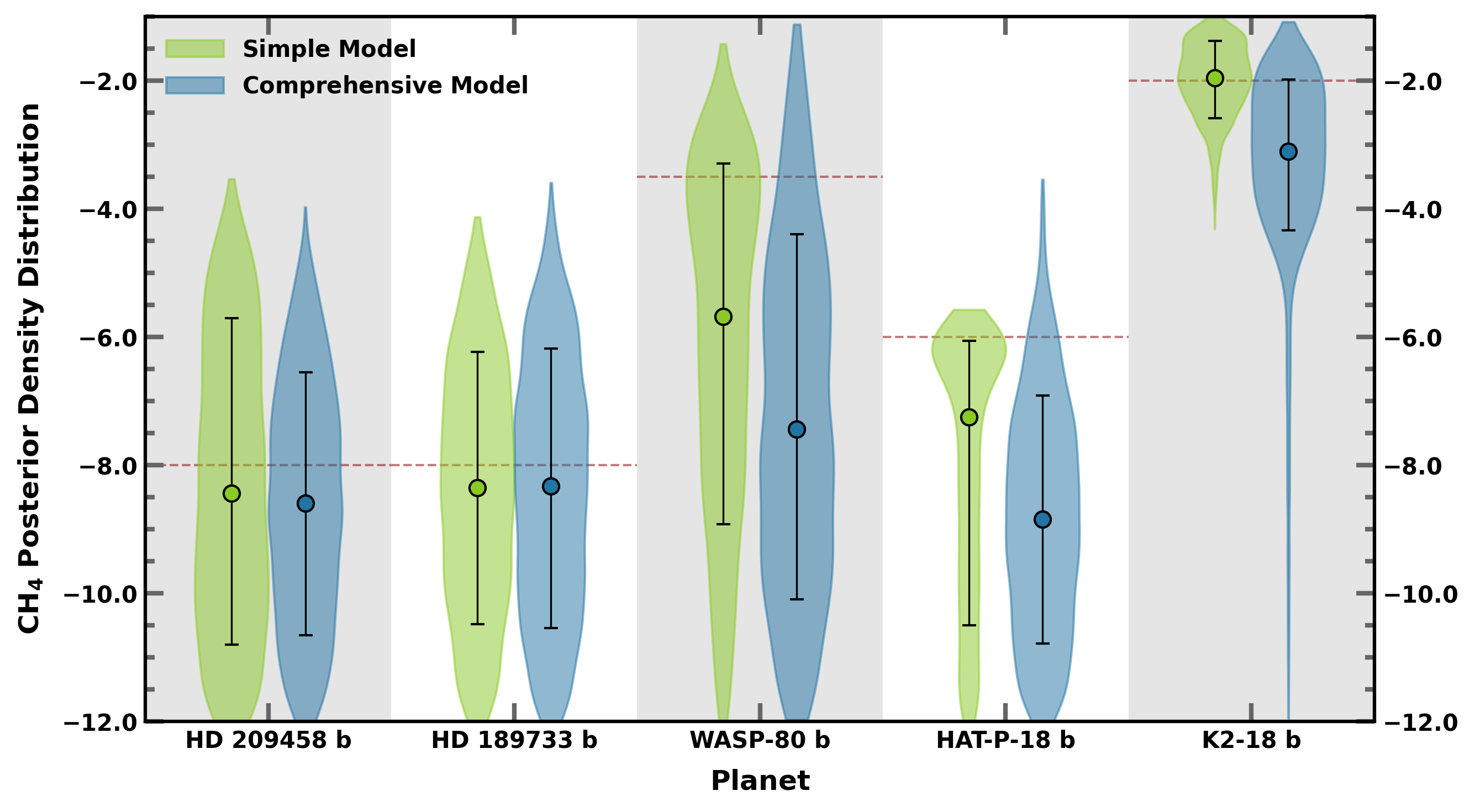}
    \end{subfigure}
    \hfill
    \begin{subfigure}[t]{0.48\linewidth}
        \centering
        \includegraphics[width=\linewidth]{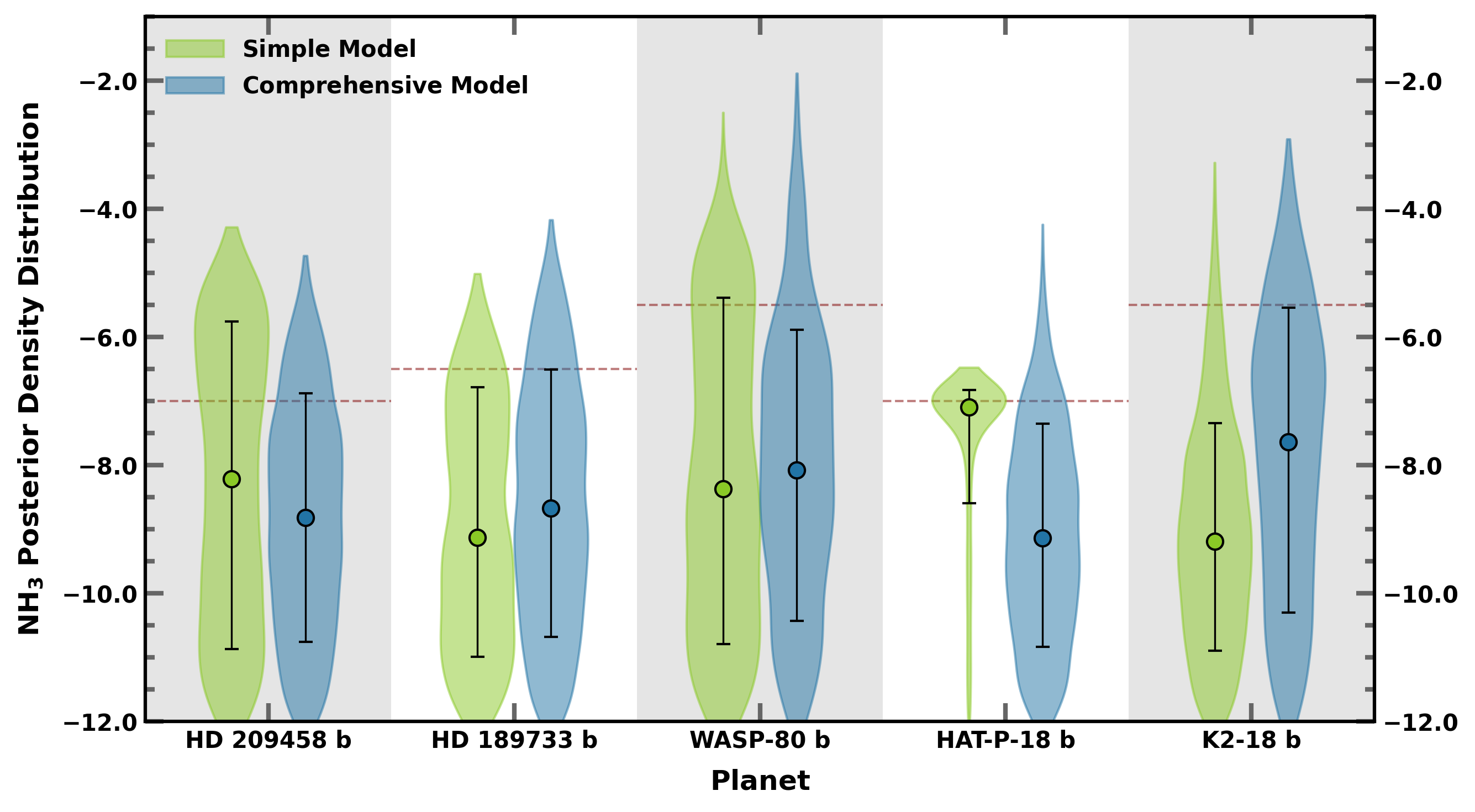}
    \end{subfigure}
    \caption{Same as Figure~\ref{fig:pandora_posteriors}, but showing the CH$_4$ (left) and NH$_3$ (right) abundance constraints from the simple (green) and comprehensive (blue) models.}
    \label{fig:pandora_posteriors_CH4_NH3}
\end{figure}




    

\section{\edit{H$_2$O, CH$_4$, CO, and CO$_2$} Constraints from Pandora and JWST}\label{appendix:jwst_posteriors}
\renewcommand{\thefigure}{C\arabic{figure}}
\setcounter{figure}{0}

Here, we show the posterior distributions of \edit{H$_2$O and CH$_4$} in Figure \ref{fig:jwst_posteriors_CH4_NH3} and for CO and CO$_2$ in Figure \ref{fig:jwst_posteriors_CO_CO2} obtained with Pandora-only, JWST-only, and combined observations, using the comprehensive model for both the data simulation and retrieval. \edit{For high CH$_4$ abundances, Pandora may allow for reliable inferences of the abundance, particularly for high-SNR targets (e.g., HAT-P-18~b and K2-18~b).} Notably, despite not containing any strong CO$_2$ absorption features in the NIRDA bandpass, the inclusion of Pandora data leads to more accurate constraints on CO$_2$ abundance than JWST alone (\edit{particularly for HD~209458~b and HD~189733~b)}. Additionally, marginal improvement in the CO constraints \edit{for} the same planets can be seen despite the lack of CO features in the NIRDA bandpass.


\begin{figure*}[h]
    \centering
    \begin{subfigure}[t]{0.49\linewidth}
        \centering
        \includegraphics[width=\linewidth]{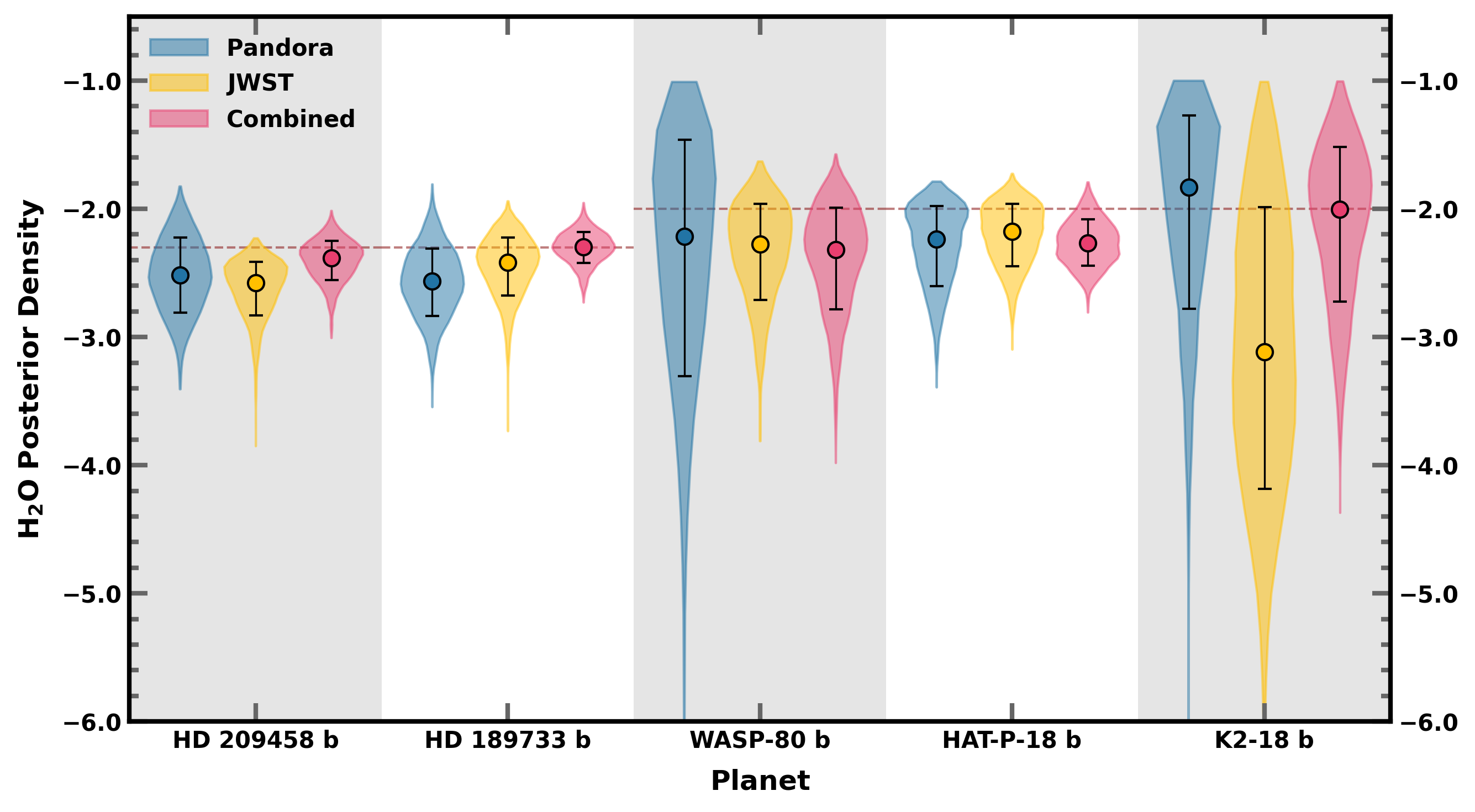}
    \end{subfigure}
    \hfill
    \begin{subfigure}[t]{0.49\linewidth}
        \centering
        \includegraphics[width=\linewidth]{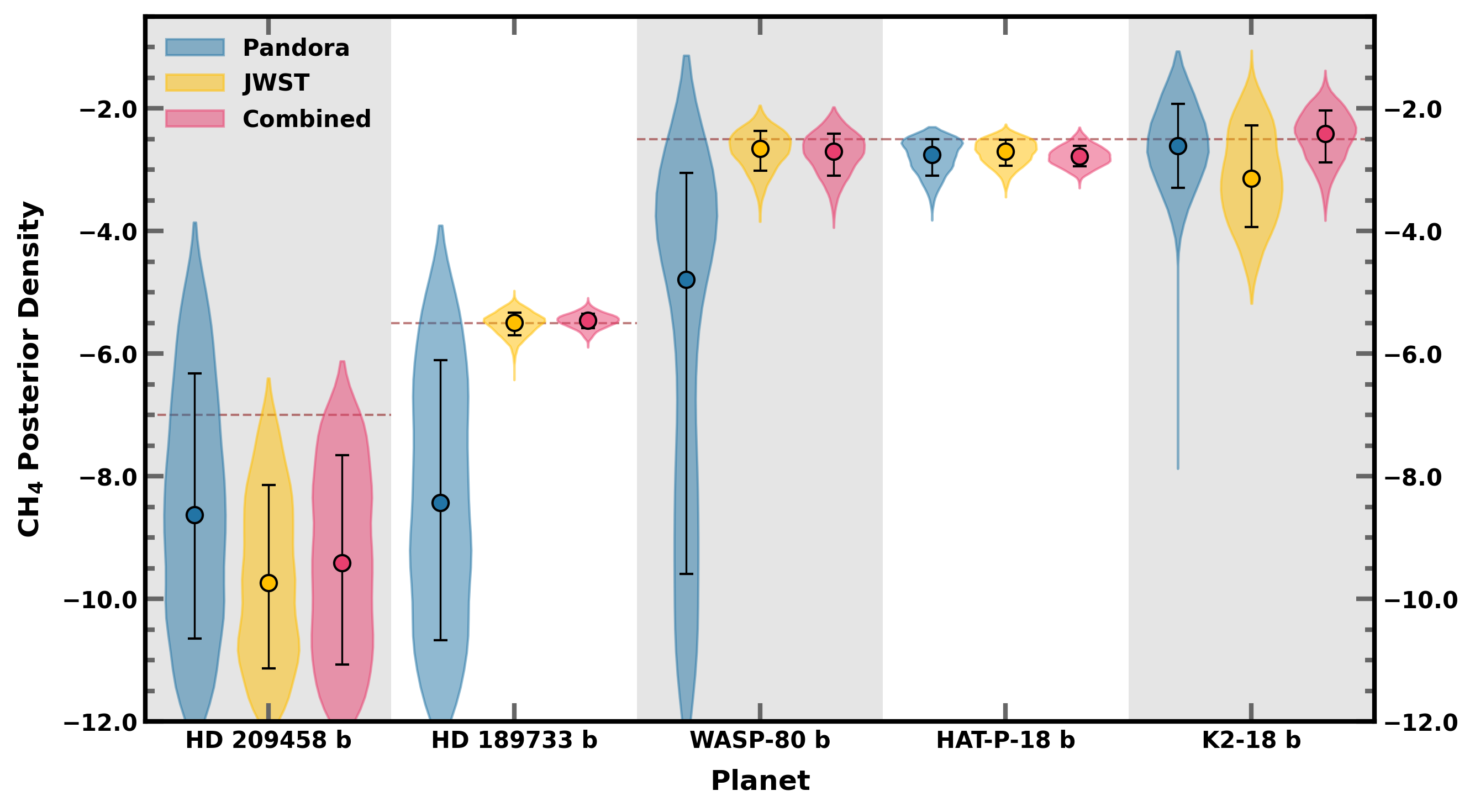}
    \end{subfigure}
    \caption{Same as Figure~\ref{fig:jwst_posteriors}, but showing the \edit{H$_2$O (left) and CH$_4$ (right)} constraints from Pandora (blue), JWST (yellow), and combined (pink) data.}
    \label{fig:jwst_posteriors_CH4_NH3}
\end{figure*}

\begin{figure*}[h]
    \centering
    \begin{subfigure}[t]{0.48\linewidth}
        \centering
        \includegraphics[width=\linewidth]{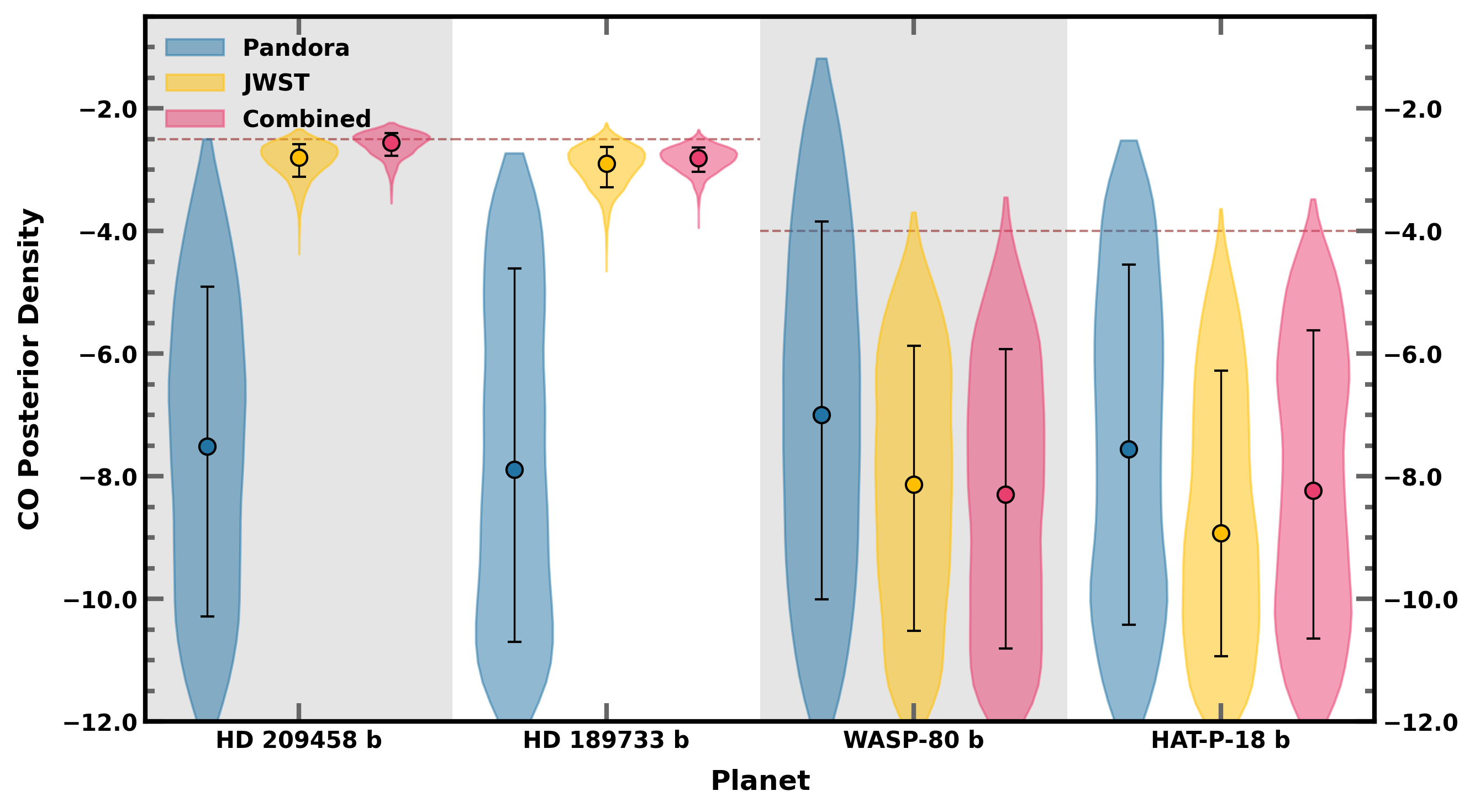}
    \end{subfigure}
    \hfill
    \begin{subfigure}[t]{0.48\linewidth}
        \centering
        \includegraphics[width=\linewidth]{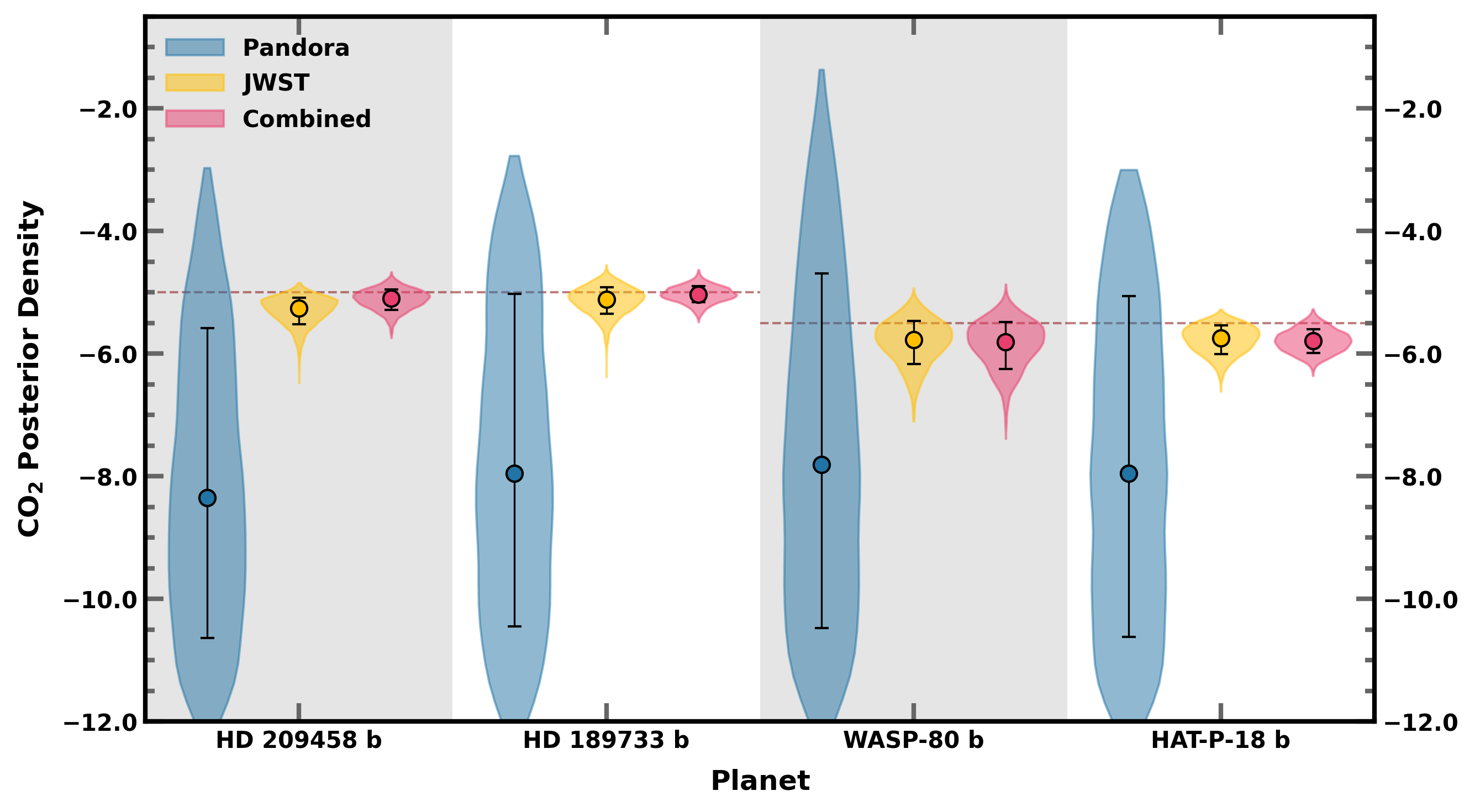}
    \end{subfigure}
    \caption{Same as Figure~\ref{fig:jwst_posteriors}, but showing the CO (left) and CO$_2$ (right) abundance constraints from Pandora (blue), JWST (yellow), and combined (pink) data.}
    \label{fig:jwst_posteriors_CO_CO2}
\end{figure*}




\end{document}